\documentclass{bmcart}

\usepackage[utf8]{inputenc} 
\usepackage{placeins} 
\usepackage{multirow} 
\usepackage{enumitem} 

\usepackage[nonumberlist]{glossaries}

\makeglossaries
\newacronym{3d}{3D}{three-dimensional}
\newacronym{alida}{Alida}{Advanced Library for Integrated Development of data analysis Applications}
\newacronym{api}{API}{Application Programming Interface}
\newacronym{awt}{AWT}{Abstract Windowing Toolkit}
\newacronym{bar}{BAR}{Broadly Applicable Routines}
\newacronym{beakerx}{BeakerX}{Beaker Extensions for Jupyter}
\newacronym{bisque}{BisQue}{Bio-Image Semantic Query User Environment}
\newacronym{bmc}{BMC}{BioMed Central}
\newacronym{bsd}{BSD}{Berkeley Software Distribution}
\newacronym{ci}{CI}{continuous integration}
\newacronym{cpu}{CPU}{Central Processing Unit}
\newacronym{cuda}{CUDA}{Compute Unified Device Architecture}
\newacronym{di}{DI}{dependency injection}
\newacronym{dna}{DNA}{deoxyribonucleic acid}
\newacronym{dry}{DRY}{Don't Repeat Yourself}
\newacronym{epfl}{EPFL}{École Polytechnique Fédérale de Lausanne}
\newacronym{fiji}{Fiji}{Fiji Is Just ImageJ}
\newacronym{gimp}{GIMP}{GNU Image Manipulation Program}
\newacronym{gnu}{GNU}{GNU's Not Unix}
\newacronym{gpu}{GPU}{Graphics Processing Unit}
\newacronym{gtk}{GTK}{GIMP ToolKit}
\newacronym{http}{HTTP}{Hypertext Transfer Protocol}
\newacronym{io}{I/O}{Input/Output}
\newacronym{ide}{IDE}{Integrated Development Environment}
\newacronym{il1}{IL-1}{Interleukin-1}
\newacronym{ioc}{IoC}{inversion of control}
\newacronym{itk}{ITK}{Insight ToolKit}
\newacronym{jar}{JAR}{Java ARchive}
\newacronym{jit}{JIT}{Just-In-Time compiler}
\newacronym{jvm}{JVM}{Java Virtual Machine}
\newacronym{knime}{KNIME}{KoNstanz Information MinEr}
\newacronym{knip}{KNIP}{KNIME Image Processing}
\newacronym{laf}{L\&F}{Look \& Feel}
\newacronym{list}{LIST}{Luxembourg Institute of Science and Technology}
\newacronym{loci}{LOCI}{Laboratory for Optical and Computational Instrumentation}
\newacronym{lut}{LUT}{color lookup table}
\newacronym{macos}{macOS}{Macintosh Operating System}
\newacronym{mamut}{MaMuT}{Massive Multi-view Tracker}
\newacronym{matlab}{MATLAB}{MATrix LABoratory}
\newacronym{mitobo}{MiToBo}{Microscopy image analysis ToolBox}
\newacronym{moma}{MoMA}{MotherMachine Analyzer}
\newacronym{nemo}{NEMO}{NF-$\kappa$B essential modulator}
\newacronym{nigms}{NIGMS}{National Institute of General Medical Sciences}
\newacronym{nih}{NIH}{National Institutes of Health}
\newacronym{ome}{OME}{Open Microscopy Environment}
\newacronym{omero}{OMERO}{OME Remote Objects}
\newacronym{opencl}{OpenCL}{Open Computing Language}
\newacronym{opencv}{OpenCV}{Open source Computer Vision library}
\newacronym{plga}{PLGA}{poly (lactic-co-glycolic acid)}
\newacronym{ram}{RAM}{Random-Access Memory}
\newacronym{rest}{REST}{REpresentational State Transfer}
\newacronym{rgb}{RGB}{Red+Green+Blue color model}
\newacronym[longplural={regions of interest}]{roi}{ROI}{region of interest}
\newacronym{scifio}{SCIFIO}{SCientific Image Format Input and Output}
\newacronym{scp}{SCP}{Secure CoPy}
\newacronym{sftp}{SFTP}{Secure File Transfer Protocol}
\newacronym{ssh}{SSH}{Secure SHell}
\newacronym{snt}{SNT}{Simple Neurite Tracer}
\newacronym{swt}{SWT}{Standard Widget Toolkit}
\newacronym{ui}{UI}{user interface}
\newacronym{uri}{URI}{Uniform Resource Identifier}
\newacronym{url}{URL}{Uniform Resource Locator}
\newacronym{webdav}{WebDAV}{Web Distributed Authoring and Versioning}
\newacronym{xml}{XML}{eXtensible Markup Language}

\usepackage{graphicx} 

\startlocaldefs
\endlocaldefs

\renewcommand{\theenumi}{\arabic{enumi}.} 

\begin{document}

\begin{frontmatter}

\begin{fmbox}
\dochead{Software}


\title{ImageJ2: ImageJ for the next generation of scientific image data}


\author[
   addressref={aff1}
]{\inits{CTR}\fnm{Curtis T} \snm{Rueden}}
\author[
   addressref={aff1,aff2}
]{\inits{JS}\fnm{Johannes} \snm{Schindelin}}
\author[
   addressref={aff1}
]{\inits{MCH}\fnm{Mark C} \snm{Hiner}}
\author[
   addressref={aff1}
]{\inits{BED}\fnm{Barry E} \snm{DeZonia}}
\author[
   addressref={aff1,aff2}
]{\inits{AEW}\fnm{Alison E} \snm{Walter}}
\author[
   addressref={aff1,aff2}
]{\inits{ETA}\fnm{Ellen T} \snm{Arena}}
\author[
   addressref={aff1,aff2},
   email={eliceiri@wisc.edu}
]{\inits{KWE}\fnm{Kevin W} \snm{Eliceiri}}


\address[id=aff1]{
  \orgname{Laboratory for Optical and Computational Instrumentation, University of Wisconsin at Madison},
  \city{Madison},
  \state{Wisconsin},
  \cny{USA}
}
\address[id=aff2]{%
  \orgname{Morgridge Institute for Research},
  \city{Madison},
  \state{Wisconsin},
  \cny{USA}
}

\end{fmbox}



\begin{abstractbox}

\begin{abstract} 
\parttitle{Background}
  ImageJ is an image analysis program extensively used in the biological
  sciences and beyond. Due to its ease of use, recordable macro language, and
  extensible plug-in architecture, ImageJ enjoys contributions from
  non-programmers, amateur programmers, and professional developers alike.
  Enabling such a diversity of contributors has resulted in a large community
  that spans the biological and physical sciences.
  However, a rapidly growing user base, diverging plugin suites, and technical
  limitations have revealed a clear need for a concerted software engineering
  effort to support emerging imaging paradigms, to ensure the software's
  ability to handle the requirements of modern science.

\parttitle{Results}
  We rewrote the entire ImageJ codebase, engineering a redesigned plugin
  mechanism intended to facilitate extensibility at every level, with the goal
  of creating a more powerful tool that continues to serve the existing
  community while addressing a wider range of scientific requirements. This
  next-generation ImageJ, called ``ImageJ2'' in places where the distinction
  matters, provides a host of new functionality. It separates concerns, fully
  decoupling the data model from the user interface. It emphasizes integration
  with external applications to maximize interoperability. Its robust new
  plugin framework allows everything from image formats, to scripting
  languages, to visualization to be extended by the community. The redesigned
  data model supports arbitrarily large, N-dimensional datasets, which are
  increasingly common in modern image acquisition. Despite the scope of these
  changes, backwards compatibility is maintained such that this new
  functionality can be seamlessly integrated with the classic ImageJ interface,
  allowing users and developers to migrate to these new methods at their own
  pace.

\parttitle{Conclusions}
  Scientific imaging benefits from open-source programs that advance new method
  development and deployment to a diverse audience. ImageJ has continuously
  evolved with this idea in mind; however, new and emerging scientific
  requirements have posed corresponding challenges for ImageJ's development.
  The described improvements provide a framework engineered for flexibility,
  intended to support these requirements as well as accommodate future needs.
  Future efforts will focus on implementing new algorithms in this framework
  and expanding collaborations with other popular scientific software suites.
\end{abstract}




\end{abstractbox}
%

\end{frontmatter}




\section*{Background}
ImageJ \cite{imagej_history} is a powerful, oft-referenced platform for image
processing, developed by Wayne Rasband at the \acrfull{nih}. Since its initial
release in 1997, ImageJ has proven paramount in many scientific endeavors and
projects, particularly those within the life sciences \cite{imagej_review}.
Over the past twenty years, the program has evolved far beyond its originally
intended scope. After such an extended period of sustained growth, any software
project benefits from a subsequent period of scrutiny and refactoring; ImageJ
is no exception. Such restructuring helps the program to remain accessible to
newcomers, powerful enough for experts, and relevant to the demands of its
ever-growing community. As such, we have developed ImageJ2: a total redesign of
the previous incarnation (hereafter ``ImageJ 1.x''), which builds on the
original's successful qualities while improving its core architecture to
encompass the scientific demands of the decades to come. Key motivations for
the development of ImageJ2 include:

\begin{enumerate}
  \item \textbf{Supporting the next generation of image data.} Over time, the
    infrastructure of image acquisition has grown in sophistication and
    complexity. For example, in the field of microscopy we were once limited to
    single image planes. However, with modern techniques we can record much
    more information: physical location in time and space (X, Y, Z, time),
    lifetime histograms across a range of spectral emission channels,
    polarization state of light, phase and frequency, angles of rotation (e.g.,
    in light sheet fluorescence microscopy), and high-throughput screens, just
    to name a few. The ImageJ infrastructure needed improvement to work
    effectively with these new modes of image data.

  \item \textbf{Enabling new software collaborations.} The field of software
    engineering has seen an explosion of available development tools and
    infrastructure, and it is no longer realistic to expect a single standalone
    application to remain universally relevant. We wanted to improve ImageJ's
    modularity to facilitate its use as a software library, the creation of
    additional user interfaces, and integration and interoperability with
    external software suites.

  \item \textbf{Broadening the ImageJ community.} Though initially developed
    for the life sciences, ImageJ is used in various other scientific
    disciplines as well. It has the potential to be a powerful tool for any
    field that benefits from image visualization, processing, and analysis:
    earth sciences, astronomy, fluid dynamics, computer vision, signal
    processing, etc. We wanted to enhance ImageJ's impact in the greater
    scientific community by adopting software engineering best practices,
    generalizing the codebase, and providing unified, comprehensive,
    consistently structured, community-editable online resources.
\end{enumerate}

From these motivations emerge the six pillars of the ImageJ2 mission
statement:

\begin{itemize}
  \item \textbf{Design} the next generation of ImageJ, driven by the needs of
    the community.
  \item \textbf{Collaborate} across organizations, fostering open development
    through sharing and reuse.
  \item \textbf{Broaden} ImageJ's usefulness and relevance across many
    disciplines of the scientific community.
  \item \textbf{Maintain} backwards compatibility with existing ImageJ
    functionality.
  \item \textbf{Unify} online resources to a central location for the ImageJ
    community.
  \item \textbf{Lead} ImageJ development with a clear vision.
\end{itemize}

It is important to stress that this mission is, and always will be, informed
by pragmatism. For instance, much of ImageJ's existing user community is
centered in the biosciences and related life science fields, and the core
ImageJ developers and contributors are part of bioimaging laboratories as
principal investigators, staff, students, consultants, etc.
\cite{imagej_contributors}. As such, ImageJ's current development directions
tend toward addressing problems in bioimaging. However, most image processing
algorithms are generally applicable, and there are users of ImageJ in other
scientific fields as well. Hence, we wish to avoid pigeonholing the software
as a tool for bioimage analysis only, which would implicitly preclude it from
being adopted for other purposes. One of our explicit goals is to exploit
commonality across scientific disciplines, leaving the door open for others
to collaborate and improve ImageJ in cases where doing so is useful.

\subsection*{Why ImageJ?}
Any time a development effort of this scale is undertaken on an existing tool,
it is worth evaluating its impact and the decision to invest such resources.
The bioimage informatics field \cite{bioimage_informatics} is fortunate to have
a wide range of software tools available in both commercial and open source
arenas \cite{bioimaging_software_review}. Open-source tools are especially
important in science due to their transparency and inherent ability for sharing
and extensibility \cite{bioimaging_cell_biology}. This need and ability for
method sharing has resulted in a plethora of open-source solutions in bioimage
informatics, ranging from image acquisition tools such as $\mu$Manager
\cite{micro_manager_2010, micro_manager_2014}; databases such as
\acrfull{bisque} \cite{bisque} and \acrfull{omero} \cite{omero}; image analysis
suites such as Icy \cite{icy} and BioImageXD \cite{bioimagexd}; scientific
workflow and pipeline tools such as CellProfiler \cite{cellprofiler,
cellprofiler_2011}, \acrfull{knime} \cite{knime, knip} and Pipeline Pilot
\cite{workflow_systems}; and \acrfull{3d} rendering applications such as
FluoRender \cite{fluorender} and Vaa3D \cite{vaa3d}. There are many other open,
bioimaging-oriented software packages besides these, including solutions
written in powerful scripting platforms such as R, Python and MATLAB. With such
an extensive array of tools, does it make sense to invest in an updated ImageJ
platform, rather than building on some combination of more recent tools?

The ImageJ2 project aims to do both, by rearchitecting ImageJ as a shared
platform for integration and interoperability across many bioimaging software
packages. ImageJ has a unique niche in that it is not a monolithic or
single-purpose application, but rather a platform for discovery
where the bench biologist can adapt and deploy new image analysis methods.
Historically, ImageJ 1.x has been popular due to not only pre-designed tools
developed for a single purpose and regularly maintained and updated, but also
its powerful yet approachable plugin and macro environments that have enabled
hundreds of groups to generate results through the development of thousands of
customized plugins and scripts \cite{imagej_review, imagej_ecosystem,
imagej_list_of_update_sites}. It is this ability for sharing, and the desire to
engage the professional and amateur developer alike, that drove the development
for ImageJ2. The new version of ImageJ is a platform for extensibility and
cross-application cooperation, broadening the scope of ImageJ into a new effort
called SciJava \cite{scijava}: a collaboration of projects striving to
cooperate and build on one another both socially and technically. It is our
intent that with the developments detailed in this paper, the synergy between
these tools, which include ImageJ, \acrshort{knime}, CellProfiler,
\acrshort{omero} and others, will only increase as each tool continues to
evolve along with current avenues of scientific inquiry, benefiting not only existing users, but new users and communities as
well. See Table 1 in the "Results and Discussion" section for a detailed
breakdown of software that has been successfully integrated with ImageJ.


\subsection*{Design Goals}
The central technical design goals of ImageJ2 can be divided into seven key
categories: functionality, extensibility, reproducibility, usability,
performance, compatibility and community. In this section, we discuss the goals
of ImageJ2 from its outset; for how these goals have been met in practice, see
the subsequent sections.

\subsubsection*{Functionality}
The overriding principle of ImageJ2 is to create \textbf{\textit{powerful}}
software, capable of meeting the expanding requirements of an ever-more-complex
landscape of scientific image processing and analysis for the foreseeable
future. As such, ImageJ needs to be more than a desktop application: it must be
a
\textbf{\textit{modular}}, multi-layered set of functions with each layer
encapsulated and building upon lower layers. In computer science terminology,
ImageJ2 strives to have a proper \textbf{\textit{separation of concerns}}
between data model and display thereof, enabling use within a wide variety of
scenarios, such as headless operation---i.e., running remotely on a server,
cluster or cloud without a graphical \acrfull{ui}.

At its core, ImageJ2 aims to provide robust support for
\textbf{\textit{N-dimensional}} image data, to support domains with dimensions
beyond time and space. Examples include: multispectral and hyperspectral
images, fluorescence lifetime measured in the time or frequency domains,
multi-angle data from acquisition modalities such as light sheet fluorescence
microscopy, multi-position data from High Content Screens, and experiments
using polarized light. In general, the design must be robust enough to express
any newly emerging modalities within its infrastructure.

Finally, it is not sufficient to provide a modular framework---ImageJ2 must
also provide \textbf{\textit{built-in routines}} as default behavior for
standard tasks in image processing and analysis. These core plugins must span a
wealth of algorithms for image processing and analysis, image visualization,
and image file import and export. Such built-in features ensure users have an
application they can apply out-of-the-box.

\subsubsection*{Extensibility}
According to a survey of ImageJ users, the greatest strength of ImageJ is its
\textbf{\textit{extensibility}} \cite{imagej_survey}. From its inception
\cite{imagej_history},
ImageJ 1.x has had a mechanism by which users can develop their own plugins and
macros to extend its capabilities. Two decades later, a plethora of such
plugins and macros have been shared and published \cite{imagej_ecosystem}. It
is paramount that ImageJ2 maintains this ease of modification and extension by
its user community, and furthermore leverages its improved separation of
concerns to actually make user extension easier and more powerful; e.g., if
image processing plugins are agnostic to user interface, new interfaces can be
developed without a loss of functionality.

A related preeminent concern is \textbf{\textit{interoperability}}. There is no
silver bullet in image processing. No matter how powerful ImageJ becomes or how
many extensions exist, there will always be powerful and useful alternative
tools available. Users benefit most when information can easily be exchanged
between such tools. One of ImageJ2's primary motivations is to enable usage of
ImageJ code from other applications and toolkits, and vice versa, and to
support open standards for data storage and exchange.

\subsubsection*{Reproducibility}
For ImageJ to be truly useful to the scientific community, it must be not only
technically feasible to extend, but also socially feasible, without legal
obstacles or other restrictions preventing the free exchange of scientific
ideas. To that end, ImageJ must be not only open source, but offer full
\textbf{\textit{reproducibility}}, following an \textbf{\textit{open
development process}} which we believe is an optimal fit for open scientific
inquiry \cite{software_usability}. We want to enable the community to not just
use ImageJ, but also to build upon it, with all project resources---revision
history, project roadmap, community contribution process, etc.---publicly
accessible, and development discussions taking place in public, archived
communication channels so that interested parties can remain informed of and
contribute to the project's future directions. Such transparency also
facilitates sensible, defensible software development processes and fosters
responsibility amongst those involved in the ImageJ project. In particular,
the project must be well covered by automated tests, to validate that it
produces consistent results on reference data sets.

\subsubsection*{Usability}
Modular systems composed of many components often have a corresponding increase
in conceptual complexity, making them harder to understand and use. To avoid
this pitfall, ImageJ2 employs the idea of complexity minimization: seeking
\textbf{\textit{sensible defaults}} that make simple things easy, but difficult
things still possible. The lowest-level software layers should define the
program's full power, while each subsequent layer reduces visible complexity by
choosing default parameters suitable for common tasks. The highest levels
should provide users with the simplicity of a ``big green button,'' performing
the most commonly desired tasks with ease---the powerful inner machinery
remaining unseen, yet accessible when needed.

To bridge the gap between extensibility and usability, there must be a painless
process of installing new functionality: a built-in, configurable
\textbf{\textit{automatic update mechanism}} to manage extensions and keep the
software up-to-date. This update mechanism must be scalable and distributed,
such that software developers can publish their own extensions on their own
websites, without needing to obtain permission from a central authority.

\subsubsection*{Performance}
N-dimensional images and the ever-expanding size of datasets increase the
computation requirements placed on analysis routines. For ImageJ2 to succeed,
it must accomplish its goals without negatively impacting performance
\textbf{\textit{efficiency}} in time---e.g., \acrfull{cpu} and
\acrfull{gpu}---or space---e.g., \acrfull{ram} and disk. Furthermore, to ensure
ImageJ2 meets performance needs for a wide variety of use cases, it should
offer choices surrounding usage of available resources, as well as sensible
defaults for balancing performance in common scenarios.

Another key consideration for performance is \textbf{\textit{scalability}}:
ImageJ must be capable of operating on increasingly huge datasets. In cloud
computing, this requirement is often met via elasticity: the ability to
transparently provision additional computing resources---i.e., throw more
computers at the problem \cite{hardware_is_cheap}. We are at the dawn of the
``Big Data'' era of computing, where both computation and storage are scalable
resources which can be purchased from remote server farms. Software like ImageJ
which hopes to remain effective for serious scientific inquiry into the coming
decades must be architected so that its algorithms scale well to increasingly
large data processed in parallel across increasingly large numbers of
\acrshort{cpu} and \acrshort{gpu} cores.

\subsubsection*{Compatibility}
There are a vast number of existing extensions---plugins, macros, and
scripts---for the original ImageJ 1.x application which have proven extremely
useful to the user community \cite{imagej_ecosystem}. ImageJ2 must continue to
support these extensions as faithfully as possible, while also providing a
clear incremental migration path to take advantage of the new framework.

\subsubsection*{Community}
The principal non-technical goal of ImageJ2 is to serve the ImageJ community as
it evolves and grows; to that end, several community-oriented technical goals
naturally follow. The ImageJ project must provide \textbf{\textit{unified
online resources}} including a central community-editable website, discussion
forum, and online technical resources for managing community extensions of
ImageJ. And the ImageJ application itself must work in concert with these
resources---e.g., users should be able to report bugs directly to online issue
tracking systems when something goes wrong.


\section*{Implementation}
Broadly speaking, ImageJ2 components are classified into one of four domains:

\begin{itemize}
  \item \textbf{SciJava.} The most fundamental layers of ImageJ2 are
    independent from image processing, but rather provide needed functionality
    common to many applications. On a technical level, the SciJava core
    components are a set of standard Java libraries for managing extensible
    applications. Socially, the SciJava initiative is a pledge among
    cooperating organizations to work together, reuse code, and synergize
    wherever possible \cite{imagej_scijava}.
  \item \textbf{ImgLib2.} To ensure generality of image analysis, ImageJ2 is
    built on the flexible ImgLib2 container model \cite{imglib2}. Decoupling the
    elements of image representation, ImgLib2 components enable general image
    processing algorithms to be reused, regardless of image type, source, or
    organization.
  \item \textbf{\acrfull{scifio}.} \acrshort{scifio} components define
    standards for reading, writing, and translating between image formats
    \cite{scifio}. These libraries ensure a broad spectrum of image data can be
    interpreted in a consistent manner across all SciJava applications.
  \item \textbf{ImageJ.} Low-level components establish image metadata and algorithm
    patterns, built on the SciJava and ImgLib2 layers. High-level
    components focus on ``end user'' tools for working with image data,
    and include user interfaces, user-facing commands, and the top-level
    ImageJ application \cite{imagej_web_site}.
\end{itemize}

These layers, taken as a whole, form the \textbf{ImageJ software stack}
\cite{imagej_architecture}, the core set of components upon which ImageJ-based
applications are built.

Each domain is itself divided into many individual libraries, each of which
targets a particular function. This separation of concerns provides a logical
organization which allows targeted reuse and extension of any given
functionality of interest.

The following sections describe, in order from lowest to highest level, the
essential backbone libraries of ImageJ2. Note that this is not an exhaustive
list of components, as many components across these domains provide secondary
functions---e.g.: script languages, build management, \acrshort{ui} elements,
or targeted implementations of specific features.

\subsection*{SciJava Common}
The ground floor of the ImageJ software stack is the SciJava Common library
\cite{imagej_sjc}, providing the core framework for creating extensible
applications. The heart of SciJava Common is its \textbf{application
container}, the \texttt{Context} class. Each \texttt{Context} encapsulates
runtime application state: available extensions, open images and documents,
user settings, etc. The application container paradigm allows multiple
independently configured instances of SciJava applications to run concurrently
within the same \acrfull{jvm}.

\subsubsection*{Service framework}
The application container consists of a collection of \textbf{services}, which
are initialized dynamically at runtime. These services provide methods which
operate on the system in various ways, such as opening data, manipulating
images, or displaying user interface elements on screen. Taken as a whole,
these service methods constitute the bulk of the \acrfull{api} of ImageJ.
Software developers are free to extend the system with new, needed services
and/or override any aspect of behavior provided by existing services. This
approach is in contrast to the most common naive design of many software
projects, which use global ``static'' state and functions, whose behavior is
difficult or impossible to override or enhance in downstream code.

The SciJava Common library itself provides the most fundamental of these
services, such as:

\begin{itemize}
  \item A \textbf{plugin service}, which dynamically discovers available
    plugins using an index generated at compile time by a Java annotation
    processor. This plugin index is used to bootstrap the application context,
    since services are themselves a type of plugin.
  \item An \textbf{event service}, which provides a hierarchical
    publish/subscribe model for event handling.
  \item A \textbf{log service}, for environment-agnostic data logging.
  \item An \textbf{object service}, which keeps a central typed index of
    available objects.
  \item A \textbf{thread service}, which manages a thread pool and dispatch
    thread(s) for code execution.
  \item An \textbf{\acrfull{io} service}, for reading and writing of data.
  \item A \textbf{preference service}, for saving and restoring user-specific
    preferences.
\end{itemize}

In principle, SciJava Common is similar to frameworks such as Spring
\cite{spring}, offering standard software engineering patterns such as
\acrfull{di} \cite{dependency_injection} and \acrfull{ioc} \cite{ioc}, but
tailored to the needs of collaborative scientific projects like ImageJ. For
example, SciJava Common provides a generalized \acrshort{io} mechanism for
opening data from any source, but the library itself has no specific knowledge
of how to open \acrfull{xml} documents, microscopy image formats, or
spreadsheets of numerical results---such functionality is provided by
downstream components as SciJava plugins (see next section).

\subsubsection*{Plugin framework}
SciJava Common provides a unified mechanism for defining \textbf{plugins}:
extensions which add new features or behavior to the software, and/or modify
existing behavior. Plugins are discovered by the system at runtime, and ordered
according to assigned priorities and types, forming type hierarchies:
structural trees that define how each individual plugin fits into the system.
The typical pattern for a desired sort of functionality is to define a
dedicated plugin type, then implement plugins that fulfill that operation in
various ways. SciJava Common is designed to make virtually any aspect of an
application extensible. Some of the most critical plugin categories and types
include:

\paragraph*{Core extensibility}
\begin{itemize}
  \item \textbf{\texttt{Service}} -- A collection of related functionality,
    expressed as an \acrshort{api}. SciJava services are singletons with
    respect to each application context. For example, each instance of ImageJ2
    has exactly one \texttt{AnimationService} responsible for managing
    animations, with methods to start and stop animations, select the dimension
    over which to animate, adjust frame rate, and other options. Note that
    while the behavior of services can certainly be modified by extensions,
    doing so is primarily the domain of advanced developers looking to
    radically alter behavior of the system.
  \item \textbf{\texttt{IOPlugin}} -- A plugin that reads data from and/or
    writes data to a location, such as a file on disk. For example, the SciJava
    layer provides \acrshort{io} plugins for common text formats such as
    Markdown \cite{markdown}, while the \acrshort{scifio} layer provides an
    \acrshort{io} plugin for image formats.
\end{itemize}

\paragraph*{Modules}
\begin{itemize}
  \item \textbf{\texttt{Command}} -- An operation, more generally known as a
    SciJava \textbf{\textit{module}}, with typed inputs and outputs. These
    modules typically appear in the menu system of the application's user
    interface, but can be exposed via interoperability mechanisms in many other
    ways, such as nodes in \acrshort{knime} or modules in CellProfiler
    \cite{cellprofiler}. When ImageJ users talk about ``writing a plugin'' they
    usually mean a \texttt{Command}. See ``Module framework'' below for more on
    SciJava modules.
  \item \textbf{\texttt{ScriptLanguage}} -- A programming language for SciJava
    scripts. Each script language plugin provides the logic necessary to
    execute scripts written in that language (e.g., JavaScript or Python) as
    SciJava modules with typed inputs and outputs, in a similar way to
    commands. It also makes it possible to express operations as code snippets
    that can be reused in scripts to repeat those operations.
  \item \textbf{\texttt{Converter}} -- A plugin which transforms data from one
    type of object to a different type of object. Converters greatly extend the
    concept of type conversion from what Java provides out of the box to
    provide automatic conversion in a wide and extensible set of circumstances.
    For example, it becomes possible for an algorithm to accept a string in
    place of a floating point numerical value, as long as that string can be
    parsed to such a value---or to transparently convert between
    normally-incompatible image data structures from different image processing
    ecosystems.
  \item \textbf{\texttt{ModulePreprocessor}} -- A ``meta-module'' which
    prepares modules to run. For example, the \texttt{LoadInputsPreprocessor}
    populates a module's inputs with the last-used values as defaults, which
    can save the user a lot of time. Preprocessor plugins are executed in
    priority order as part of the module ``preprocessing chain'' before the
    module is actually executed.
  \item \textbf{\texttt{ModulePostprocessor}} -- A ``meta-module'' which does
    something with a module after it has run. For example, the
    \texttt{DisplayPostprocessor} takes care of displaying the outputs of a
    module after it has completed execution. Postprocessor plugins are executed
    in priority order as part of the module ``postprocessing chain'' after the
    module is actually executed.
\end{itemize}

\paragraph*{User interface}
\begin{itemize}
  \item \textbf{\texttt{UserInterface}} -- A plugin providing an application
    \acrshort{ui}. These plugins include functionality for creating and showing
    windows and dialogs. ImageJ2 includes a user interface written in Java's
    Swing toolkit which is modeled closely after the ImageJ 1.x design, as well
    as a \texttt{UserInterface} plugin that wraps ImageJ 1.x itself. But other
    \acrshortpl{ui} are equally possible; since a \acrshort{ui} is simply a
    type of plugin, anyone can develop their own SciJava \acrshort{ui} without
    any code changes to the core system. The system is even flexible enough to
    display multiple \acrshortpl{ui} simultaneously.
  \item \textbf{\texttt{Platform}} -- A plugin which enables customization of
    behavior based on machine-specific criteria, such as specific flavor of
    operating system or Java language, including type, architecture, or
    version. For example, on \acrfull{macos}, the menu bar appears at the top
    of the screen, with the About, Preferences, and Quit commands relocated to
    the Application menu.
  \item \textbf{\texttt{InputWidget}} -- A user interface element for
    harvesting typed inputs. Typically, these widgets are presented as part of
    a form in a dialog box which prompts the user to fill in input values of a
    module. In principle, the widgets can be used for anything requiring typed
    input from the user. For example, a \texttt{FileWidget} allows the user to
    select a file (\texttt{java.io.File}) on disk, while a ToggleWidget
    provides a boolean toggle (typically rendered as a checkbox). The SciJava
    layer provides \acrshort{ui}-agnostic interfaces to the common widget
    types, along with widget implementations corresponding to each supported
    \texttt{UserInterface} plugin. However, an extension to the system can not
    only implement its own data structure classes which it uses as inputs to
    its modules; it can also provide corresponding widgets for those
    structures, allowing the user to populate them from the user interface in
    innovative ways.
  \item \textbf{\texttt{Display}} -- A plugin for visualizing data. For
    example, an ImageJ2 \texttt{ImageDisplay} can show two-dimensional planes
    of N-dimensional image data in a window with sliders for controlling which
    plane is visible. However, the framework imposes no limits on the sorts of
    objects that can be visualized; other examples include the
    \texttt{TextDisplay}, which shows strings, and the \texttt{TableDisplay},
    which shows tabular data as a spreadsheet. These plugins are typically used
    to display a module's typed outputs (i.e., its results).
  \item \textbf{\texttt{Tool}} -- A collection of rules binding user input
    (e.g., keyboard and mouse events) to display and data manipulation actions.
    For example, ImageJ2's \texttt{PanTool} pans a display when the mouse is
    dragged or arrow key is pressed; the \texttt{PencilTool} draws hard lines
    on the data within an image display. Many user interfaces render them as
    icons in the application toolbar.
  \item \textbf{\texttt{ConsoleArgument}} -- A plugin that handles arguments
    passed to the application as command line parameters. This plugin type
    makes the application's command line parameter handling extensible---a
    feature especially important for headless operation sans user interface.
\end{itemize}

This encapsulation of functionality, coupled with a plugin prioritization
mechanic, allows SciJava-based software to be fully customized or extended at
any point. An application such as ImageJ is then simply a collection of plugins
and services built on top of the SciJava Common framework. For instance, the
ImageJ Common \cite{imagej_common} component introduces new services
specifically for opening and displaying images, specializing the functions
defined in the lower-level components. Assigning these specialized functions a
higher plugin priority creates a natural, flexible ordering of operations.
Given that everything from user interfaces to file formats uses the SciJava
plugin mechanism, the path for overriding any behavior is clear and consistent.

Finally, to keep the plugin development process as simple as possible, great
care is taken throughout the codebase to adhere to interface-driven design with
default method implementations whenever possible. This strategy minimizes the
amount of code developers are responsible for writing, lowering the barrier to
entry for creating and modifying plugins.

\subsubsection*{Module framework}
To successfully interoperate with other scientific software, ImageJ algorithms
must be decoupled from the various user interfaces and applications which might
want to expose them to end users.

The key concept SciJava employs is that of \textbf{\textit{parameterized
modules}}: executable routines with declared input and output parameters of
specific types. These modules can take the form of \texttt{Command} plugins or
be expressed as scripts written in any supported scripting language (via
available \texttt{ScriptLanguage} plugins; see ``Plugin framework'' above). For
example, a user might write the following parameterized Groovy \cite{groovy}
script:

\begin{quote}
  \small
  \begin{verbatim}
  #@INPUT String name
  #@INPUT int age
  #@OUTPUT String greeting
  greeting = "Hello, " + name + ". You are " + age + " years old."\end{verbatim}
\end{quote}

This script accepts two parameters as input---a name and an age---and outputs a
greeting based on the input values. Note the typing: the name can be any string
of characters, but the age must be an integer value; the greeting is also a
string of characters. Note also that this script makes no assumptions about
user interface; it is the responsibility of the framework to: A) prompt the
user for the input values in the most appropriate way, B) execute the module
code itself, and finally, C) process and/or display the output values in the
most appropriate way.

As such, this scheme has great potential for reuse across a wide variety of
contexts. For example, when running the above script from the ImageJ user
interface, a Swing dialog box will pop up allowing the user to enter the name
and age values; and after OK is pressed, the greeting will be displayed in a
new window. However, when running the script headless from the command line
interface, the input values can be passed as command line arguments and the
output values echoed to the standard output stream. See Supplemental Figure 1
for an illustration. Since many computational tools have this concept of
parameterized modules, SciJava developers need only create some general adapter
code to integrate the SciJava module framework with a given tool---and then all
SciJava modules become automatically available within that tool's paradigm. We
have already implemented such integration for several other tools in the
SciJava ecosystem, including CellProfiler, \acrshort{knime} \cite{knip}, and
the \acrshort{omero} image server \cite{omero}.

SciJava Common has an important mechanism which facilitates the extensible and
configurable execution of modules: module pre- and post-processing. Developers
can write \texttt{ModulePreprocessor} and \texttt{ModulePostprocessor} plugins
to extend what happens when a module is executed (see ``Plugin framework''
above). Moreover, there are also two plugin types built on this module
processing mechanism which make it easy to customize and extend how modules
behave:

\begin{enumerate}
  \item {The process of collecting module inputs is known as \textit{input
    harvesting}. The \texttt{InputWidget} plugin type lets developers create
    widgets to harvest specific types of inputs from the user. In particular,
    the SciJava project provides Swing widgets for several data types
    (Supplemental Table 1).

    Some inputs are also automatically populated via
    \texttt{ModulePreprocessor} code. For example, when a single image
    parameter is declared, an ``active image preprocessor'' detects the
    situation, populating the value with the currently active image. In this
    way, the user does not have to explicitly select an image upon which to
    operate in the common case, but the module still has semantic knowledge
    that an image is one of the routine's input parameters.}
  \item The process of dealing with outputs after a module executes is known as
    \textit{displaying}. The \texttt{Display} plugin type lets developers
    visualize specific types of outputs in appropriate ways. The SciJava layer
    provides a basic display plugin for text outputs, which shows the text in a
    dedicated window, while the ImageJ layer provides additional similar
    display plugins for image and tabular data.
\end{enumerate}

One final SciJava subsystem of note is the \textit{conversion framework}, which
provides a general way of transforming data from one type to another. The
\texttt{Converter} plugin type lets developers extend SciJava's conversion
capabilities, allowing objects of one type to be used as module inputs of a
different type, in cases where the two types are conceptually analogous. For
example, data stored in memory as a \acrfull{matlab} matrix can be expressed as
an ImageJ image object, even though \acrshort{matlab} matrices are not natively
ImageJ images \cite{imagej_matlab}. When a suitable converter plugin is
present, modules capable of operating only on \acrshort{matlab} matrices become
transparently capable of accepting ImageJ images as inputs, thanks to the
framework's auto-conversion. Similarly, a converter between ImageJ and the
\acrfull{itk} \cite{itk} images greatly streamlines use of \acrshort{itk}-based
algorithms within ImageJ \cite{imagej_itk}.

\subsection*{ImageJ Common}

Meeting the needs of contemporary scientific image analysis requires a flexible
and extensible data model, including support for arbitrary dimensions, data
types and image sizes. To this end, we have chosen to model ImageJ2 images
using the ImgLib2 library, which itself provides an extensible,
interface-driven design that supports numeric (8-bit unsigned integer, 32-bit
floating point, etc.) and non-numeric data types. It also provides great
flexibility regarding the source and structure of data. Out of the box, ImgLib2
provides several data sources and sample organizations, including use of a
single primitive array (``array image''), one array per plane (``planar
image''), and block-based ``cell image.'' However, the library remains general
enough that alternative structures are also feasible. To quote the ImgLib2
article \cite{imglib2}:

\begin{quote}
  The core paradigm [of ImgLib2] is a clean separation of pixel algebra (how
  sample values are manipulated), data access (how sample coordinates are
  traversed), and data representation (how the samples are stored, laid out in
  memory, or paged to disc). ImgLib2 relies on virtual access to both sample
  values and coordinates, facilitating parallelizability and extensibility.
\end{quote}

ImageJ Common provides a unification of the type and storage-independence of
ImgLib2 with the SciJava Common plugin framework (described above). A
\texttt{Dataset} interface provides the fundamental representation of ImageJ
images, collections of images, and corresponding metadata: \acrfullpl{roi},
visualization settings, sample coordinates and physical calibrations, and much
more. Also provided are plugins and services for working with these
\texttt{Dataset} objects. Together, these classes form the access points for
higher-level components to open, save, generate and process these images.

Note that as of this writing, elements of the ImageJ Common data model and
corresponding services are still stabilizing. As such, we do not describe these
structures in technical detail here.

\subsection*{\acrshort{scifio}} An essential goal of ImageJ2 is to establish
universal image analysis routines, with no limits on application; however, the
proliferation of proprietary image formats from scientific instruments creates
a major obstacle to this ambition. To overcome this issue, the
\acrshort{scifio} core library establishes a common framework for reading,
writing and translating image data to and from the ImageJ Common data model, as
well as between domain-specific standard metadata models. \acrshort{scifio}
builds on the services provided in SciJava Common and ImageJ Common, defining
image \texttt{Format} and metadata \texttt{Translator} plugin types to
encapsulate the operations necessary to take an image source and standardize it
as an ImageJ \texttt{Dataset}.

\acrshort{scifio} builds upon SciJava Common's core \acrshort{io}
infrastructure, which allows it to operate on most data locations independent
of their nature. SciJava Common provides a \texttt{Location} interface which
acts as a data descriptor, similar to a \acrfull{uri}. This \texttt{Location}
interface is specialized according to the nature of the data; for example, a
\texttt{URLLocation} identifies data served by a remote \acrfull{url}, while
an \texttt{OMEROLocation} (part of the ImageJ-\acrshort{omero} integration
\cite{imagej_omero}) identifies an image from an \acrshort{omero} server. For
data locations whose raw bytes can be accessed randomly and/or sequentially
(e.g., remote \acrshortpl{url}, but not \acrshort{omero} images), SciJava
Common provides a \texttt{DataHandle} plugin type which enables such access.
The core library provides \texttt{DataHandle} plugins for several kinds of data
locations, including files on disk, remote \acrshortpl{url}, and arrays of
bytes in local computer memory. Developers can easily create new
\texttt{DataHandle} plugins which provide random access into additional sorts
of locations, and \acrshort{scifio} will be able to use them transparently
without any changes to existing \texttt{Format} or \texttt{Translator} plugins.

The \texttt{Format} plugin \acrshort{api} is architected to support reading and
writing of image data in chunks, which provides scalability. It is no longer
necessary to have a large quantity of computer \acrshort{ram} to work with
large images---\acrshort{scifio} reads the data from the source location on
demand, paging it into and out of memory as needed. \acrshort{scifio}'s caching
mechanism persists any changes made to image pixels, even when chunks leave
memory, by using temporary storage on disk.

\acrshort{scifio} \texttt{Translator} plugins provide the means to translate
not only between image formats, but between common metadata models of various
scientific disciplines. For example, the \acrfull{ome} defines a data model
called \acrshort{ome}-\acrshort{xml} \cite{ome_xml}, for which the
\acrshort{scifio}-\acrshort{ome}-\acrshort{xml} component provides a suite of
translators to and from ImageJ Common data structures. In this way,
\acrshort{scifio} has the potential to bridge interoperability gaps across
various discipline-specific scientific software packages.

Further details about \acrshort{scifio} can be found in the \acrfull{bmc}
Bioinformatics software article ``\acrshort{scifio}: an extensible framework to
support scientific image formats'' \cite{scifio}.

\subsection*{ImageJ Ops}

ImageJ's ultimate purpose is image processing and analysis. To that end, we
have crafted the ImageJ Ops component: ImageJ2's shared, extensible library of
reusable image processing operations. As of version 0.33.0, the core Ops
library provides 788 \texttt{Op} plugins across nearly 350 types of ops in
more than 20 namespaces, covering functionality such as: image arithmetic,
trigonometry, Fourier transformations, deconvolution, global and local
thresholding, image statistics, image filtering, binary morphological
operations, type conversion, image transformations (scaling, rotation,
etc.)---even 2- and 3-dimensional geometric operations such as marching cubes
\acrshort{3d} mesh generation (see Figure 1 for examples). A thorough
treatment of available ops can be found in the ImageJ Ops tutorial notebook
\cite{imagej_notebooks}.

  \begin{figure}[h]
    \caption{Examples of image processing algorithms available in ImageJ Ops.}
    \includegraphics[width=4.75in]{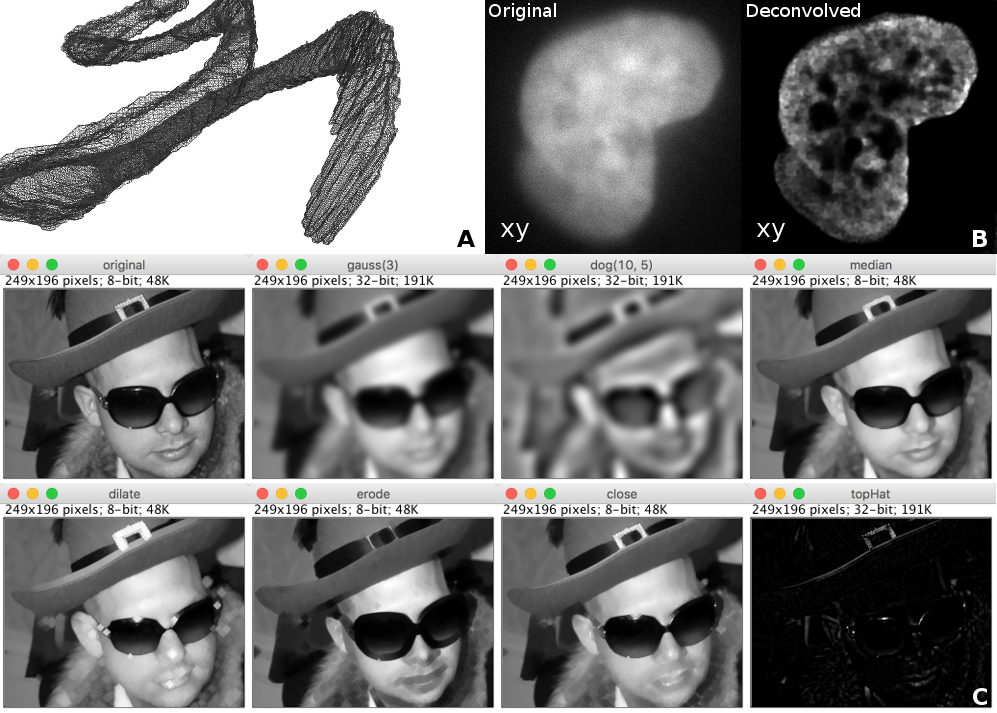}
    \begin{flushleft}
      \footnotesize
      Panel A (top left): \acrshort{3d} wireframe mesh of ImageJ's Bat Cochlea
      Volume sample dataset \cite{bat_cochlea_volume}, computed by the
      \texttt{geom.marchingCubes} op, an implementation of the marching cubes
      algorithm \cite{marching_cubes}, visualized using MeshLab \cite{meshlab}.
      Credit to Kyle Harrington for the figure, Tim-Oliver Buchholz for
      authoring the op, and Art Keating for the dataset. Panel B (top right):
      Richardson-Lucy Total Variation deconvolution \cite{richardson_lucy} of
      the Stellaris FISH dataset \#1 \cite{stellaris_fish}, computed by the
      \texttt{deconvolve.richardsonLucyTV} op. Credit to Brian Northan for
      authoring the op and figure \cite{bnorthan_ops_decon}, and George
      McNamara for the dataset. Panel C (bottom): Grayscale morphology and
      neighborhood filter operations on Fiji's New Lenna sample image, using a
      diamond-shaped structuring element with radius 3. Credit to Jean-Yves
      Tinevez, Jonathan Hale and Leon Yang for authoring the ops.
    \end{flushleft}
  \end{figure}

ImageJ Ops was conceived with three major design goals: 1) easy to use and
extend; 2) powerful and general; and 3) high performance. To achieve all three
of these goals, Ops utilizes a plugin-based design enabling ``extensible case
logic.'' Ops defines a new plugin type, \texttt{Op}, each of which has a name
and a list of typed parameters, analogous to a function definition in most
programming languages. When invoking an op, callers typically do not specify
the exact \texttt{Op} plugin to use, but instead specify the operation's name
and arguments; the Ops framework then performs a \textit{matching} process,
finding the optimal fit for the given request. For example, calling
\texttt{math.add} with a planar image and a 64-bit floating point number leads
to a match of \texttt{net.imagej.ops.math.ConstantToPlanarImage.AddDouble},
which adds a constant value to each element of an image, whereas calling
\texttt{math.add} with two planar images results in a match of
\texttt{net.imagej.ops.math.IIToIIOutputII.Add}, which adds two images
element-wise.

This scheme is similar to---but more powerful than---the method
overloading capabilities of many programming languages: op behavior can be
further specialized by tailoring \texttt{Op} implementations for specific
subclasses, generic parameters, and \texttt{Converter} substitutions (see
``SciJava Common'' above). Consider an op sqrt(image), which computes the
element-wise square root of an image. If we implement this op as
\texttt{sqrt(Dataset)}, we miss out on performance optimizations for
\texttt{ArrayImg}, an ImgLib2 container type where the entire collection of
image samples is stored in a single Java primitive array. However, if we only
implement \texttt{sqrt(ArrayImg)}, we are restricted in supported data types,
since not all images can be stored in such a manner. The power of the Ops
matching approach is that both of these and more can coexist simultaneously and
extensibly, and the most specific will always be selected at runtime.

Furthermore, as algorithm implementations increasingly become available for the
\acrshort{gpu} via libraries such as \acrfull{opencl} \cite{opencl} and
Nvidia's \acrfull{cuda} \cite{cuda}, as well as for clusters via libraries such
as Apache Spark \cite{apache_spark}, such implementations could also be
expressed as ops so that they can be selected automatically based on the
currently available hardware environment.

The \texttt{Op} plugin type extends SciJava's \texttt{Command}, and therefore
all ops are SciJava parameterized modules, usable anywhere SciJava modules are
supported---see the ``module framework'' section in ``SciJava Common'' above.
Like standard modules, an op declares typed inputs and outputs. However, unlike
modules in general, an op must be a ``pure function'' with a fixed number of
parameters and no side effects; i.e., it must be deterministic in its behavior,
operating only on the given inputs, and populating or mutating only the given
outputs. These restrictions provide some very useful guarantees which allow the
system to reason about an op's use and behavior; e.g., after computing an op
with particular arguments once, the result can be cached to dramatically
improve subsequent time performance at the potential expense of additional
space. Properly constructed ops will also always be usable headless because
they do not rely on the existence of \acrshort{ui} elements.

\paragraph*{Op chaining and special ops}
It is often the case in image processing that an algorithm can be expressed as
a composition of lower level algorithms. For example, a simple difference of
Gaussians (``DoG'') operation is merely two Gaussian blur operations along with
a subtraction:

\begin{quote}
  $dog(image, \sigma_1, \sigma_2) =
  sub(gauss(image, \sigma_1), gauss(image, \sigma_2))$
\end{quote}

For users calling into the Ops framework via scripting, the core library
provides an \texttt{eval} op backed by SciJava's expression parser library,
which enables executing such sequences of ops via standard mathematical
expressions, including use of unary and infix binary operators.

For developers, the Ops library provides a mechanism for efficient
\textit{chaining} of ops calls. An op may declare other ops as inputs,
resulting in a tree of ops which are resolved when an op is matched; the
matched op instance can then be reused across any number of input values. In
this way, very general operations can be created to address a broad range of
use cases---e.g., the \texttt{map} operation provides a unified way of
executing an op such as \texttt{math.sqrt(number)} element-wise on a collection
(e.g., an image) whose elements are numbers. Indeed, in the case of DoG, the
Ops library's baseline implementation takes an image as input, along with two
\texttt{filter.gauss} ops and a \texttt{math.sub} op, and then invokes them on
the input image. The baseline \texttt{stats.mean} implementation is similar,
built on the \texttt{stats.sum}, \texttt{stats.size} and \texttt{math.div} ops.
Higher level DoG ops provide sensible defaults, enabling calls like
\texttt{dog(image, sigma1, sigma2)} to work, making common operations simple,
while leaving the door open for additional customization as needed.

To facilitate type-safe and efficient chaining of ops, the Ops library has a
subsystem known as \textit{special ops}. Such special ops are specifically
intended to be called repeatedly from other ops, without needing to invoke the
op matching algorithm every time. This repeat usage is achieved in a type-safe
and efficient way by explicitly declaring the types of the op's primary
inputs---i.e., the inputs whose values can be efficiently varied across
invocations of the op---as well as the type of the op's primary output.

Special ops have two major characteristics beyond regular ops. First, each
special op has a declared \textit{arity}, indicating the number of primary
inputs, which are explicitly typed via Java generics and can thus efficiently
vary across invocations of the op. Three arities are currently implemented:
\textit{nullary} for no inputs, \textit{unary} for one input, and
\textit{binary} for two inputs. It is important to note that unlike a formal
mathematical function, a unary special op may have more than one input
parameter---the ``unary'' in this case refers to the number of explicitly typed
parameters intended to vary when calling an instance of the op multiple times.
For instance, in the DoG example above, the baseline DoG is declared as a unary
op, so that the input image can vary efficiently while the sigmas etc. are held
constant in value.

Secondly, every special op is one of three kinds:

\begin{itemize}
  \item A \textit{function} operates on inputs, producing outputs, in a way
    consistent with the functional programming paradigm. Inputs are immutable,
    and outputs are generated during computation and subsequently also
    immutable. Functions are very useful for parallel processing since they are
    fully thread-safe even when object references overlap---but this safety
    comes at the expense of space, since they offer no way to reuse
    preallocated output buffers.
  \item A \textit{computer} is similar to a function, but populates a
    preallocated output object instead of generating a new object every time.
    Computers have many of the same advantages of functions, but provide the
    ability to reuse preallocated output buffers to improve efficiency in space
    and time.
  \item An \textit{inplace} op mutates its input(s) in place---i.e., its input
    and output are the same object. Inplace ops are highly space efficient, but
    lack the mathematical guarantees of functions and computers, since they
    destroy the original input data.
\end{itemize}

Some ops are implemented as \textit{hybrids}, offering a choice between two or
more of the function, computer and inplace computation mechanisms. Users of the
ops library---even advanced users---will rarely if ever need to know about this
implementation detail, but for developers crafting new ops, it is convenient to
have unifying interfaces which provide common logic for combining these
paradigms. See Supplemental Table 2 for a complete breakdown of the special op
kinds and arities.

\subsection*{ImageJ Legacy}
To maximize backwards compatibility with ImageJ 1.x, ImageJ2 must continue to
provide access to the complete existing \acrshort{ui} and \acrshort{api} with
which ImageJ users are familiar, while also making all new ImageJ2 features
available for exploration and use. Furthermore, to bridge the gap, ImageJ2 must
provide improved functionality transparently when possible, as well as support
seamless ``mixing and matching'' of the two respective \acrshortpl{api}. In
this way, ImageJ2 can enable gradual migration to the more powerful
capabilities of ImageJ2, while empowering developers' contributions to the
framework to be immediately effective. To achieve this goal, we identified the
major functional pathways of ImageJ 1.x and reworked them to delegate first to
ImageJ2 equivalents, falling back on the old behavior if needed.

There are two ImageJ components dedicated to maintaining backwards
compatibility with ImageJ 1.x. The lower level of the two is the IJ1-patcher:
using a tool called Javassist \cite{javassist} to perform an advanced Java
technique known as bytecode manipulation, ImageJ 1.x code is modified at
runtime to expose callback hooks at critical locations---e.g.: when opening
images with \textit{File $\triangleright$ Open\ldots}, closing the ImageJ
application, or displaying \acrshort{ui} components. These hooks are built
using the SciJava plugin infrastructure, allowing new behavior to be injected
into ImageJ 1.x despite the fact that it was not designed to support such
extensibility. In essence, ImageJ2 ``rewrites'' portions of ImageJ 1.x at
runtime to make integration possible. This approach is necessary because
altering ImageJ 1.x directly to enable such hooks would break backwards
compatibility with existing macros and plugins, ruining established scientific
workflows which have otherwise remained functional across many years.

By default, these hooks are exploited to inject ImageJ2 functionality in the
second compatibility layer: ImageJ Legacy. ImageJ2 intercepts an ImageJ 1.x
request and attempts to delegate to its own routines. For example, in our
implementation of the \textit{File $\triangleright$ Open\ldots} hook, we use
the SciJava \acrshort{io} service, which provides extensible support for data
types via SciJava \acrshort{io} plugins. This allows the full power of
\acrshort{scifio} to be called automatically by \textit{File $\triangleright$
Open\ldots} without requiring users to select individual loader plugins. In
this way, ImageJ2 exposes new ``seams'' which provide extensibility points not
available in the standalone ImageJ 1.x project \cite{legacy_code}.

A second major function of ImageJ Legacy is to provide a wrapping legacy
\acrshort{ui}: an ImageJ2 \texttt{UserInterface} plugin that reuses the ImageJ
1.x \acrshort{ui}, but maintains synchronization between respective data
structures. For example, consider the \texttt{ImagePlus} structure in ImageJ
1.x and its equivalent, the \texttt{Dataset}, in ImageJ2. By default, an
\texttt{ImagePlus} and \texttt{Dataset} could not be interchanged; they have
different Java class hierarchies, and with ImageJ2's expanded data model, a
\texttt{Dataset} is more expressive than an \texttt{ImagePlus}. However,
requiring plugins to ``select one'' would impose a technical barrier, even if
both structures are available in the same application. Thus, the legacy
\acrshort{ui} notes when either an \texttt{ImagePlus} or a \texttt{Dataset} is
created and ensures a complementary instance is mapped, via SciJava
\texttt{Converter} plugins. This brings the ImageJ 1.x and ImageJ2 worlds
closer together: when an image is opened, it can be used by plugins that would
take an \texttt{ImagePlus} or a \texttt{Dataset} regardless of whether that
image was opened via an ImageJ 1.x or ImageJ2 mechanism. Furthermore, because
conversion is handled in the ImageJ Legacy layer, individual plugins do not
require knowledge of the synchronization.

Shared image data structures are but one aspect of the legacy \acrshort{ui}'s
synchronization. Others include logging, notification, and status
events---essentially all \acrshort{ui} events are mapped across paradigms.
Whenever possible, these conversions are achieved using an adapter class that
implements a common interface (e.g., \texttt{Dataset}), which wraps the object
of interest (e.g., \texttt{ImagePlus}) by reference. This approach enables
information to be translated between ImageJ 1.x and ImageJ2 structures on
demand, while minimizing the performance impact. Wrapping by reference also
mitigates the burden of updates; once synchronization is established, changes
to the underlying object are automatically reflected in the wrapper.

\subsection*{ImageJ Updater}
The ImageJ Updater is the mechanism by which the available and installed
components of ImageJ are managed. At its core, the Updater is a flexible
component for tracking ImageJ update sites: endpoints containing versioned
collections of files. Users can pick and choose which update sites they wish to
enable, with ImageJ's core functionality offered on the base ``ImageJ'' update
site, which is on by default. Distributions of ImageJ such as \acrfull{fiji}
\cite{fiji} extend this base with additional functionality (Figure 2) in the
form of more plugins, scripts, sample images, \acrfullpl{lut}, etc., leveraging
the ability to override ImageJ's base behavior using SciJava's plugin priority
mechanism (see ``SciJava Common'' above for details).

  \begin{figure}[h]
    \caption{ImageJ update sites provide additional functionality to ImageJ.}
    \includegraphics[width=4.75in]{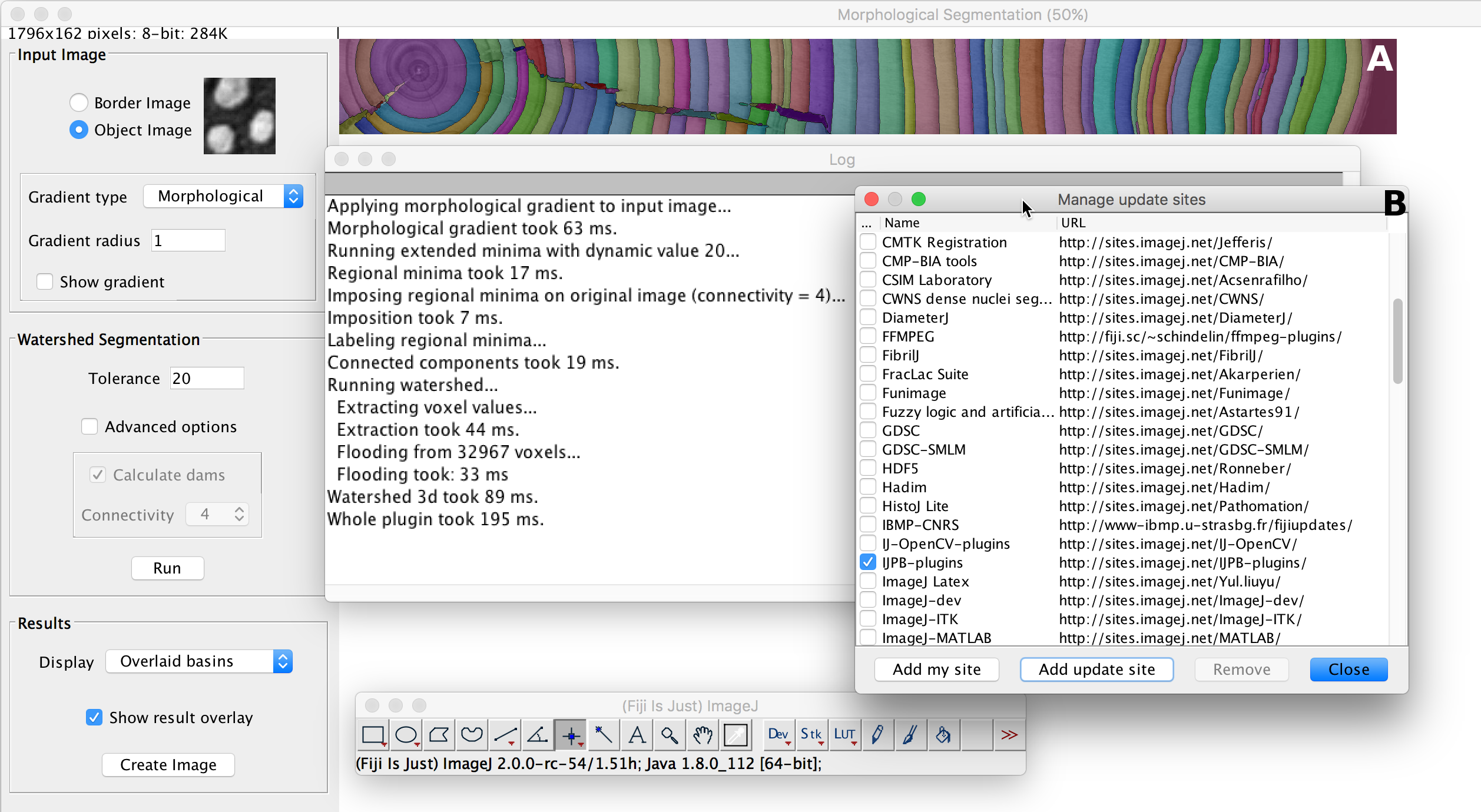}
    \begin{flushleft}
      \footnotesize
      The Morphological Segmentation plugin, part of the MorphoLibJ plugin
      collection \cite{morpholibj}, easily segments the rings of ImageJ's Tree
      Rings sample dataset (panel A). The MorphoLibJ plugins are installed into
      the \acrshort{fiji} distribution of ImageJ by enabling the IJPB-plugins
      update site (panel B). Credit to David Legland and Ignacio
      Arganda-Carreras for authoring the plugins.
    \end{flushleft}
  \end{figure}

The Updater stores metadata in a \texttt{db.xml.gz} file in the root of each
update site, which describes the files that are part of that update site,
including checksums and timestamps for all previous versions. In this way, the
Updater can tell whether each local file is: A) an up-to-date tracked file; B)
an old version of a tracked file; C) a locally modified version of a tracked
file; or D) an untracked file. Update sites are served to users over
\acrfull{http}. Developers may upload files to an update site via an extensible
set of protocols, as defined by Uploader plugins. The core ImageJ distribution
includes plugins for \acrfull{ssh}, \acrfull{scp}, \acrfull{sftp} and
\acrfull{webdav}, but in principle, the Updater makes no assumptions about how
files are uploaded.

The \texttt{db.xml.gz} structure was originally designed for use with the
\acrshort{fiji} Updater, the ImageJ Updater's predecessor. The logic of the
\acrshort{fiji} Updater was migrated into the core of ImageJ2, with backwards
compatibility preserved for existing \acrshort{fiji} installations. As part of
that migration, the Updater was heavily refactored to be \acrshort{ui}
agnostic, such that additional user interface plugins for the Updater could be
created which leverage the same core. Out of the box, ImageJ provides two
different user interfaces for the Updater: a command-line tool intended for
power users and developers, and a Swing \acrshort{ui} intended to make updating
easy for end users. When ImageJ is first launched, it automatically runs the
``Up-to-date check'' command, which then displays the Updater \acrshort{ui} if
updates are available from any of the currently enabled update sites.


\section*{Results and Discussion}
ImageJ has transformed from a single-user, single-bench application to a
versatile framework of extensible, reusable operations. In the following
sections, we discuss how each core aspect of ImageJ2 has impacted community
usage and how we expect these qualities to shape future developments.

\subsection*{Functionality}

The architecture of ImageJ2 enables it to meet current and future demands in
image analysis.

\textbf{Dimensions.} Using ImgLib2 opens up caching options for operating on
extremely large images, an area in which ImageJ 1.x has previously struggled.
ImageJ 1.x is inherently limited to five dimensions (X, Y, Z, time, and
channel) with fewer than $2^{31}$ pixels per XY plane, e.g. a $50,000 \times
50,000$ plane being too large to represent. ImageJ 1.x allows composite images,
but is constrained to a maximum of seven composited channels. ImageJ2's
N-dimensional data model supports up to $2^{31} - 1$ dimensions, each with up
to $2^{63} - 1$ elements, and composite rendering over any dimension of
interest regardless of length. There are several preset dimensional axis types,
and new types can also be defined as needed. When visualizing multi-channel
data, each channel can now have its own \acrshort{lut} without constraint.

  \begin{figure}[h]
    \caption{ImageJ 1.x ~case logic compared to a unified ImgLib2
    implementation.}
    \includegraphics[width=4.75in]{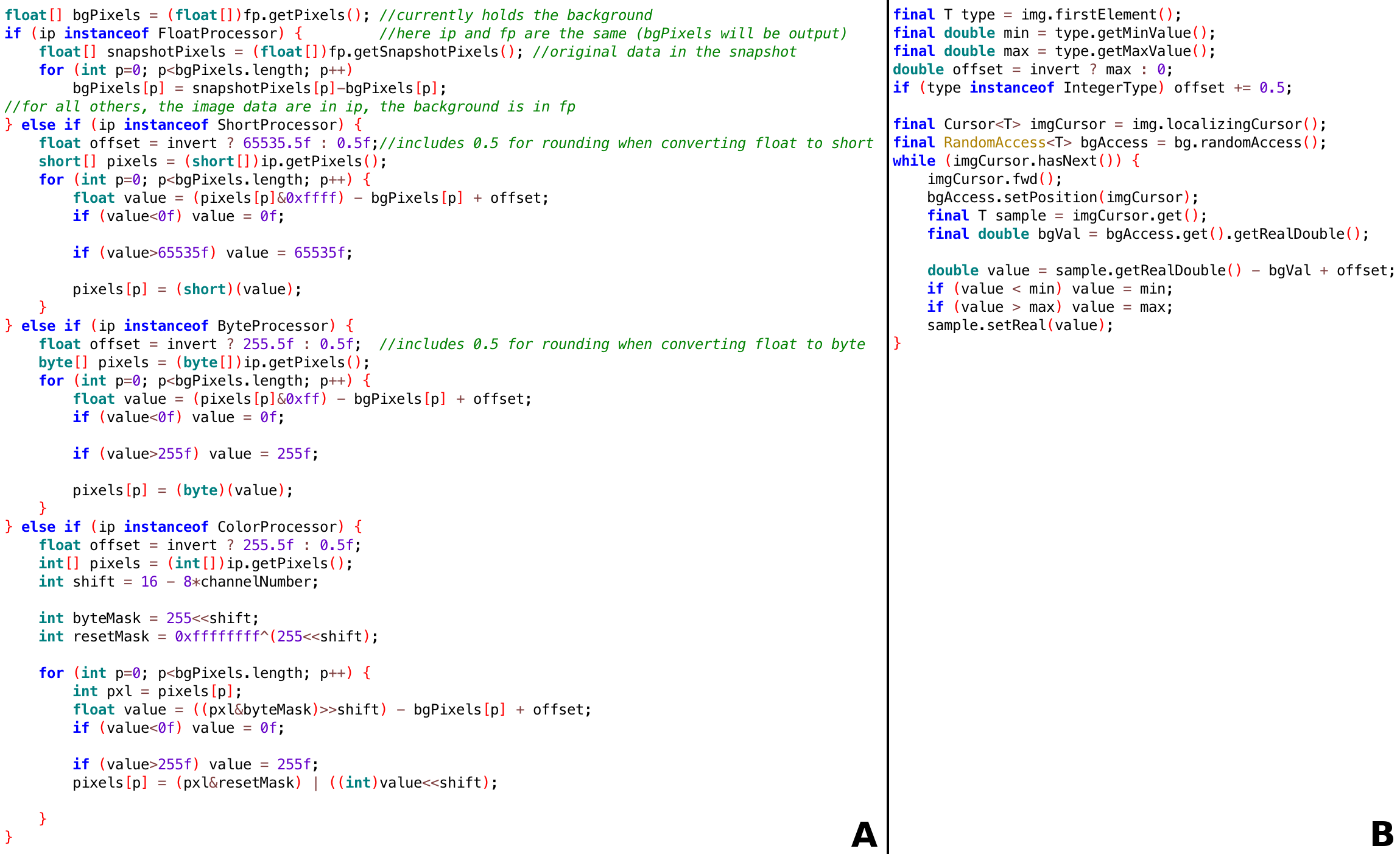}
    \begin{flushleft}
      \footnotesize
      Panel A (left) shows the ImageJ 1.x implementation of a rolling ball
      background subtraction method, part of the
      \texttt{ij.plugin.filter.BackgroundSubtracter} class. Panel B (right)
      shows an equivalent implementation using ImgLib2, without the need for
      extensive case logic.
    \end{flushleft}
  \end{figure}

\textbf{Types.} ImageJ 1.x supports only five image types for representing
sample values: 8-bit unsigned integer grayscale, 8-bit with a color lookup
table, 16-bit unsigned integer grayscale, 32-bit floating point, and a 32-bit
integer-packed color type representing three 8-bit unsigned color channels:
red, green, and blue. Furthermore, this support is highly static, sometimes
requiring case logic for algorithms to properly handle each of desired image
types independently. In contrast, ImgLib2 is explicitly designed to facilitate
algorithms developed agnostic of image type (Figure 3). ImageJ2 already
supports over twenty different image types (Supplemental Table 3), including
arbitrary precision integer and decimal types, and further types are definable
using SciJava Common's flexible plugin framework. SciJava \texttt{Converter}
plugins also extend the reach of ImageJ2-based algorithms even further into
additional data structures, such as \acrshort{matlab} matrices
\cite{imagej_matlab} and \acrshort{itk} images \cite{imagej_itk}.

\textbf{Storage.} The prime example of an alternate storage source in ImageJ
1.x is the virtual stack, allowing image slices to be read on demand---e.g., if
the image would not normally fit in memory. However, ImageJ 1.x commands must
explicitly account for whether or not they can operate on a virtual stack,
requiring a proliferation of case logic and complexity. ImageJ2 takes advantage
of ImgLib2's extensible container system, which enables data to be stored
flexibly: as files on disk, remote \acrshortpl{url}, within a database,
generated on-the-fly, etc. Such routines can even be used with pixel and
storage types implemented well after their creation without having to change
the original implementation. As image acquisition sizes increase, we expect
virtualized image data to be particularly critical to the future of image
analysis. The \acrshort{scifio} library already provides an ImgLib2 image type
(``\acrshort{scifio} cell image'') that supports block-based read/write caching
from disk, effectively behaving as a writeable virtual stack.

\textbf{\Acrfullpl{roi}.} Like ImageJ 1.x, ImageJ2 provides support for
\textit{\acrshortpl{roi}}, which are functions that identify samples upon which
to operate, as well as \textit{overlays}, which are visuals (e.g., text)
superimposed for visualization. ImageJ2 builds upon the \acrshort{roi}
interfaces of ImgLib2, allowing for any number of simultaneous \acrshortpl{roi}
and overlays to be associated with a particular image without the need for
additional tools like ImageJ 1.x's global \acrshort{roi} Manager window.

Because \acrshortpl{roi} are part of the core ImgLib2 library, it is possible
to process subsets of images identified by one or more \acrshortpl{roi} using
an ImgLib2-based algorithm, and the Ops library can process data within a
\acrshort{roi} as a single functional operation. This continues ImageJ2's
migration towards image processing algorithms that need not add explicit case
logic---e.g., to handle \acrshortpl{roi} separately---but instead simply
provide a pixelwise function, or iterate using ImgLib2's generic iteration
mechanism. In this way, we continue to reduce the effort and complexity of
ImageJ2 plugins, while increasing their utility and application.

\textbf{Modularity.} ImageJ 1.x was developed with a ``single computer, single
user, single operation'' in mind. Although ImageJ 1.x can be used as a library,
it will always be a single unit that cannot be decoupled from its dependencies,
which are implicit in its source code. ImageJ2 has succeeded in building a
cohesive application from encapsulated, modular components unified by the
SciJava plugin framework. Each component is independently deployed and
accessible via the build automation tool Maven \cite{apache_maven}, allowing
developers to pick and choose the individual pieces relevant to them---be it
the ImageJ application, a particular scripting language, image format, or the
SciJava Common core. SciJava-based projects inherit a ``bill of materials''
which enables components to be combined at versions known to be compatible with
each other \cite{imagej_architecture}. We have already seen the benefits of
this modularity---for example, the use of the \acrshort{scifio} library in
\acrfull{knip} to produce images compatible with ImageJ2 commands.

\textbf{Ops.} The ImageJ Ops library is the centerpiece of ImageJ2, bringing
Java's mantra of ``write once, run anywhere'' to image processing algorithms.
Ops provides a wealth of image processing algorithms to users, accessible in a
unified way that empowers developers to transparently extend and enhance the
behavior and capabilities of each operation. It is critical to appreciate that
each type of op (more than 350 different operations as of version 0.33.0)
represents a potential extension point for optimizing existing parameters, or
supporting new ones. In contrast to algorithms coded using ImageJ 1.x data
structures, all ops work without modification on all image types (Supplemental
Table 3) and containers, including those not yet in existence. As the Ops
project is a very active collaboration across several institutions including
the University of Konstanz, University of Wisconsin-Madison and others, we
expect the core library to continue to grow in both available functionality and
use within the community.

\subsection*{Extensibility}

\textbf{Plugins.} The ImageJ2 plugin framework, built on top of SciJava Common,
is a modular and extensible infrastructure for adding features. Plugins can now
take many forms, including image processing operations, new tools, and even
completely new displays. In ImageJ 1.x there are three kinds of plugins: 1) the
standard \texttt{PlugIn}, which provides a freeform \texttt{run(String arg)}
method; 2) \texttt{PlugInFilter}, which processes images one plane at a time;
and 3) \texttt{PlugInTool}, which adds a function to the toolbar. In ImageJ2,
these ideas are expressed in the form of \texttt{Command}, \texttt{Op} and
\texttt{Tool} plugins, respectively---although these plugin types have many
advantages over their ImageJ 1.x analogues: type-safe chaining of operations,
dynamic selection of ops based on arguments, \acrshort{ui} agnosticism, etc.
Furthermore, many other types of plugins are available as well, and the
flexibility of the SciJava plugin framework also allows for additional new
types of plugins to be defined.

\textbf{Modules.} The ImageJ application's menu structure is made up of SciJava
modules---most commonly commands and scripts. Thus, scripts and
\texttt{Command} plugins are probably the most common points of extension for
developers exploring the ImageJ2 architecture. Writing such extensions in
ImageJ2 is much simpler than in ImageJ 1.x, which requires each extension to
explicitly create its own dialog box to collect user input. In ImageJ2, the use
of parameters results in more declarative extensions, freeing software
developers from the need to explicitly ask the user for input values in the
vast majority of cases, and substantially reducing boilerplate and
\acrshort{ui}-specific code, making commands shorter and easier to understand
(see Figure 5 in ``Usability'' below). Moreover, this mechanism makes ImageJ2
modules truly independent of the user interface, allowing them to work with any
\acrshort{ui} or headlessly. The module simply declares its inputs and outputs
using the appropriate parameter syntax, and lets ImageJ do the rest.

\textbf{Formats.} In an open source image analysis program like ImageJ, an
extensive collection of supported image formats is necessary to maximize
relevance and impact across the community. ImageJ 1.x provided a central
\texttt{HandleExtraFileTypes} class to enable extensibility, but required
direct modification of this class to do so, resulting in many third parties
each shipping their own modified version. Only one modification could ``win,''
effectively breaking any other supported formats. To fill this role in ImageJ2,
the \acrshort{scifio} library provides extensible image format support tailored
to the ImageJ Common data model.

As of version 0.29.0, the core \acrshort{scifio} library provides a collection
of more than 30 open formats, and also includes a wrapping of the Bio-Formats
library \cite{bio_formats}, which enables a wide variety of supported images
throughout all ImageJ operations. Furthermore, \acrshort{scifio} enables
developers to create their own \texttt{Format} plugins to smoothly integrate
support for new proprietary formats and metadata standards without modification
of core functions or proliferation of one-off format commands.

\textbf{Image processing.} ImageJ's main purpose is effective and extensible
image processing; therefore, ImageJ's extension mechanism for image processing
algorithms must be one of its central features. ImageJ2's op
matching subsystem offers extensible case logic: an \texttt{Op} plugin can be
written to add a new algorithm, to extend an existing algorithm to support new
data structures, or to make an algorithm more efficient for specific data
types, all without impacting previously written code. As the Ops library
matures, we expect to see new \texttt{Op} implementations along all of these
lines in existing third-party suites, conveniently shipped to users via ImageJ
update sites. Hence, unlike in ImageJ 1.x, existing user scripts using the Ops
library will automatically benefit from new performance-enhancing ops.

\textbf{User interface.} ImageJ 1.x's user interface is written in Java
\acrfull{awt} with many assumptions throughout the codebase relying on this
fact. Hence, ImageJ 1.x is only limitedly usable in a headless way (e.g., for
image processing on a server cluster). Normally, ImageJ 1.x cannot be used
headless at all: some lynchpin ImageJ 1.x classes---notably \texttt{ij.ImageJ}
and \texttt{ij.gui.GenericDialog}---derive from \texttt{java.awt.Window}, and
such classes cannot be instantiated when running in headless mode. Fortunately,
the ImageJ Legacy layer's runtime patcher rewrites affected ImageJ 1.x classes
to derive from non-\acrshort{awt} window classes when in headless mode; as
such, ImageJ2 makes headless execution of ImageJ 1.x scripts feasible.

Furthermore, ImageJ 1.x's reliance on \acrshort{awt} also limits its ability to
be embedded into other applications using different \acrshort{ui} frameworks,
such as Swing or Eclipse \acrfull{swt}. While some applications have succeeded
in doing so \cite{bio7}, the amount of work required in response to each ImageJ
1.x update is considerable, since many changes must be made to the ImageJ 1.x
core source code.

In contrast, ImageJ2's separation between the underlying data model and the
user interface enables it to run headless or within a variety of different user
interface paradigms with no changes to the core. Developers can create their
own plugins to provide alternative user interfaces. ImageJ2 is even capable of
displaying multiple \acrshortpl{ui} simultaneously in the same Java runtime.
Adding support for a new \acrshort{ui} now only requires writing a new
\texttt{UserInterface} plugin and corresponding display and widget plugins.
As one of the most common questions about ImageJ from software developers is
how to customize the ImageJ \acrshort{ui}, we believe that this improved user
interface framework will yield substantial future dividends.

While the current flagship user interface for ImageJ2 is still the ImageJ 1.x
\acrshort{ui} via the ImageJ Legacy component, ImageJ2 also has a Swing user
interface modeled after the ImageJ 1.x \acrshort{ui}, but which stands alone
with no dependence on ImageJ 1.x code. We have been successful in
``reskinning'' this Swing \acrshort{ui} with various Java \acrfullpl{laf},
including the Metal, Motif, Nimbus, Aqua, Windows and \acrfull{gtk}
\acrshortpl{laf}. Furthermore, we explored proof-of-concept \acrshort{ui}
implementations in other frameworks, such as Eclipse's \acrshort{swt}, Java
\acrshort{awt} sans Swing, and Apache Pivot. There is also a JavaFX
\acrshort{ui} for ImageJ2 developed by Cyril Mongis at the University of
Heidelberg \cite{imagejfx}, as well as integrations of ImageJ into other
powerful end-user applications such as \acrshort{knime} and CellProfiler. See
Figure 4 for a side-by-side illustration of \acrshortpl{ui}.

  \begin{figure}[h]
    \caption{Side-by-side comparison of ImageJ2-based user interfaces
    and integrations.}
    \includegraphics[width=4.75in]{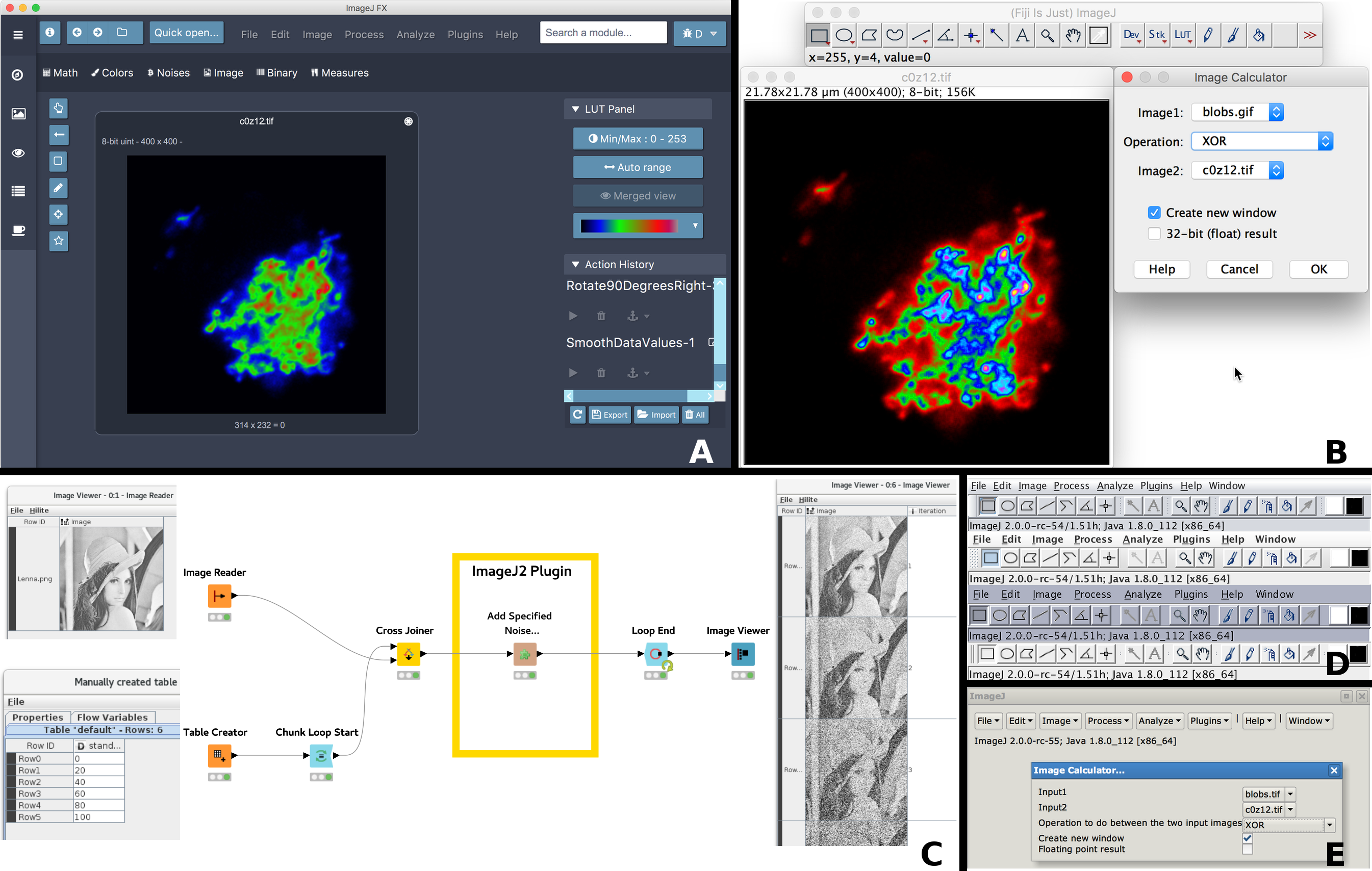}
    \begin{flushleft}
      \footnotesize
      Panel A (top left): ImageJFX, a JavaFX-based user interface built on
      ImageJ2. Panel B (top right): ImageJ2's default user interface, the
      ImageJ Legacy \acrshort{ui}, which wraps ImageJ 1.x. Panel C (bottom
      left): Example \acrshort{knime} workflow utilizing ImageJ2 image
      processing nodes. Panel D (middle right): Swing \acrshort{ui} prototype,
      closely modeled after ImageJ 1.x so that it remains familiar to existing
      users, in various Java ``\acrlong{laf}'' modes. Panel E (bottom right): A
      proof-of-concept Apache Pivot user interface. The ImageJFX and ImageJ
      Legacy \acrshortpl{ui} display an XY slice of ImageJ's Confocal Series
      sample dataset (dataset courtesy of Joel Sheffield), which has been
      rotated, smoothed and colorized.
    \end{flushleft}
  \end{figure}

\textbf{Interoperability.} There is no one-size-fits-all tool for scientific
image processing. A diversity of tools benefits users, even more so when they
can interoperate. ImageJ 1.x was designed to be run by a single user on a
single desktop computer. Many aspects of the program are structured as
singletons: one macro interpreter, one \texttt{WindowManager}, one active
image, one \texttt{PlugInClassLoader}, one active \acrshort{roi} per image, one
set of overlays, one active \acrshort{roi} in the \acrshort{roi} manager, etc.
This structure imposes many limitations: for example, multiple macros cannot
run concurrently, and it is not possible to operate more than one instance of
ImageJ 1.x in the same \acrshort{jvm} simultaneously---e.g., on a single web
page as applets.

ImageJ2 is structured as an application container, avoiding the static
singleton pattern and hence many of ImageJ 1.x's limitations. Multiple
instances of ImageJ2 can run in the same \acrshort{jvm}, each with multiple (or
no) user interfaces and multiple concurrent operations. Our primary goal is to
make each encapsulated component of ImageJ2 usable in other software projects.
There are several examples of other projects leaning on this generality to
expose SciJava modules in interesting ways: for example, automated conversion
to nodes in a \acrshort{knime} workflow. Continuing efforts are underway at
the \acrfull{loci} and elsewhere to integrate ImageJ with many other software
projects, script languages and paradigms (Table 1).

  \begin{table}[h!]
    \caption{ImageJ software integrations}
    \begin{tabular}{| l | l | l |}
      \hline
      \textbf{Software}                                                  & \textbf{Integration project}                                         & \textbf{Supporting technologies}                                           \\ \hline
      Apache Groovy                               \cite{groovy}          & SciJava Scripting: Groovy             \cite{scripting_groovy}        & -                                                                          \\ \hline
      BeanShell                                   \cite{beanshell}       & SciJava Scripting: BeanShell          \cite{scripting_beanshell}     & -                                                                          \\ \hline
      Bio-Formats                                 \cite{bio_formats}     & \acrshort{scifio}-Bio-Formats         \cite{scifio_bf_compat}        & \acrshort{scifio}-\acrshort{ome}-\acrshort{xml}     \cite{scifio_ome_xml}  \\ \hline
      Bio7 (R + ImageJ 1.x)                       \cite{bio7}            & -                                                                    & Eclipse                                             \cite{eclipse}         \\ \hline
      CellProfiler                                \cite{cellprofiler}    & ImageJ Server*                        \cite{imagej_server}           & -                                                                          \\ \hline
      ImageJ 1.x                                  \cite{imagej_history}  & ImageJ Legacy                         \cite{imagej_legacy}           & ImageJ 1.x Patcher \cite{ij1_patcher}, Javassist    \cite{javassist}       \\ \hline
      \acrshort{itk}                              \cite{itk}             & ImageJ-\acrshort{itk}                 \cite{imagej_itk}              & SimpleITK                                           \cite{simpleitk}       \\ \hline
      JavaScript                                  \cite{javascript}      & SciJava Scripting: JavaScript         \cite{scripting_javascript}    & Nashorn \cite{nashorn}, Rhino                       \cite{rhino}           \\ \hline
      Jupyter Notebook                            \cite{jupyter}         & SciJava Jupyter Kernel                \cite{scijava_jupyter_kernel}  & \acrshort{beakerx}$^\ddagger$                       \cite{beakerx}         \\ \hline
      \acrshort{knime}                            \cite{knime}           & \acrshort{knime} Image Processing     \cite{knip}                    & -                                                                          \\ \hline
      Kotlin                                      \cite{kotlin}          & SciJava Scripting: Kotlin             \cite{scripting_kotlin}        & -                                                                          \\ \hline
      Lisp (\acrshort{jvm})                       \cite{lisp}            & SciJava Scripting: Clojure            \cite{scripting_clojure}       & Clojure                                             \cite{clojure}         \\ \hline
      \acrshort{matlab}                           \cite{matlab}          & SciJava Scripting: \acrshort{matlab}  \cite{scripting_matlab}        & matlabcontrol                                       \cite{matlabcontrol}   \\ \hline
      \acrshort{matlab}                                                  & ImageJ-\acrshort{matlab}              \cite{imagej_matlab}           & SciJava Scripting: \acrshort{matlab}                                       \\ \hline
      \acrshort{mitobo}$^\mathsection$            \cite{mitobo}          & -                                                                    & \acrshort{alida}$^\dagger$                          \cite{alida}           \\ \hline
      \acrshort{omero}                            \cite{omero}           & ImageJ-\acrshort{omero}               \cite{imagej_omero}            & -                                                                          \\ \hline
      \acrshort{opencv}$^\mathparagraph$          \cite{opencv}          & IJ-OpenCV                             \cite{ij_opencv}               & JavaCV                                              \cite{javacv}          \\ \hline
      Python (CPython or \acrshort{jvm})          \cite{python}          & imglib2-imglyb                        \cite{imglib2_imglyb}          & pyJNIus \cite{pyjnius}, Jython \cite{jython}, JyNI  \cite{jyni}            \\ \hline
      Python (CPython)                                                   & imagey                                \cite{imagey}                  & imglib2-imglyb                                                             \\ \hline
      Python (CPython)                                                   & SciJava Scripting: CPython            \cite{scripting_cpython}       & javabridge                                          \cite{javabridge}      \\ \hline
      Python (\acrshort{jvm})                                            & SciJava Scripting: Jython             \cite{scripting_jython}        & Jython, JyNI                                                               \\ \hline
      R (\acrshort{jvm})                          \cite{r}               & SciJava Scripting: Renjin             \cite{scripting_renjin}        & Renjin                                              \cite{renjin}          \\ \hline
      \acrshort{rest}$^\|$                        \cite{rest}            & ImageJ Server*                        \cite{imagej_server}           & Dropwizard                                          \cite{dropwizard}      \\ \hline
      Ruby (\acrshort{jvm})                       \cite{ruby}            & SciJava Scripting: JRuby              \cite{scripting_jruby}         & Ruby                                                \cite{ruby}            \\ \hline
      Scala                                       \cite{scala}           & SciJava Scripting: Scala              \cite{scripting_scala}         & -                                                                          \\ \hline
      TensorFlow                                  \cite{tensorflow}      & ImageJ-TensorFlow                     \cite{imagej_tensorflow}       & -                                                                          \\ \hline
    \end{tabular}
    \begin{flushleft}
      * Provides cross-language interprocess integration with JavaScript, Python and others.

      $^\dagger$ \acrfull{alida}.

      $^\ddagger$ \acrfull{beakerx}.

      $^\mathsection$ \acrfull{mitobo}.

      $^\mathparagraph$ \acrfull{opencv}.

      $^\|$ \acrfull{rest}.
    \end{flushleft}
  \end{table}

\subsection*{Reproducibility}

In the interest of transparency and reproducibility---especially in the context
of open science---the ImageJ2 project strives to be accessible. Ultimately, we
want to spur the community to improve ImageJ in a collaborative way, by
providing open access to resources. Of course, we recognize the need for
responsive, reliable maintainers to coordinate and facilitate contributions.
However, with the Internet's modern software infrastructure, it is now quite
feasible to push ImageJ development in a more collaborative and
community-driven direction, embracing the ``GitHub effect''
\cite{github_effect} of worldwide, distributed development.

All ImageJ2 source code is open and publicly available on GitHub
\cite{imagej_source_code}, and all core components are permissively licensed
\cite{imagej_licensing} to avoid any ambiguity over how the code can be used.
But visibility alone is not sufficient to keep a project open; each line of
code adds complexity, making the project harder to understand and maintain.
Modular, encapsulated design, the application of the ``\acrfull{dry}'' concept,
and avoidance of overly ``clever'' code keeps ImageJ2 well-organized and easier
to understand. Extensive online documentation \cite{imagej_web_site} and
Javadoc \cite{imagej_javadoc} provide further insight, while effective unit
testing and dedicated tutorial components \cite{imagej_tutorials}
illustrate concrete use cases. By
keeping a clean, well-organized and well-documented codebase, we facilitate
community contributions, as well as continued maintenance of ImageJ into the
future.

ImageJ2's open development process provides many benefits over the centralized
process of ImageJ 1.x. The use of Maven makes dependency management human
readable and enables the use of a ``bill of materials'' to unambiguously
determine which versions of each ImageJ component are compatible. The use of
Git has evolved revision control to a new level of documentation, clearly
communicating why changes are made and encouraging atomic, easily understood
changes. Furthermore, ImageJ2 minimizes the barrier to community contributions
via an open issue tracking system \cite{imagej_issues} and open patch
submission process \cite{imagej_contributing}.

Finally, it is critical for reproducibility that algorithms produce consistent
results over time. We utilize the public Travis \acrfull{ci} infrastructure
\cite{travis_ci} to run automated regression tests whenever modifications to
ImageJ's source code are published. Such tests help to avoid and detect
regression bugs so that core functionality and behavior will continue to work
reliably as the program evolves. We especially prioritize test coverage for
the crucial base levels of ImageJ: as of this writing, there are approximately
500 tests for SciJava Common, 1200 tests for ImageJ Ops, and 1200 tests for
ImageJ 1.x. We also plan to integrate automated test coverage analysis, to
measure the percentage of code which is exercised by the tests, which should be
straightforward thanks to the project's use of Maven.

\subsection*{Usability}

The ImageJ community includes both end users---who use ImageJ as an application
and want it to ``just work''---and software developers---who want to customize
and invoke parts of ImageJ as a software library from their own programs.
However, these roles are not rigid; many users write scripts and macros to
facilitate their image analysis, and many developers also use ImageJ as an
application. ImageJ2 includes a powerful Script Editor with many tools to aid
users as they transition into the realm of development. This tool removes much
of the complexity of traditional software development, allowing users to focus
on coding without the added burden of applying compilers, \acrfullpl{ide}, or
the command line.

In addition, the SciJava parameterized scripting mechanism makes it easier for
users to write scripts whose inputs and outputs are declared in a clear and
straightforward manner. SciJava parameters reduce the boilerplate code
historically needed to define a script's input values, in some cases by 50\% or
more (Figure 5). Leveraging SciJava annotations also frees the plugin from the
Java \acrshort{awt} dependencies of ImageJ 1.x's \texttt{GenericDialog},
allowing the plugin to be used headlessly, in future \acrshortpl{ui}, and even
in other applications.

  \begin{figure}[h]
    \caption{Comparison of pure ImageJ 1.x command with one using SciJava
    declarative syntax.}
    \includegraphics[width=4.75in]{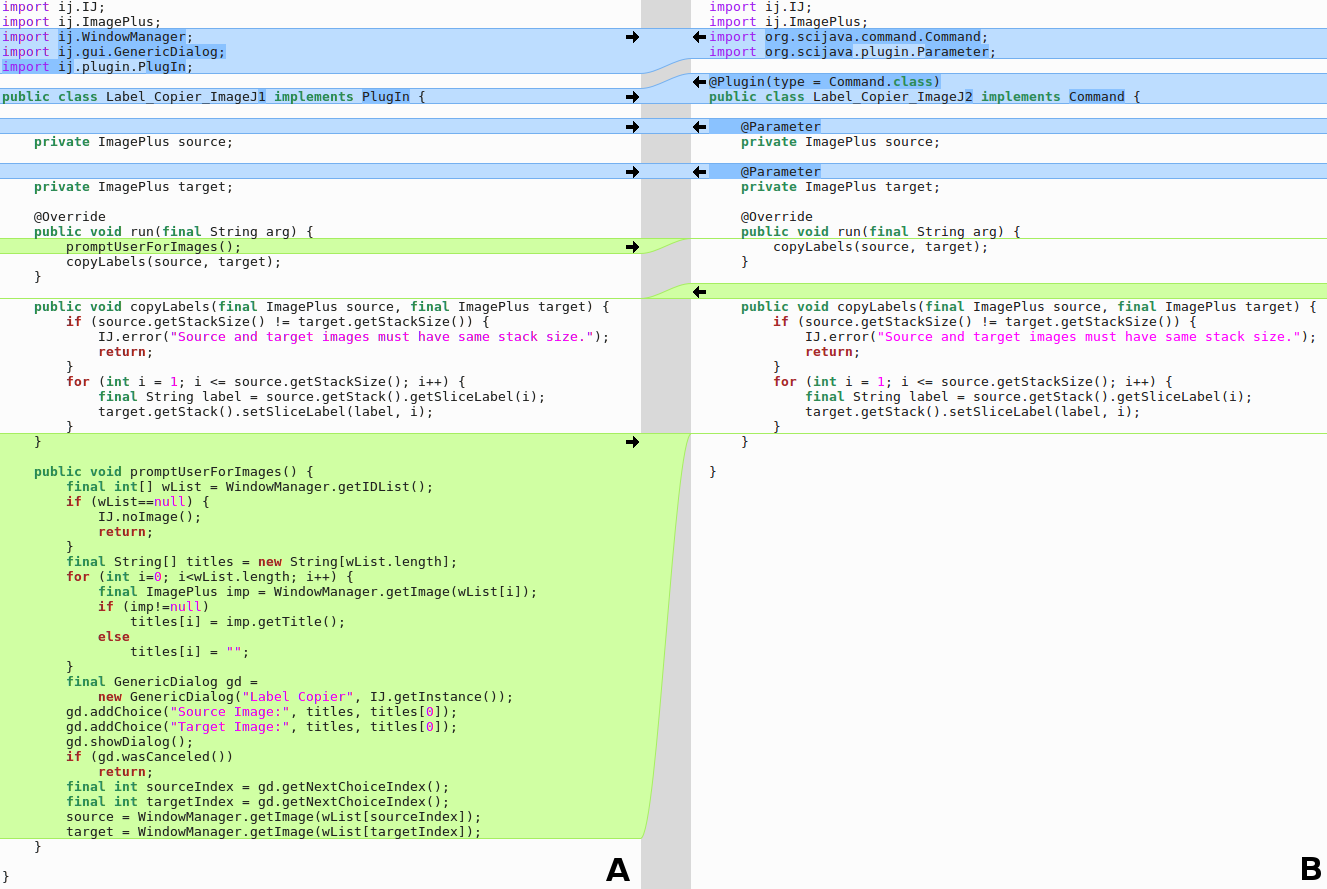}
    \begin{flushleft}
      \footnotesize
      Panel A (left) shows an ImageJ 1.x implementation of a plugin that copies
      slice labels from one image to another, as chosen by the user. Panel B
      (right) shows the same plugin written using the SciJava declarative
      command syntax. Changed lines are highlighted in blue, new lines in
      green. The actual operation (the \texttt{copyLabels} method) is
      identical, but the routine for selecting which images to process is no
      longer necessary.
    \end{flushleft}
  \end{figure}

\textbf{Sensible defaults.} A key example of reasonable default behavior is the
SciJava conversion framework with its specialized \texttt{Converter} plugin
type. \texttt{Converter} plugins define useful type substitutions that would
not normally be allowed by the Java type hierarchy. For example, conversion of
ImageJ \texttt{Dataset} objects to and from other paradigms (\acrshort{matlab}
arrays, \acrshort{itk} images, etc.) is facilitated by \texttt{Converter}
plugins which encapsulate the logic for each conversion case. The framework
then uses the converters automatically when modules are executed, so when a
user says e.g. ``run this \acrshort{itk} algorithm on this dataset I opened''
everything ``just works'' without the user needing to perform an explicit
manual conversion. From a software development perspective, this scheme lets
ImageJ2 retain the advantages of strong typing while escaping the corresponding
shackles that often accompany it.

The ImageJ Ops library provides another illustration of ImageJ2's sensible
defaults structured in layers. While every op in the system is a dynamically
callable plugin, the core Ops library also organizes its built-in operations
into a centrally accessible collection of type-safe namespaces in a standard
Java \acrshort{api} explorable from \acrshortpl{ide} like Eclipse e.g. via its
Content Assist functions. This structure also provides an elegant and
easy-to-read syntax for calling ops in script-driven workflows (see Figure 7 in
``Compatibility'' below).

\textbf{Automatic updates.} In ImageJ 1.x, plugin installation requires users
to download a \acrfull{jar} or Java class file and place it within the ImageJ
plugins folder. Updating an installed plugin essentially requires repeating the
manual installation steps, which is both tedious and error-prone. Developers
have to manually manage their plugin's dependencies, which in practice leads to
users receiving cryptic error messages when multiple plugins require
incompatible component versions. Even worse, some developers then make
suboptimal design decisions to work around this difficulty, such as
reimplementing functionality already provided by well-tested third party
libraries, and/or creating so-called ``uber-\acrshortpl{jar}'' which lump
together the dependencies into intractable bundles \cite{imagej_uber_jar}.

The ImageJ Updater vastly simplifies this process by informing users
automatically when a new plugin version is available and enabling single-click
upgrades to the latest version of all components. On the development side, the
use of Maven by the ImageJ2 and \acrshort{fiji} projects provides a clear best
practice for managing dependencies in a consistent way, which reduces the
chance of broken end-user installations.

ImageJ's support for multiple update sites makes it feasible for community
developers and third parties to create their own update sites from which users
can pick and choose, without them needing to become a part of the core ImageJ
or \acrshort{fiji} distribution. This distributed model of update sites fits in
very well with the community-driven aspects of ImageJ, dramatically lowering
the barrier for sharing effort. This capability is made even more potent by the
Personal Update Sites feature, which lets users link their ImageJ wiki account
to their own personal web space. Furthermore, the Updater derives its initial
list of available update sites from the ``List of update sites'' wiki page of
the ImageJ website \cite{imagej_list_of_update_sites}---plugin developers can
edit this wiki page in the same way as the rest of the ImageJ website, in order
to make their site automatically available to all users of ImageJ. Editing this
page is not mandatory, however; users can also manually edit their ImageJ
installation's list of available update sites---e.g., if their organization
offers an internally managed update site for plugins specific to their
institute.

Although manual plugin installation is still supported in ImageJ2, many
organizations have already publicized their own update sites, and thanks to the
Updater together with the ImageJ Legacy layer, all of the plugins served from
those sites are easily accessible within ImageJ2. This has helped to focus the
\acrshort{fiji} project on its original goal of being a curated collection of
plugins facilitating scientific image analysis. In addition to \acrshort{fiji},
hundreds of third-party update sites are available, such as \acrshort{loci},
\acrfull{bar}, BioVoxxel, the Stowers Institute, and the BioImaging and Optics
platform of the \acrfull{epfl}, most of which are served from the centrally
managed Personal Update Sites server \cite{imagej_sites}.

\subsubsection*{Performance}

ImageJ2 is engineered to accommodate the growing size and complexity of image
data. Although performance has been a serious design consideration, we believe
in aggressive performance optimization only on an as-needed basis as software
components stabilize and mature \cite{premature_optimization}. By designing a
robust framework that allows for specialization at every level, we avoid
compromising design for the sake of incremental performance gains, while
empowering developers to optimize when necessary. Furthermore, the fact that
ImageJ2 maintains 100\% backwards compatibility with ImageJ 1.x (see
``Compatibility'' below) means that existing high-performance image processing
approaches continue to work as is, even if they do not benefit from ImageJ2's
architectural improvements.

\textbf{Efficiency.} The time performance of ImageJ2 data structures is
generally consistent with those of ImageJ 1.x. The core of performance in
ImageJ2 hinges on the efficiency of the various ImgLib2 containers. We have run
benchmarks comparing the time performance of iteration and access on ImgLib2
image structures with that of ImageJ 1.x images, as well as compared to raw
array access \cite{imglib2_benchmarks}. We found that thanks to Java's
\acrfull{jit}, ImgLib2 is highly comparable to ImageJ 1.x in these regards
(Supplemental Figure 2).

When time performance is dominated by the overhead of looping itself, some
ImgLib2 container types such as cell images may be measurably slower to iterate
and access. However, this loop overhead is generally very small, and for most
container types (e.g., array and planar images) the \acrshort{jit} quickly
optimizes the code to equal the speed of raw array access. Hence, for
non-trivial image processing operations which take significant time to compute
per sample, overall time performance converges across all data structures and
container types, ImageJ 1.x and ImageJ2 alike.

The advantages of ImgLib2's type- and container-agnostic algorithm development
outweigh any minor differences in time performance, saving developer time and
effort via simpler, more maintainable code. Furthermore, ImgLib2's more
comprehensive set of image types (Supplemental Table 3) make it easier to
optimize for space efficiency. For example, an image sequence recorded using a
12-bit detector requires 16 bits per sample in ImageJ 1.x, whereas ImageJ2 can
pack that data without wasted bits using ImgLib2's \texttt{uint12} data type,
resulting in a 33\% increase in space efficiency.

Relatedly, the design of the ImageJ Ops library realizes these same efficiency
advantages. End user scripts invoke ops by name and arguments, and the Ops
matching algorithm takes care of selecting the implementation optimized for
those arguments. This scheme enables image processing algorithms to be written
once, then automatically benefit from future performance optimizations without
explicit case logic.

  \begin{figure}[h]
    \caption{Benchmarks of a simple addition operation with ImageJ Ops and
    ImageJ 1.x.}
    \includegraphics[width=4.75in]{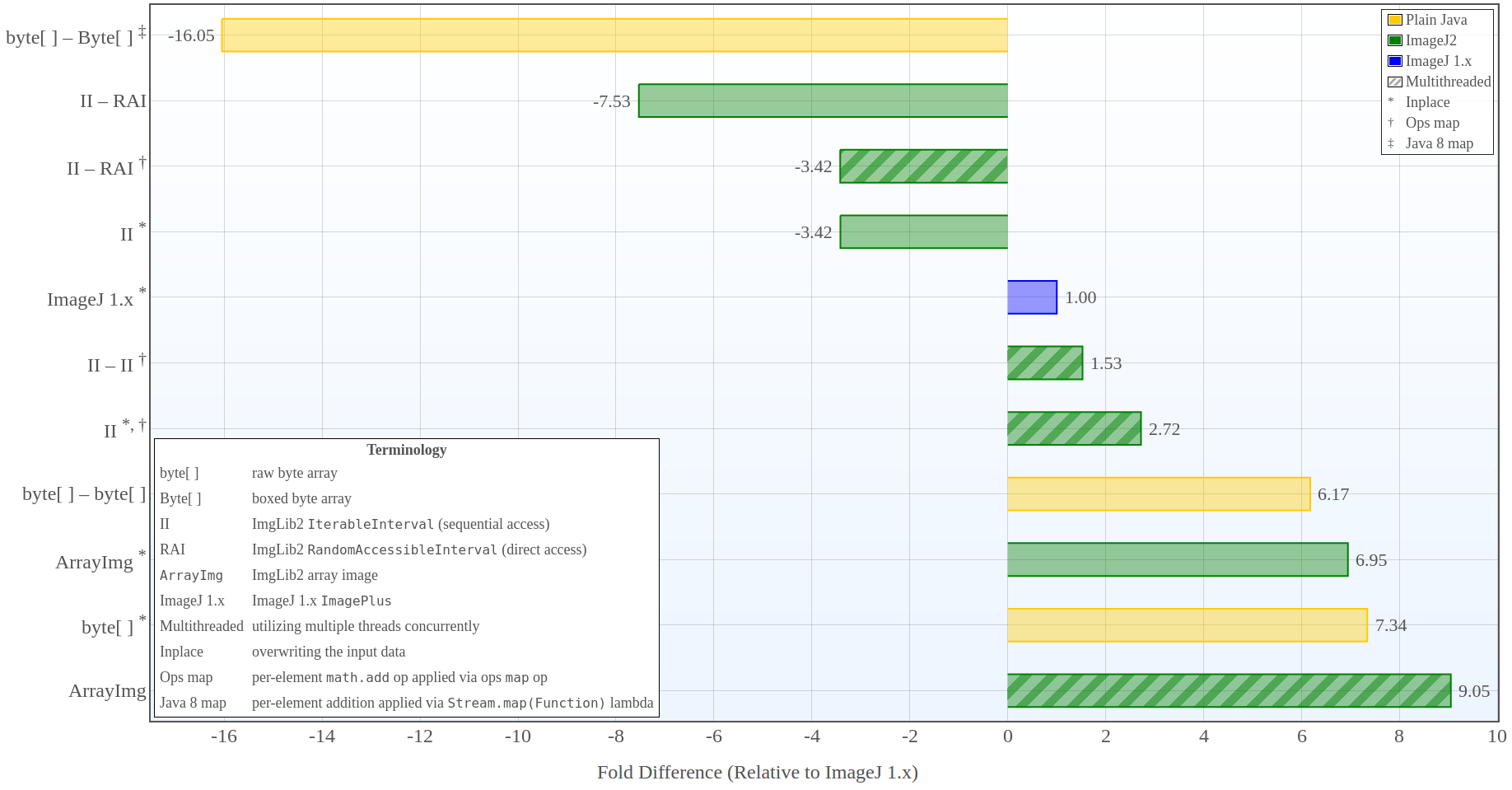}
    \begin{flushleft}
      \footnotesize
      Time performance comparison of simple addition operations between raw
      Java array manipulation, various \texttt{math.add} operations of ImageJ
      Ops, and ImageJ 1.x's \textit{Process $\triangleright$ Math
      $\triangleright$ Add\ldots} command. Benchmarks were run for 20 rounds on
      randomly generated \texttt{uint8} noise images dimensioned $15,000 \times
      15,000$, using the JUnit Benchmarks framework, on a MacBook Pro (Retina,
      15-inch, Mid 2015) running \acrshort{macos} Sierra 10.12 with 2.5 GHz
      Intel Core i7 processor and 16 GB 1600 MHz DDR3 memory. Positive numbers
      are fold faster, negative numbers are fold slower. The routines which
      produced these results can be found in the ImageJ Ops test code, in the
      \texttt{AddOpBenchmarkTest} class of the
      \texttt{net.imagej.ops.benchmark} package.
    \end{flushleft}
  \end{figure}

To validate this approach, we benchmarked the core Ops library's
\texttt{math.add} operations which add a constant value to each element of an
image (Figure 6). As evidenced by the results, the inplace ImageJ 1.x version
of this operation (the \textit{Add\ldots} command under Math in the Process
menu) performs much better than some generalized op implementations
(\texttt{II} source to \texttt{RAI} destination) which work on all image
types---but the optimized ops outperform it, with the single-threaded inplace
\texttt{ArrayImg} op finishing 7 times faster, and the multithreaded version
finishing 9 times faster. While some of this gain is likely due to the expense
of ImageJ 1.x's bounds checking, it is also evident from the results that the
optimized ops are comparable in efficiency to operations on raw primitive
arrays.

\textbf{Scalability.} As discussed in ``Functionality'' above, ImageJ 1.x is
fundamentally limited to XY image planes of less than $2^{31}$ pixels due to
its use of one Java primitive array per plane, and to the size of available
computer memory for many of its image processing operations. In contrast,
ImageJ2 has been engineered at every level to support scalable image processing
using cell images which are cached to and from mass storage on demand.
ImageJ2's careful separation of concerns and enhanced command line parameter
handling enable ImageJ to run headless on remote servers, opening up a wide
array of possibilities for scalable computation. The \acrshort{scifio} library
enables direct access into image data samples, so that image data many orders
of magnitude larger than available computer memory can be systematically
processed on an individual workstation or using a cluster. And thanks to
visualization tools like the BigDataViewer plugin \cite{bigdataviewer}, which
is also built on ImgLib2 cell images, it is now realistic to quickly visualize
and explore such massive datasets.

\subsection*{Compatibility}
The ImageJ2 project, by design, enables software developers to use a
combination of ImageJ 1.x and ImageJ2 features. Many ImageJ-based tools and
plugins continue to rely on ImageJ 1.x data structures, with varying levels of
dependence on ImageJ2 and the SciJava framework (Supplemental Figure 3). A few
examples include:

\begin{itemize}
  \item \textbf{TrackMate} \cite{trackmate}, a plugin for object
    identification and tracking, has been used to achieve a semi-automated
    workflow for \acrfull{dna} double-strand break-induced telomere mobility
    quantitative analysis \cite{science_trackmate}, as well as for
    \textit{Caenorhabditis-elegans} lineage analysis during light-induced
    damage, recruitment of \acrfull{nemo} clusters in fibroblasts after
    \acrfull{il1} stimulation, and clathrin-mediated endocytosis analysis in
    plant cells \cite{trackmate}.
  \item \textbf{\acrfull{mamut}}, a tool for the annotation of massive,
    multi-view data, has been used for reconstruction of the complete cell
    lineage of an outgrowing thoracic limb of the crustacean \textit{Parhyale
    hawaiensis}, with single-cell resolution \cite{science_mamut}.
  \item \textbf{Multiview Reconstruction} \cite{multiview_2010,
    multiview_2014}, a pipeline for registering multi-angle \acrshort{3d}
    volumes and visualizing them using the BigDataViewer \cite{bigdataviewer},
    is commonly part of experimental protocols for light sheet fluorescence
    microscopy, and has been used to analyze zebrafish embryo eye development
    \cite{science_multiview_1}, as well as directional movement of
    cerebrospinal fluid in zebrafish larvae during developmental neurogenesis
    \cite{science_multiview_2}.
  \item \textbf{Sholl Analysis} \cite{sholl_analysis} and
    \textbf{\acrfull{snt}} \cite{simple_neurite_tracer}, plugins for
    quantifying traced structures such as neurites, have been used to analyze
    dendritic morphology of the amygdala and hippocampus in
    conventionally-colonized versus germ-free mice \cite{science_sholl_1},
    morphologies of retinal ganglion cells from neural retina on
    \acrfull{plga} scaffold \cite{science_sholl_2}, as well as dendritic
    complexity and arborization in absence of $\alpha$2-chimaerin, a key
    regulator of Rac1-dependent signaling \cite{science_sholl_3}.
\end{itemize}

It is thanks to the ImageJ Legacy and IJ1-patcher components that the community
can blend the usage of ImageJ 1.x and ImageJ2 functionality, cherry-picking the
best from each world to accomplish their image analysis tasks. For example,
parameterized ImageJ commands and scripts may continue to use ImageJ 1.x data
structures and plugins as needed, while taking advantage of ImageJ2
functionality as appropriate (Figure 7), and declaring and populating input
values with less boilerplate code (see Figure 5 in ``Usability'' above).

  \begin{figure}[h]
    \caption{A mixed-world ImageJ 1.x + ImageJ2 script.}
    \includegraphics[width=4.75in]{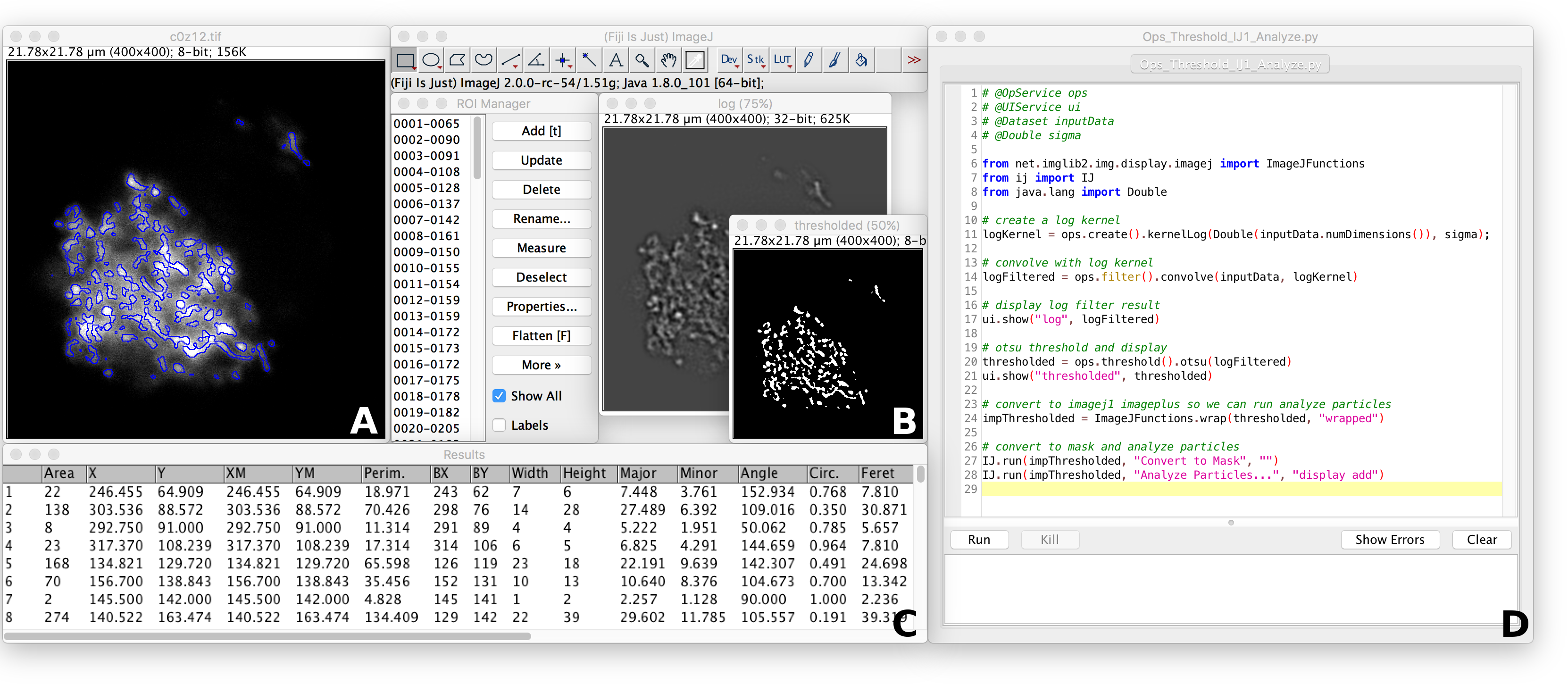}
    \begin{flushleft}
      \footnotesize
      This example Python script (panel D) uses ImageJ Ops to preprocess a
      confocal image and perform an automatic thresholding (panel B). ImageJ
      1.x's Analyze Particles routine is then called to isolate (panel A) and
      measure (panel C) foreground objects. Script contributed by Brian
      Northan, True North Intelligent Algorithms LLC. This script is available
      within ImageJ as a sample from the Tutorials submenu of the Script
      Editor's Templates menu.
    \end{flushleft}
  \end{figure}

As ImageJ2 continues to mature, usage of ImageJ 1.x functionality will be
increasingly replaced with more powerful ImageJ2 equivalents: image processing
algorithms built on ImageJ Ops, data format plugins built on \acrshort{scifio},
block-based cell images using ImgLib2, N-dimensional ROIs, metadata-rich
images, nonlinear registration transforms, etc. However, this process is both
lengthy and necessarily incomplete: migrating ImageJ 1.x core functionality
alone is a years-long process as the ImageJ2 \acrshortpl{api} continue to
evolve, mature and stabilize---and there are countless other useful plugins and
scripts in the wild, some of which will never be updated to the new
\acrshortpl{api}. Meanwhile, development of ImageJ 1.x also progresses, with
users continuing to request bug fixes and new features within its intended
scope. As such, the importance of a robust transitional strategy for migrating
from ImageJ 1.x to ImageJ2 cannot be overstated.

Although the development of ImageJ2 has necessitated reimplementation of
ImageJ 1.x functionality, maintaining backwards compatibility with ImageJ 1.x
will remain a fundamental goal. Abandoning or ignoring ImageJ 1.x would have
been a significant disservice to the community, causing a rift to the detriment
of all parties. Our efforts to enable incremental migration from ImageJ 1.x to
ImageJ2 allow the two projects to continue developing in tandem, with new
features in each reaching a unified ImageJ community.

\subsection*{Community}

Ultimately, the goal of ImageJ is to enable scientific collaboration and
achievement, which requires community management as much as code management.
ImageJ 1.x leverages open-source code, a public web site and a mailing list to
support discussion and contributions from people across the globe. However, it
follows a centralized ``cathedral'' development model, rather than a
community-driven ``bazaar'' style model \cite{cathedral_bazaar}, with its
primary resources and scalability fundamentally limited by a single
``gatekeeper.'' For ImageJ's continued success and growth, it is critical to
renew its focus on partnership and communication with the community
\cite{imagej_communication}.

\textbf{Online resources.} The centrally organized documentation of ImageJ2
takes the form of a collaborative wiki \cite{imagej_web_site} with over 800
articles: a ``world-writable'' location for both users and developers to learn
about and contribute to ImageJ. The wiki is complemented by the ImageJ Forum
\cite{imagej_forum}, a powerful, friendly and universally accessible discussion
channel driven by the excellent Discourse software, which is engineered to
encourage civil interaction \cite{discourse}. Finally, ImageJ's source code and
issue tracking via GitHub completes the community-centric approach for managing
and discussing changes and improvements.

These resources together enable project developers to clearly communicate the
expectations and norms surrounding plugin development, contribution,
maintenance and support, empowering users to easily see who is responsible for
each plugin as well as its support and development status \cite{imagej_team},
outstanding bugs \cite{imagej_issues} and future plans \cite{imagej_roadmap}.
This is a critical service for the community: it is not enough to provide a
convenient way for people to publish, share and consume extensions---we have
learned from experience that there must be a social framework in place for
managing and understanding the software development lifecycle of the myriad
community efforts.

\subsection*{Future directions}
ImageJ is more than a single application: it is a living ecosystem of
scientific exchange. As acquisition technology continues to advance, there will
always be a need for new development and maintenance within ImageJ. There are
still key technical tasks remaining for ImageJ2 to achieve stability, as well
as new directions made possible by the ImageJ2 platform which we are excited to
explore:

\begin{itemize}
  \item Finalize the ImageJ Common data model to support extensible attachment
    of metadata, including spatial metadata, that respond robustly to image
    processing operations such as transformation.
  \item Extend ImgLib2's N-dimensional \acrshort{roi} interfaces to cover all
    needed cases, including all \acrshort{roi} types supported by ImageJ 1.x,
    all \acrshort{roi} types supported by \acrshort{omero}, and any additional
    \acrshort{roi} types available in other image-oriented software packages
    for which integration with ImageJ is pursued.
  \item Update the core SciJava \acrshort{io} mechanism to be plugin-driven
    for improved extensibility of data source location.
  \item Generalize \acrshort{scifio}'s planar model to operate on arbitrary
    ``blocks'' at a fundamental level.
  \item Retire the custom C-based ImageJ desktop application launcher,
    migrating to the industry-standard application bundling of JavaFX.
  \item Complete our ongoing effort to automate the documentation regarding
    development and maintenance responsibility for every core component of the
    ImageJ ecosystem, including \acrshort{fiji} components \cite{imagej_team}.
  \item Develop our ImageJ-based \acrshort{rest} image server prototype toward
    production use, providing a common language- and implementation-independent
    \acrshort{api}.
  \item Implement a web \acrshort{ui} built on the \acrshort{rest} image
    server.
  \item Improve the ImageJ Updater user interface to be more user friendly, so
    that users can more easily cherry-pick extensions of interest from each
    update site.
  \item Expand the list of built-in ImageJ Ops with additional image processing
    and analysis routines, including Deep Learning approaches
    \cite{deep_learning} and novel algorithms from computer vision and
    statistics.
  \item Continue building bridges between ImageJ and other image processing
    frameworks such as \acrshort{opencv} \cite{opencv} and scikit-image
    \cite{scikit_image}.
  \item Integrate support for cloud computing frameworks such as Apache Spark
    \cite{apache_spark} running on platforms such as Amazon Web Services
    \cite{aws}.
  \item Continue supporting community requests for bug fixes, new features and
    image analysis advice.
  \item Continue migrating ImageJ resources into the main ImageJ wiki website,
    including the ImageJ User Guide \cite{imagej_user_guide}, ImageJ 1.x
    documentation \cite{imagej1_docs} and \acrfull{list}'s ImageJ Information
    and Documentation Portal \cite{imagej_docu}.
  \item Redesign ImageJ's bug submission system such that users can
    automatically submit an issue report to the correct location online
    whenever something goes wrong in the software.
\item Continue listening to, and working with, the user and developer
  community.
\end{itemize}

A detailed breakdown and discussion of each specific issue can be found on
GitHub, searchable from the unified ImageJ Search portal \cite{imagej_search}.


\section*{Conclusions}
Based on feedback from the existing ImageJ community, we have over the last
several years been designing and implementing ImageJ2, a radically improved
application that employs best practices and proven software engineering
approaches. ImageJ2 directly addresses two major needs, supporting applications
where: 1) the underlying ImageJ data engine was not sufficient to analyze
modern datasets; and 2) the lack of an underlying robust software design
impeded the addition of new functionality. This overhaul of ImageJ transforms
it into not only a powerful and flexible image processing and analysis
application in its own right, but also a framework for interoperability between
a plethora of external image visualization and analysis programs. ImageJ2
strengthens ImageJ's utility as a platform for scientific image analysis by: 1)
generalizing the ImageJ data model; 2) introducing a robust architecture
instrumental in building bridges across a range of other image processing
tools; 3) remaining open source and cross-platform with permissive licensing,
enabling continued widespread adoption and extension; 4) building on the huge
collection of existing ImageJ plugins while enabling the creation of new
plugins with more powerful features; and 5) leveraging a correspondingly large
and diverse community to foster a collaborative and interdisciplinary project
that facilitates the collective advancement of science.


\printglossary[title=List of abbreviations,type=\acronymtype,style=long]


\begin{backmatter}

\section*{Declarations}

\subsection*{Ethics approval and consent to participate}
Not applicable.

\subsection*{Consent for publication}
Not applicable.

\subsection*{Availability of data and material}

\textbf{Project name:} ImageJ\\
\textbf{Project home page:} https://imagej.net/\\
\textbf{Archived version:} net.imagej:imagej:2.0.0-rc-55\\
\textbf{Operating system(s):} Platform independent\\
\textbf{Programming language:} Java\\
\textbf{Other requirements:} Java 8 or higher\\
\textbf{License:} Simplified (2-clause) \acrfull{bsd}\\
\textbf{Any restrictions to use by non-academics:} None

All data generated or analyzed during this study are included in this
published article.

\subsection*{Competing interests}
  The authors declare that they have no competing interests.

\subsection*{Funding}
  ImageJ2 was funded from 2010 through 2012 by the \acrfull{nigms} through the
  American Recovery and Reinvestment Act of 2009 \acrshort{nih} Research and
  Research Infrastructure ``Grand Opportunities'' Grant, ``Hardening'' of
  Biomedical Informatics/Computing Software for Robustness and Dissemination
  (Ref: RC2 GM092519-01), as well as a Wellcome Trust Strategic Award (Ref:
  095931). \acrshort{scifio} was funded by the National Science Foundation,
  award number 1148362. ImageJ2 projects were also funded by internal funding
  from the \acrlong{loci}, and the Morgridge Institute for Research.

\subsection*{Authors' contributions}
  CTR acted as the technical lead of the ImageJ2 project and primary architect
  of ImageJ2's software architecture. JS migrated key portions of
  \acrshort{fiji} into ImageJ2, including the Launcher and Updater components,
  and advised and improved upon many architectural aspects of ImageJ2,
  particularly the legacy layer. MCH served as the lead \acrshort{scifio}
  developer and contributed to all layers of the ImageJ software stack. BED
  developed substantial portions of the ImageJ2 codebase, including much of the
  legacy layer for backwards compatibility, prototype versions of ImageJ Ops
  for numerical processing, and many command implementations. AEW contributed
  to ImageJ Ops and the ImageJ-\acrshort{omero} integration layer. ETA made
  extensive edits and improvements to the manuscript. Lastly, as the primary
  principal investigator of ImageJ2, KWE directed and advised on all aspects of
  the project, including development directions and priorities. All authors
  contributed to, read, and approved the final manuscript.

\subsection*{Acknowledgements}
  Many people have contributed to the development of ImageJ2 on both technical
  and leadership levels. In particular, the authors gratefully thank and
  acknowledge the efforts of (in alphabetical order): Ignacio Arganda-Carreras,
  Michael Berthold, Tim-Oliver Buchholz, Jean-Marie Burel, Albert Cardona, Anne
  Carpenter, Christian Dietz, Richard Domander, Jan Eglinger, Gabriel Einsdorf,
  Adam Fraser, Aivar Grislis, Ulrik Günther, Robert Haase, Jonathan Hale, Kyle
  Harrington, Grant Harris, Stefan Helfrich, Martin Horn, Florian Jug, Lee
  Kamentsky, Gabriel Landini, Rick Lentz, Melissa Linkert, Mark Longair, Kevin
  Mader, Hadrien Mary, Kota Miura, Birgit Möller, Cyril Mongis, Josh Moore,
  Alec Neevel, Brian Northan, Rudolf Oldenbourg, Aparna Pal, Tobias Pietzsch,
  Stefan Posch, Stephan Preibisch, Loïc Royer, Stephan Saalfeld, Benjamin
  Schmid, Daniel Seebacher, Jason Swedlow, Jean-Yves Tinevez, Pavel Tomancak,
  Jay Warrick, Leon Yang, Yili Zhao and Michael Zinsmaier. We also thank the
  entire ImageJ community, especially those who contributed patch submissions,
  use cases, feature requests, and bug reports. A special thanks to Wayne
  Rasband for his tireless work on, and continuing maintenance of, ImageJ 1.x
  for these many years. Finally, our deep thanks to the \acrshort{nih}, whose
  initial funding of ImageJ2 in 2009 was instrumental in launching the project,
  as well as to all funding agencies and organizations who have supported the
  project's continued development \cite{imagej_funding}.

\section*{Endnotes}
  None.


\bibliographystyle{bmc-mathphys} 
\bibliography{imagej2}      


\begin{thebibliography}{159}
\ifx \bisbn   \undefined \def \bisbn  #1{ISBN #1}\fi
\ifx \binits  \undefined \def \binits#1{#1}\fi
\ifx \bauthor  \undefined \def \bauthor#1{#1}\fi
\ifx \batitle  \undefined \def \batitle#1{#1}\fi
\ifx \bjtitle  \undefined \def \bjtitle#1{#1}\fi
\ifx \bvolume  \undefined \def \bvolume#1{\textbf{#1}}\fi
\ifx \byear  \undefined \def \byear#1{#1}\fi
\ifx \bissue  \undefined \def \bissue#1{#1}\fi
\ifx \bfpage  \undefined \def \bfpage#1{#1}\fi
\ifx \blpage  \undefined \def \blpage #1{#1}\fi
\ifx \burl  \undefined \def \burl#1{\textsf{#1}}\fi
\ifx \doiurl  \undefined \def \doiurl#1{\textsf{#1}}\fi
\ifx \betal  \undefined \def \betal{\textit{et al.}}\fi
\ifx \binstitute  \undefined \def \binstitute#1{#1}\fi
\ifx \binstitutionaled  \undefined \def \binstitutionaled#1{#1}\fi
\ifx \bctitle  \undefined \def \bctitle#1{#1}\fi
\ifx \beditor  \undefined \def \beditor#1{#1}\fi
\ifx \bpublisher  \undefined \def \bpublisher#1{#1}\fi
\ifx \bbtitle  \undefined \def \bbtitle#1{#1}\fi
\ifx \bedition  \undefined \def \bedition#1{#1}\fi
\ifx \bseriesno  \undefined \def \bseriesno#1{#1}\fi
\ifx \blocation  \undefined \def \blocation#1{#1}\fi
\ifx \bsertitle  \undefined \def \bsertitle#1{#1}\fi
\ifx \bsnm \undefined \def \bsnm#1{#1}\fi
\ifx \bsuffix \undefined \def \bsuffix#1{#1}\fi
\ifx \bparticle \undefined \def \bparticle#1{#1}\fi
\ifx \barticle \undefined \def \barticle#1{#1}\fi
\ifx \bconfdate \undefined \def \bconfdate #1{#1}\fi
\ifx \botherref \undefined \def \botherref #1{#1}\fi
\ifx \url \undefined \def \url#1{\textsf{#1}}\fi
\ifx \bchapter \undefined \def \bchapter#1{#1}\fi
\ifx \bbook \undefined \def \bbook#1{#1}\fi
\ifx \bcomment \undefined \def \bcomment#1{#1}\fi
\ifx \oauthor \undefined \def \oauthor#1{#1}\fi
\ifx \citeauthoryear \undefined \def \citeauthoryear#1{#1}\fi
\ifx \endbibitem  \undefined \def \endbibitem {}\fi
\ifx \bconflocation  \undefined \def \bconflocation#1{#1}\fi
\ifx \arxivurl  \undefined \def \arxivurl#1{\textsf{#1}}\fi
\csname PreBibitemsHook\endcsname

\bibitem{imagej_history}
\begin{barticle}
\bauthor{\bsnm{Schneider}, \binits{C.A.}},
\bauthor{\bsnm{Rasband}, \binits{W.S.}},
\bauthor{\bsnm{Eliceiri}, \binits{K.W.}}, \betal:
\batitle{Nih image to imagej: 25 years of image analysis}.
\bjtitle{Nat methods}
\bvolume{9}(\bissue{7}),
\bfpage{671}--\blpage{675}
(\byear{2012})
\end{barticle}
\endbibitem

\bibitem{imagej_review}
\begin{botherref}
\oauthor{\bsnm{Arena}, \binits{E.T.}},
\oauthor{\bsnm{Rueden}, \binits{C.T.}},
\oauthor{\bsnm{Hiner}, \binits{M.C.}},
\oauthor{\bsnm{Wang}, \binits{S.}},
\oauthor{\bsnm{Yuan}, \binits{M.}},
\oauthor{\bsnm{Eliceiri}, \binits{K.W.}}:
Quantitating the cell: turning images into numbers with imagej.
Wiley Interdisciplinary Reviews: Developmental Biology
(2016)
\end{botherref}
\endbibitem

\bibitem{imagej_contributors}
\begin{botherref}
ImageJ Contributors.
\url{https://imagej.net/Contributors}
\end{botherref}
\endbibitem

\bibitem{bioimage_informatics}
\begin{barticle}
\bauthor{\bsnm{Peng}, \binits{H.}}:
\batitle{Bioimage informatics: a new area of engineering biology}.
\bjtitle{Bioinformatics}
\bvolume{24}(\bissue{17}),
\bfpage{1827}--\blpage{1836}
(\byear{2008})
\end{barticle}
\endbibitem

\bibitem{bioimaging_software_review}
\begin{barticle}
\bauthor{\bsnm{Eliceiri}, \binits{K.W.}},
\bauthor{\bsnm{Berthold}, \binits{M.R.}},
\bauthor{\bsnm{Goldberg}, \binits{I.G.}},
\bauthor{\bsnm{Ib{\'a}{\~n}ez}, \binits{L.}},
\bauthor{\bsnm{Manjunath}, \binits{B.S.}},
\bauthor{\bsnm{Martone}, \binits{M.E.}},
\bauthor{\bsnm{Murphy}, \binits{R.F.}},
\bauthor{\bsnm{Peng}, \binits{H.}},
\bauthor{\bsnm{Plant}, \binits{A.L.}},
\bauthor{\bsnm{Roysam}, \binits{B.}}, \betal:
\batitle{Biological imaging software tools}.
\bjtitle{Nature methods}
\bvolume{9}(\bissue{7}),
\bfpage{697}--\blpage{710}
(\byear{2012})
\end{barticle}
\endbibitem

\bibitem{bioimaging_cell_biology}
\begin{barticle}
\bauthor{\bsnm{Swedlow}, \binits{J.R.}},
\bauthor{\bsnm{Eliceiri}, \binits{K.W.}}:
\batitle{Open source bioimage informatics for cell biology}.
\bjtitle{Trends in cell biology}
\bvolume{19}(\bissue{11}),
\bfpage{656}--\blpage{660}
(\byear{2009})
\end{barticle}
\endbibitem

\bibitem{micro_manager_2010}
\begin{botherref}
\oauthor{\bsnm{Edelstein}, \binits{A.}},
\oauthor{\bsnm{Amodaj}, \binits{N.}},
\oauthor{\bsnm{Hoover}, \binits{K.}},
\oauthor{\bsnm{Vale}, \binits{R.}},
\oauthor{\bsnm{Stuurman}, \binits{N.}}:
Computer control of microscopes using $\mu$manager.
Current protocols in molecular biology,
14--20
(2010)
\end{botherref}
\endbibitem

\bibitem{micro_manager_2014}
\begin{botherref}
\oauthor{\bsnm{Edelstein}, \binits{A.D.}},
\oauthor{\bsnm{Tsuchida}, \binits{M.A.}},
\oauthor{\bsnm{Amodaj}, \binits{N.}},
\oauthor{\bsnm{Pinkard}, \binits{H.}},
\oauthor{\bsnm{Vale}, \binits{R.D.}},
\oauthor{\bsnm{Stuurman}, \binits{N.}}:
Advanced methods of microscope control using $\mu$manager software.
Journal of biological methods
\textbf{1}(2)
(2014)
\end{botherref}
\endbibitem

\bibitem{bisque}
\begin{barticle}
\bauthor{\bsnm{Kvilekval}, \binits{K.}},
\bauthor{\bsnm{Fedorov}, \binits{D.}},
\bauthor{\bsnm{Obara}, \binits{B.}},
\bauthor{\bsnm{Singh}, \binits{A.}},
\bauthor{\bsnm{Manjunath}, \binits{B.}}:
\batitle{Bisque: a platform for bioimage analysis and management}.
\bjtitle{Bioinformatics}
\bvolume{26}(\bissue{4}),
\bfpage{544}--\blpage{552}
(\byear{2010})
\end{barticle}
\endbibitem

\bibitem{omero}
\begin{barticle}
\bauthor{\bsnm{Allan}, \binits{C.}},
\bauthor{\bsnm{Burel}, \binits{J.-M.}},
\bauthor{\bsnm{Moore}, \binits{J.}},
\bauthor{\bsnm{Blackburn}, \binits{C.}},
\bauthor{\bsnm{Linkert}, \binits{M.}},
\bauthor{\bsnm{Loynton}, \binits{S.}},
\bauthor{\bsnm{MacDonald}, \binits{D.}},
\bauthor{\bsnm{Moore}, \binits{W.J.}},
\bauthor{\bsnm{Neves}, \binits{C.}},
\bauthor{\bsnm{Patterson}, \binits{A.}}, \betal:
\batitle{Omero: flexible, model-driven data management for experimental
  biology}.
\bjtitle{Nature methods}
\bvolume{9}(\bissue{3}),
\bfpage{245}--\blpage{253}
(\byear{2012})
\end{barticle}
\endbibitem

\bibitem{icy}
\begin{barticle}
\bauthor{\bsnm{De~Chaumont}, \binits{F.}},
\bauthor{\bsnm{Dallongeville}, \binits{S.}},
\bauthor{\bsnm{Chenouard}, \binits{N.}},
\bauthor{\bsnm{Herv{\'e}}, \binits{N.}},
\bauthor{\bsnm{Pop}, \binits{S.}},
\bauthor{\bsnm{Provoost}, \binits{T.}},
\bauthor{\bsnm{Meas-Yedid}, \binits{V.}},
\bauthor{\bsnm{Pankajakshan}, \binits{P.}},
\bauthor{\bsnm{Lecomte}, \binits{T.}},
\bauthor{\bsnm{Le~Montagner}, \binits{Y.}}, \betal:
\batitle{Icy: an open bioimage informatics platform for extended reproducible
  research}.
\bjtitle{Nature methods}
\bvolume{9}(\bissue{7}),
\bfpage{690}--\blpage{696}
(\byear{2012})
\end{barticle}
\endbibitem

\bibitem{bioimagexd}
\begin{barticle}
\bauthor{\bsnm{Kankaanp{\"a}{\"a}}, \binits{P.}},
\bauthor{\bsnm{Paavolainen}, \binits{L.}},
\bauthor{\bsnm{Tiitta}, \binits{S.}},
\bauthor{\bsnm{Karjalainen}, \binits{M.}},
\bauthor{\bsnm{P{\"a}iv{\"a}rinne}, \binits{J.}},
\bauthor{\bsnm{Nieminen}, \binits{J.}},
\bauthor{\bsnm{Marjom{\"a}ki}, \binits{V.}},
\bauthor{\bsnm{Heino}, \binits{J.}},
\bauthor{\bsnm{White}, \binits{D.J.}}:
\batitle{Bioimagexd: an open, general-purpose and high-throughput
  image-processing platform}.
\bjtitle{Nature methods}
\bvolume{9}(\bissue{7}),
\bfpage{683}--\blpage{689}
(\byear{2012})
\end{barticle}
\endbibitem

\bibitem{cellprofiler}
\begin{barticle}
\bauthor{\bsnm{Carpenter}, \binits{A.E.}},
\bauthor{\bsnm{Jones}, \binits{T.R.}},
\bauthor{\bsnm{Lamprecht}, \binits{M.R.}},
\bauthor{\bsnm{Clarke}, \binits{C.}},
\bauthor{\bsnm{Kang}, \binits{I.H.}},
\bauthor{\bsnm{Friman}, \binits{O.}},
\bauthor{\bsnm{Guertin}, \binits{D.A.}},
\bauthor{\bsnm{Chang}, \binits{J.H.}},
\bauthor{\bsnm{Lindquist}, \binits{R.A.}},
\bauthor{\bsnm{Moffat}, \binits{J.}}, \betal:
\batitle{Cellprofiler: image analysis software for identifying and quantifying
  cell phenotypes}.
\bjtitle{Genome biology}
\bvolume{7}(\bissue{10}),
\bfpage{100}
(\byear{2006})
\end{barticle}
\endbibitem

\bibitem{cellprofiler_2011}
\begin{barticle}
\bauthor{\bsnm{Kamentsky}, \binits{L.}},
\bauthor{\bsnm{Jones}, \binits{T.R.}},
\bauthor{\bsnm{Fraser}, \binits{A.}},
\bauthor{\bsnm{Bray}, \binits{M.-A.}},
\bauthor{\bsnm{Logan}, \binits{D.J.}},
\bauthor{\bsnm{Madden}, \binits{K.L.}},
\bauthor{\bsnm{Ljosa}, \binits{V.}},
\bauthor{\bsnm{Rueden}, \binits{C.}},
\bauthor{\bsnm{Eliceiri}, \binits{K.W.}},
\bauthor{\bsnm{Carpenter}, \binits{A.E.}}:
\batitle{Improved structure, function and compatibility for cellprofiler:
  modular high-throughput image analysis software}.
\bjtitle{Bioinformatics}
\bvolume{27}(\bissue{8}),
\bfpage{1179}--\blpage{1180}
(\byear{2011})
\end{barticle}
\endbibitem

\bibitem{knime}
\begin{botherref}
\oauthor{\bsnm{Michael}, \binits{B.}},
\oauthor{\bsnm{Nicolas}, \binits{C.}},
\oauthor{\bsnm{Fabian}, \binits{D.}},
\oauthor{\bsnm{Thomas}, \binits{G.}},
\oauthor{\bsnm{Tobias}, \binits{O.}},
\oauthor{\bsnm{Thorsten}, \binits{M.}},
\oauthor{\bsnm{Peter}, \binits{O.}},
\oauthor{\bsnm{Christoph}, \binits{S.}},
\oauthor{\bsnm{Kilian}, \binits{T.}},
\oauthor{\bsnm{Bernd}, \binits{W.}}:
Knime: The konstanz information miner.
Studies in Classification, Data Analysis, and Knowledge Organization. GfKL
(2007)
\end{botherref}
\endbibitem

\bibitem{knip}
\begin{bchapter}
\bauthor{\bsnm{Dietz}, \binits{C.}},
\bauthor{\bsnm{Berthold}, \binits{M.R.}}:
\bctitle{Knime for open-source bioimage analysis: A tutorial}.
In: \bbtitle{Focus on Bio-Image Informatics},
pp. \bfpage{179}--\blpage{197}.
\bpublisher{Springer}, \blocation{???}
(\byear{2016})
\end{bchapter}
\endbibitem

\bibitem{workflow_systems}
\begin{barticle}
\bauthor{\bsnm{Warr}, \binits{W.A.}}:
\batitle{Scientific workflow systems: Pipeline pilot and knime}.
\bjtitle{Journal of computer-aided molecular design}
\bvolume{26}(\bissue{7}),
\bfpage{801}--\blpage{804}
(\byear{2012})
\end{barticle}
\endbibitem

\bibitem{fluorender}
\begin{bchapter}
\bauthor{\bsnm{Wan}, \binits{Y.}},
\bauthor{\bsnm{Otsuna}, \binits{H.}},
\bauthor{\bsnm{Chien}, \binits{C.-B.}},
\bauthor{\bsnm{Hansen}, \binits{C.}}:
\bctitle{Fluorender: an application of 2d image space methods for 3d and 4d
  confocal microscopy data visualization in neurobiology research}.
In: \bbtitle{Pacific Visualization Symposium (PacificVis), 2012 IEEE},
pp. \bfpage{201}--\blpage{208}
(\byear{2012}).
\bcomment{IEEE}
\end{bchapter}
\endbibitem

\bibitem{vaa3d}
\begin{barticle}
\bauthor{\bsnm{Peng}, \binits{H.}},
\bauthor{\bsnm{Bria}, \binits{A.}},
\bauthor{\bsnm{Zhou}, \binits{Z.}},
\bauthor{\bsnm{Iannello}, \binits{G.}},
\bauthor{\bsnm{Long}, \binits{F.}}:
\batitle{Extensible visualization and analysis for multidimensional images
  using vaa3d}.
\bjtitle{Nature protocols}
\bvolume{9}(\bissue{1}),
\bfpage{193}--\blpage{208}
(\byear{2014})
\end{barticle}
\endbibitem

\bibitem{imagej_ecosystem}
\begin{barticle}
\bauthor{\bsnm{Schindelin}, \binits{J.}},
\bauthor{\bsnm{Rueden}, \binits{C.T.}},
\bauthor{\bsnm{Hiner}, \binits{M.C.}},
\bauthor{\bsnm{Eliceiri}, \binits{K.W.}}:
\batitle{The imagej ecosystem: An open platform for biomedical image analysis}.
\bjtitle{Molecular reproduction and development}
\bvolume{82}(\bissue{7-8}),
\bfpage{518}--\blpage{529}
(\byear{2015})
\end{barticle}
\endbibitem

\bibitem{imagej_list_of_update_sites}
\begin{botherref}
List of ImageJ Update Sites.
\url{https://imagej.net/List\_of\_update\_sites}
\end{botherref}
\endbibitem

\bibitem{scijava}
\begin{botherref}
SciJava.
\url{http://www.scijava.org/}
\end{botherref}
\endbibitem

\bibitem{imagej_survey}
\begin{botherref}
2015 ImageJ Conference Presentation: Survey.
\url{https://imagej.github.io/presentations/2015-09-03-imagej2-and-fiji/\#/6}
\end{botherref}
\endbibitem

\bibitem{software_usability}
\begin{barticle}
\bauthor{\bsnm{Carpenter}, \binits{A.E.}},
\bauthor{\bsnm{Kamentsky}, \binits{L.}},
\bauthor{\bsnm{Eliceiri}, \binits{K.W.}}:
\batitle{A call for bioimaging software usability}.
\bjtitle{Nature methods}
\bvolume{9}(\bissue{7}),
\bfpage{666}
(\byear{2012})
\end{barticle}
\endbibitem

\bibitem{hardware_is_cheap}
\begin{botherref}
\oauthor{\bsnm{Atwood}, \binits{J.}}:
Hardware Is Cheap, Programmers Are Expensive.
\url{https://blog.codinghorror.com/hardware-is-cheap-programmers-are-expensive/}
\end{botherref}
\endbibitem

\bibitem{imagej_scijava}
\begin{botherref}
SciJava.
\url{https://imagej.net/SciJava}
\end{botherref}
\endbibitem

\bibitem{imglib2}
\begin{barticle}
\bauthor{\bsnm{Pietzsch}, \binits{T.}},
\bauthor{\bsnm{Preibisch}, \binits{S.}},
\bauthor{\bsnm{Toman{\v{c}}{\'a}k}, \binits{P.}},
\bauthor{\bsnm{Saalfeld}, \binits{S.}}:
\batitle{Imglib2---generic image processing in java}.
\bjtitle{Bioinformatics}
\bvolume{28}(\bissue{22}),
\bfpage{3009}--\blpage{3011}
(\byear{2012})
\end{barticle}
\endbibitem

\bibitem{scifio}
\begin{barticle}
\bauthor{\bsnm{Hiner}, \binits{M.C.}},
\bauthor{\bsnm{Rueden}, \binits{C.T.}},
\bauthor{\bsnm{Eliceiri}, \binits{K.W.}}:
\batitle{Scifio: an extensible framework to support scientific image formats}.
\bjtitle{BMC bioinformatics}
\bvolume{17}(\bissue{1}),
\bfpage{521}
(\byear{2016})
\end{barticle}
\endbibitem

\bibitem{imagej_web_site}
\begin{botherref}
ImageJ.
\url{https://imagej.net/}
\end{botherref}
\endbibitem

\bibitem{imagej_architecture}
\begin{botherref}
ImageJ Architecture.
\url{https://imagej.net/Architecture}
\end{botherref}
\endbibitem

\bibitem{imagej_sjc}
\begin{botherref}
SciJava Common.
\url{https://imagej.net/SciJava\_Common}
\end{botherref}
\endbibitem

\bibitem{spring}
\begin{botherref}
Spring.
\url{https://spring.io/}
\end{botherref}
\endbibitem

\bibitem{dependency_injection}
\begin{botherref}
Dependency Injection.
\url{https://en.wikipedia.org/wiki/Dependency\_injection}
\end{botherref}
\endbibitem

\bibitem{ioc}
\begin{botherref}
Inversion of Control.
\url{https://en.wikipedia.org/wiki/Inversion\_of\_control}
\end{botherref}
\endbibitem

\bibitem{markdown}
\begin{botherref}
\oauthor{\bsnm{Gruber}, \binits{J.}}:
Daring Fireball: Markdown.
\url{https://daringfireball.net/projects/markdown/}
\end{botherref}
\endbibitem

\bibitem{imagej_common}
\begin{botherref}
ImageJ Common.
\url{https://imagej.net/ImageJ\_Common}
\end{botherref}
\endbibitem

\bibitem{groovy}
\begin{botherref}
Groovy.
\url{http://groovy-lang.org/}
\end{botherref}
\endbibitem

\bibitem{imagej_matlab}
\begin{botherref}
\oauthor{\bsnm{Hiner}, \binits{M.C.}},
\oauthor{\bsnm{Rueden}, \binits{C.T.}},
\oauthor{\bsnm{Eliceiri}, \binits{K.W.}}:
Imagej-matlab: a bidirectional framework for scientific image analysis
  interoperability.
Bioinformatics,
681
(2016)
\end{botherref}
\endbibitem

\bibitem{itk}
\begin{botherref}
\oauthor{\bsnm{Yoo}, \binits{T.S.}},
\oauthor{\bsnm{Ackerman}, \binits{M.J.}},
\oauthor{\bsnm{Lorensen}, \binits{W.E.}},
\oauthor{\bsnm{Schroeder}, \binits{W.}},
\oauthor{\bsnm{Chalana}, \binits{V.}},
\oauthor{\bsnm{Aylward}, \binits{S.}},
\oauthor{\bsnm{Metaxas}, \binits{D.}},
\oauthor{\bsnm{Whitaker}, \binits{R.}}:
Engineering and algorithm design for an image processing api: a technical
  report on itk-the insight toolkit.
Studies in health technology and informatics,
586--592
(2002)
\end{botherref}
\endbibitem

\bibitem{imagej_itk}
\begin{botherref}
ImageJ-ITK.
\url{https://imagej.net/ITK}
\end{botherref}
\endbibitem

\bibitem{imagej_omero}
\begin{botherref}
ImageJ-OMERO.
\url{https://github.com/imagej/imagej-omero}
\end{botherref}
\endbibitem

\bibitem{ome_xml}
\begin{barticle}
\bauthor{\bsnm{Goldberg}, \binits{I.G.}},
\bauthor{\bsnm{Allan}, \binits{C.}},
\bauthor{\bsnm{Burel}, \binits{J.-M.}},
\bauthor{\bsnm{Creager}, \binits{D.}},
\bauthor{\bsnm{Falconi}, \binits{A.}},
\bauthor{\bsnm{Hochheiser}, \binits{H.}},
\bauthor{\bsnm{Johnston}, \binits{J.}},
\bauthor{\bsnm{Mellen}, \binits{J.}},
\bauthor{\bsnm{Sorger}, \binits{P.K.}},
\bauthor{\bsnm{Swedlow}, \binits{J.R.}}:
\batitle{The open microscopy environment (ome) data model and xml file: open
  tools for informatics and quantitative analysis in biological imaging}.
\bjtitle{Genome biology}
\bvolume{6}(\bissue{5}),
\bfpage{47}
(\byear{2005})
\end{barticle}
\endbibitem

\bibitem{imagej_notebooks}
\begin{botherref}
ImageJ Tutorial Notebooks.
\url{https://imagej.github.io/tutorials/}
\end{botherref}
\endbibitem

\bibitem{bat_cochlea_volume}
\begin{botherref}
\oauthor{\bsnm{Keating}, \binits{A.}}:
Bat Cochlea Volume.
\url{https://imagej.net/images/bat-cochlea-volume.txt}
\end{botherref}
\endbibitem

\bibitem{marching_cubes}
\begin{bchapter}
\bauthor{\bsnm{Lorensen}, \binits{W.E.}},
\bauthor{\bsnm{Cline}, \binits{H.E.}}:
\bctitle{Marching cubes: A high resolution 3d surface construction algorithm}.
In: \bbtitle{ACM Siggraph Computer Graphics},
vol. \bseriesno{21},
pp. \bfpage{163}--\blpage{169}
(\byear{1987}).
\bcomment{ACM}
\end{bchapter}
\endbibitem

\bibitem{meshlab}
\begin{bchapter}
\bauthor{\bsnm{Cignoni}, \binits{P.}},
\bauthor{\bsnm{Callieri}, \binits{M.}},
\bauthor{\bsnm{Corsini}, \binits{M.}},
\bauthor{\bsnm{Dellepiane}, \binits{M.}},
\bauthor{\bsnm{Ganovelli}, \binits{F.}},
\bauthor{\bsnm{Ranzuglia}, \binits{G.}}:
\bctitle{Meshlab: an open-source mesh processing tool.}
In: \bbtitle{Eurographics Italian Chapter Conference},
vol. \bseriesno{2008},
pp. \bfpage{129}--\blpage{136}
(\byear{2008})
\end{bchapter}
\endbibitem

\bibitem{richardson_lucy}
\begin{barticle}
\bauthor{\bsnm{Dey}, \binits{N.}},
\bauthor{\bsnm{Blanc-Feraud}, \binits{L.}},
\bauthor{\bsnm{Zimmer}, \binits{C.}},
\bauthor{\bsnm{Roux}, \binits{P.}},
\bauthor{\bsnm{Kam}, \binits{Z.}},
\bauthor{\bsnm{Olivo-Marin}, \binits{J.-C.}},
\bauthor{\bsnm{Zerubia}, \binits{J.}}:
\batitle{Richardson--lucy algorithm with total variation regularization for 3d
  confocal microscope deconvolution}.
\bjtitle{Microscopy research and technique}
\bvolume{69}(\bissue{4}),
\bfpage{260}--\blpage{266}
(\byear{2006})
\end{barticle}
\endbibitem

\bibitem{stellaris_fish}
\begin{botherref}
\oauthor{\bsnm{McNamara}, \binits{G.}}:
Leica Microscope GPU Deconvolution Stellaris FISH Dataset \#1.
\url{https://works.bepress.com/gmcnamara/31/}
\end{botherref}
\endbibitem

\bibitem{bnorthan_ops_decon}
\begin{botherref}
\oauthor{\bsnm{Northan}, \binits{B.}}:
Flexible Deconvolution Using ImageJ Ops.
\url{https://imagej.github.io/presentations/2015-09-04-imagej2-deconvolution/}
\end{botherref}
\endbibitem

\bibitem{opencl}
\begin{botherref}
OpenCL.
\url{https://www.khronos.org/opencl/}
\end{botherref}
\endbibitem

\bibitem{cuda}
\begin{botherref}
CUDA.
\url{http://www.nvidia.com/object/cuda\_home\_new.html}
\end{botherref}
\endbibitem

\bibitem{apache_spark}
\begin{botherref}
Apache Spark.
\url{https://spark.apache.org/}
\end{botherref}
\endbibitem

\bibitem{javassist}
\begin{bchapter}
\bauthor{\bsnm{Chiba}, \binits{S.}},
\bauthor{\bsnm{Nishizawa}, \binits{M.}}:
\bctitle{An easy-to-use toolkit for efficient java bytecode translators}.
In: \bbtitle{International Conference on Generative Programming and Component
  Engineering},
pp. \bfpage{364}--\blpage{376}
(\byear{2003}).
\bcomment{Springer}
\end{bchapter}
\endbibitem

\bibitem{legacy_code}
\begin{bbook}
\bauthor{\bsnm{Feathers}, \binits{M.}}:
\bbtitle{Working Effectively with Legacy Code}.
\bpublisher{Prentice Hall Professional}, \blocation{???}
(\byear{2004})
\end{bbook}
\endbibitem

\bibitem{fiji}
\begin{barticle}
\bauthor{\bsnm{Schindelin}, \binits{J.}},
\bauthor{\bsnm{Arganda-Carreras}, \binits{I.}},
\bauthor{\bsnm{Frise}, \binits{E.}},
\bauthor{\bsnm{Kaynig}, \binits{V.}},
\bauthor{\bsnm{Longair}, \binits{M.}},
\bauthor{\bsnm{Pietzsch}, \binits{T.}},
\bauthor{\bsnm{Preibisch}, \binits{S.}},
\bauthor{\bsnm{Rueden}, \binits{C.}},
\bauthor{\bsnm{Saalfeld}, \binits{S.}},
\bauthor{\bsnm{Schmid}, \binits{B.}}, \betal:
\batitle{Fiji: an open-source platform for biological-image analysis}.
\bjtitle{Nature methods}
\bvolume{9}(\bissue{7}),
\bfpage{676}--\blpage{682}
(\byear{2012})
\end{barticle}
\endbibitem

\bibitem{morpholibj}
\begin{barticle}
\bauthor{\bsnm{Legland}, \binits{D.}},
\bauthor{\bsnm{Arganda-Carreras}, \binits{I.}},
\bauthor{\bsnm{Andrey}, \binits{P.}}:
\batitle{Morpholibj: integrated library and plugins for mathematical morphology
  with imagej}.
\bjtitle{Bioinformatics}
\bvolume{32}(\bissue{22}),
\bfpage{3532}--\blpage{3534}
(\byear{2016})
\end{barticle}
\endbibitem

\bibitem{apache_maven}
\begin{botherref}
Apache Maven.
\url{https://maven.apache.org/}
\end{botherref}
\endbibitem

\bibitem{bio_formats}
\begin{barticle}
\bauthor{\bsnm{Linkert}, \binits{M.}},
\bauthor{\bsnm{Rueden}, \binits{C.T.}},
\bauthor{\bsnm{Allan}, \binits{C.}},
\bauthor{\bsnm{Burel}, \binits{J.-M.}},
\bauthor{\bsnm{Moore}, \binits{W.}},
\bauthor{\bsnm{Patterson}, \binits{A.}},
\bauthor{\bsnm{Loranger}, \binits{B.}},
\bauthor{\bsnm{Moore}, \binits{J.}},
\bauthor{\bsnm{Neves}, \binits{C.}},
\bauthor{\bsnm{MacDonald}, \binits{D.}}, \betal:
\batitle{Metadata matters: access to image data in the real world}.
\bjtitle{The Journal of cell biology}
\bvolume{189}(\bissue{5}),
\bfpage{777}--\blpage{782}
(\byear{2010})
\end{barticle}
\endbibitem

\bibitem{bio7}
\begin{barticle}
\bauthor{\bsnm{Austenfeld}, \binits{M.}},
\bauthor{\bsnm{Beyschlag}, \binits{W.}}:
\batitle{A graphical user interface for r in a rich client platform for
  ecological modeling}.
\bjtitle{Journal of Statistical Software}
\bvolume{49}(\bissue{4}),
\bfpage{1}--\blpage{19}
(\byear{2012})
\end{barticle}
\endbibitem

\bibitem{imagejfx}
\begin{botherref}
\oauthor{\bsnm{Mongis}, \binits{C.}}:
ImageJFX - an Enhanced Interface for ImageJ.
\url{http://www.imagejfx.net/}
\end{botherref}
\endbibitem

\bibitem{scripting_groovy}
\begin{botherref}
SciJava Scripting: Groovy.
\url{https://github.com/scijava/scripting-groovy}
\end{botherref}
\endbibitem

\bibitem{beanshell}
\begin{botherref}
BeanShell: Lightweight Scripting for Java.
\url{http://beanshell.org/}
\end{botherref}
\endbibitem

\bibitem{scripting_beanshell}
\begin{botherref}
SciJava Scripting: BeanShell.
\url{https://github.com/scijava/scripting-beanshell}
\end{botherref}
\endbibitem

\bibitem{scifio_bf_compat}
\begin{botherref}
Scifio-bf-compat.
\url{https://github.com/scifio/scifio-bf-compat}
\end{botherref}
\endbibitem

\bibitem{scifio_ome_xml}
\begin{botherref}
SCIFIO OME-XML Support.
\url{https://github.com/scifio/scifio-ome-xml}
\end{botherref}
\endbibitem

\bibitem{eclipse}
\begin{botherref}
Eclipse.
\url{https://eclipse.org/}
\end{botherref}
\endbibitem

\bibitem{imagej_server}
\begin{botherref}
ImageJ Server.
\url{https://github.com/imagej/imagej-server}
\end{botherref}
\endbibitem

\bibitem{imagej_legacy}
\begin{botherref}
ImageJ Legacy.
\url{https://github.com/imagej/imagej-legacy}
\end{botherref}
\endbibitem

\bibitem{ij1_patcher}
\begin{botherref}
ImageJ 1.x Patcher.
\url{https://github.com/imagej/ij1-patcher}
\end{botherref}
\endbibitem

\bibitem{simpleitk}
\begin{botherref}
SimpleITK.
\url{https://simpleitk.org/}
\end{botherref}
\endbibitem

\bibitem{javascript}
\begin{botherref}
JavaScript.
\url{https://developer.mozilla.org/en-US/docs/Web/JavaScript}
\end{botherref}
\endbibitem

\bibitem{scripting_javascript}
\begin{botherref}
SciJava Scripting: JavaScript.
\url{https://github.com/scijava/scripting-javascript}
\end{botherref}
\endbibitem

\bibitem{nashorn}
\begin{botherref}
Project Nashorn.
\url{http://openjdk.java.net/projects/nashorn/}
\end{botherref}
\endbibitem

\bibitem{rhino}
\begin{botherref}
Rhino JavaScript Implementation.
\url{https://developer.mozilla.org/en-US/docs/Mozilla/Projects/Rhino}
\end{botherref}
\endbibitem

\bibitem{jupyter}
\begin{botherref}
Project Jupyter.
\url{https://jupyter.org/}
\end{botherref}
\endbibitem

\bibitem{scijava_jupyter_kernel}
\begin{botherref}
SciJava Jupyter Kernel.
\url{https://github.com/scijava/scijava-jupyter-kernel}
\end{botherref}
\endbibitem

\bibitem{beakerx}
\begin{botherref}
Beaker Extensions for Jupyter Notebook.
\url{https://github.com/twosigma/beakerx}
\end{botherref}
\endbibitem

\bibitem{kotlin}
\begin{botherref}
Kotlin.
\url{https://kotlinlang.org/}
\end{botherref}
\endbibitem

\bibitem{scripting_kotlin}
\begin{botherref}
SciJava Scripting: Kotlin.
\url{https://github.com/scijava/scripting-kotlin}
\end{botherref}
\endbibitem

\bibitem{lisp}
\begin{botherref}
Lisp (programming Language).
\url{https://en.wikipedia.org/wiki/Lisp\_(programming\_language)}
\end{botherref}
\endbibitem

\bibitem{scripting_clojure}
\begin{botherref}
SciJava Scripting: Clojure.
\url{https://github.com/scijava/scripting-clojure}
\end{botherref}
\endbibitem

\bibitem{clojure}
\begin{botherref}
The Clojure Programming Language.
\url{https://clojure.org/}
\end{botherref}
\endbibitem

\bibitem{matlab}
\begin{botherref}
MATLAB: The Language of Technical Computing.
\url{https://www.mathworks.com/products/matlab.html}
\end{botherref}
\endbibitem

\bibitem{scripting_matlab}
\begin{botherref}
SciJava Scripting: MATLAB.
\url{https://github.com/scijava/scripting-matlab}
\end{botherref}
\endbibitem

\bibitem{matlabcontrol}
\begin{botherref}
Matlabcontrol.
\url{https://code.google.com/archive/p/matlabcontrol/}
\end{botherref}
\endbibitem

\bibitem{mitobo}
\begin{botherref}
\oauthor{\bsnm{M{\"o}ller}, \binits{B.}},
\oauthor{\bsnm{Gla{\ss}}, \binits{M.}},
\oauthor{\bsnm{Misiak}, \binits{D.}},
\oauthor{\bsnm{Posch}, \binits{S.}}:
Mitobo-a toolbox for image processing and analysis.
Journal of Open Research Software
\textbf{4}(1)
(2016)
\end{botherref}
\endbibitem

\bibitem{alida}
\begin{botherref}
Alida.
\url{http://www.informatik.uni-halle.de/alida/}
\end{botherref}
\endbibitem

\bibitem{opencv}
\begin{botherref}
OpenCV: Open Source Computer Vision.
\url{http://opencv.org/}
\end{botherref}
\endbibitem

\bibitem{ij_opencv}
\begin{barticle}
\bauthor{\bsnm{Domínguez}, \binits{C.}},
\bauthor{\bsnm{Heras}, \binits{J.}},
\bauthor{\bsnm{Pascual}, \binits{V.}}:
\batitle{Ij-opencv: Combining imagej and opencv for processing images in
  biomedicine}.
\bjtitle{Computers in Biology and Medicine}
\bvolume{84}(\bissue{C}),
\bfpage{189}--\blpage{194}
(\byear{2017})
\end{barticle}
\endbibitem

\bibitem{javacv}
\begin{botherref}
JavaCV: Java Interface to OpenCV and More.
\url{https://github.com/bytedeco/javacv}
\end{botherref}
\endbibitem

\bibitem{python}
\begin{botherref}
Python.
\url{https://python.org/}
\end{botherref}
\endbibitem

\bibitem{imglib2_imglyb}
\begin{botherref}
Imglib2-imglyb.
\url{https://github.com/hanslovsky/imglib2-imglyb}
\end{botherref}
\endbibitem

\bibitem{pyjnius}
\begin{botherref}
PyJNIus: Access Java Classes from Python.
\url{https://github.com/kivy/pyjnius}
\end{botherref}
\endbibitem

\bibitem{jython}
\begin{botherref}
Jython: Python for the Java Platform.
\url{http://jython.org/}
\end{botherref}
\endbibitem

\bibitem{jyni}
\begin{botherref}
JyNI -- Jython Native Interface.
\url{https://jyni.org/}
\end{botherref}
\endbibitem

\bibitem{imagey}
\begin{botherref}
Imagey: ImageJ with CPython REPL.
\url{https://github.com/hanslovsky/imagey}
\end{botherref}
\endbibitem

\bibitem{scripting_cpython}
\begin{botherref}
SciJava Scripting: CPython.
\url{https://github.com/scijava/scripting-cpython}
\end{botherref}
\endbibitem

\bibitem{javabridge}
\begin{botherref}
Python-javabridge: Python Wrapper for the Java Native Interface.
\url{https://github.com/LeeKamentsky/python-javabridge}
\end{botherref}
\endbibitem

\bibitem{scripting_jython}
\begin{botherref}
SciJava Scripting: Jython.
\url{https://github.com/scijava/scripting-jython}
\end{botherref}
\endbibitem

\bibitem{r}
\begin{botherref}
The R Project for Statistical Computing.
\url{https://r-project.org/}
\end{botherref}
\endbibitem

\bibitem{scripting_renjin}
\begin{botherref}
SciJava Scripting: Renjin.
\url{https://github.com/scijava/scripting-renjin}
\end{botherref}
\endbibitem

\bibitem{renjin}
\begin{botherref}
Renjin.
\url{http://renjin.org/}
\end{botherref}
\endbibitem

\bibitem{rest}
\begin{botherref}
Representational State Transfer.
\url{https://en.wikipedia.org/wiki/Representational\_state\_transfer}
\end{botherref}
\endbibitem

\bibitem{dropwizard}
\begin{botherref}
Dropwizard.
\url{http://dropwizard.io/}
\end{botherref}
\endbibitem

\bibitem{ruby}
\begin{botherref}
Ruby Programming Language.
\url{https://www.ruby-lang.org/}
\end{botherref}
\endbibitem

\bibitem{scripting_jruby}
\begin{botherref}
SciJava Scripting: JRuby.
\url{https://github.com/scijava/scripting-jruby}
\end{botherref}
\endbibitem

\bibitem{scala}
\begin{botherref}
The Scala Programming Language.
\url{https://scala-lang.org/}
\end{botherref}
\endbibitem

\bibitem{scripting_scala}
\begin{botherref}
SciJava Scripting: Scala.
\url{https://github.com/scijava/scripting-scala}
\end{botherref}
\endbibitem

\bibitem{tensorflow}
\begin{botherref}
TensorFlow: An Open-source Software Library for Machine Intelligence.
\url{https://www.tensorflow.org/}
\end{botherref}
\endbibitem

\bibitem{imagej_tensorflow}
\begin{botherref}
ImageJ-TensorFlow.
\url{https://github.com/imagej/imagej-tensorflow}
\end{botherref}
\endbibitem

\bibitem{github_effect}
\begin{botherref}
\oauthor{\bsnm{Preston-Werner}, \binits{T.}}:
The GitHub Effect: Forking Your Way to Better Code (FOWA Vegas 2011).
\url{http://lanyrd.com/2011/fowa-vegas/sfxcw/}
\end{botherref}
\endbibitem

\bibitem{imagej_source_code}
\begin{botherref}
ImageJ Source Code.
\url{https://imagej.net/Source\_code}
\end{botherref}
\endbibitem

\bibitem{imagej_licensing}
\begin{botherref}
ImageJ Licensing.
\url{https://imagej.net/Licensing}
\end{botherref}
\endbibitem

\bibitem{imagej_javadoc}
\begin{botherref}
ImageJ Javadocs.
\url{https://javadoc.imagej.net/}
\end{botherref}
\endbibitem

\bibitem{imagej_tutorials}
\begin{botherref}
ImageJ Tutorials.
\url{https://imagej.net/Tutorials}
\end{botherref}
\endbibitem

\bibitem{imagej_issues}
\begin{botherref}
ImageJ Issue Management.
\url{https://imagej.net/Issues}
\end{botherref}
\endbibitem

\bibitem{imagej_contributing}
\begin{botherref}
Contributing to ImageJ.
\url{https://imagej.net/Contributing}
\end{botherref}
\endbibitem

\bibitem{travis_ci}
\begin{botherref}
Travis CI.
\url{https://travis-ci.org/}
\end{botherref}
\endbibitem

\bibitem{imagej_uber_jar}
\begin{botherref}
Uber-JARs.
\url{https://imagej.net/Uber-JAR}
\end{botherref}
\endbibitem

\bibitem{imagej_sites}
\begin{botherref}
Personal Update Sites.
\url{https://sites.imagej.net/}
\end{botherref}
\endbibitem

\bibitem{premature_optimization}
\begin{barticle}
\bauthor{\bsnm{Hyde}, \binits{R.}}:
\batitle{The fallacy of premature optimization}.
\bjtitle{Ubiquity}
\bvolume{2009}(\bissue{February}),
\bfpage{1}
(\byear{2009})
\end{barticle}
\endbibitem

\bibitem{imglib2_benchmarks}
\begin{botherref}
ImgLib2 Benchmarks.
\url{https://imagej.net/ImgLib2\_Benchmarks}
\end{botherref}
\endbibitem

\bibitem{bigdataviewer}
\begin{barticle}
\bauthor{\bsnm{Pietzsch}, \binits{T.}},
\bauthor{\bsnm{Saalfeld}, \binits{S.}},
\bauthor{\bsnm{Preibisch}, \binits{S.}},
\bauthor{\bsnm{Tomancak}, \binits{P.}}:
\batitle{Bigdataviewer: visualization and processing for large image data
  sets}.
\bjtitle{Nature methods}
\bvolume{12}(\bissue{6}),
\bfpage{481}--\blpage{483}
(\byear{2015})
\end{barticle}
\endbibitem

\bibitem{trackmate}
\begin{barticle}
\bauthor{\bsnm{Tinevez}, \binits{J.-Y.}},
\bauthor{\bsnm{Perry}, \binits{N.}},
\bauthor{\bsnm{Schindelin}, \binits{J.}},
\bauthor{\bsnm{Hoopes}, \binits{G.M.}},
\bauthor{\bsnm{Reynolds}, \binits{G.D.}},
\bauthor{\bsnm{Laplantine}, \binits{E.}},
\bauthor{\bsnm{Bednarek}, \binits{S.Y.}},
\bauthor{\bsnm{Shorte}, \binits{S.L.}},
\bauthor{\bsnm{Eliceiri}, \binits{K.W.}}:
\batitle{Trackmate: An open and extensible platform for single-particle
  tracking}.
\bjtitle{Methods}
\bvolume{115},
\bfpage{80}--\blpage{90}
(\byear{2017})
\end{barticle}
\endbibitem

\bibitem{science_trackmate}
\begin{barticle}
\bauthor{\bsnm{Cho}, \binits{N.W.}},
\bauthor{\bsnm{Lampson}, \binits{M.A.}},
\bauthor{\bsnm{Greenberg}, \binits{R.A.}}:
\batitle{In vivo imaging of dna double-strand break induced telomere mobility
  during alternative lengthening of telomeres}.
\bjtitle{Methods}
\bvolume{114},
\bfpage{54}--\blpage{59}
(\byear{2017})
\end{barticle}
\endbibitem

\bibitem{science_mamut}
\begin{botherref}
\oauthor{\bsnm{Wolff}, \binits{C.}},
\oauthor{\bsnm{Tinevez}, \binits{J.-Y.}},
\oauthor{\bsnm{Pietzsch}, \binits{T.}},
\oauthor{\bsnm{Stamataki}, \binits{E.}},
\oauthor{\bsnm{Harich}, \binits{B.}},
\oauthor{\bsnm{Preibisch}, \binits{S.}},
\oauthor{\bsnm{Shorte}, \binits{S.}},
\oauthor{\bsnm{Keller}, \binits{P.J.}},
\oauthor{\bsnm{Tomancak}, \binits{P.}},
\oauthor{\bsnm{Pavlopoulos}, \binits{A.}}:
Reconstruction of cell lineages and behaviors underlying arthropod limb
  outgrowth with multi-view light-sheet imaging and tracking.
bioRxiv,
112623
(2017)
\end{botherref}
\endbibitem

\bibitem{multiview_2010}
\begin{barticle}
\bauthor{\bsnm{Preibisch}, \binits{S.}},
\bauthor{\bsnm{Saalfeld}, \binits{S.}},
\bauthor{\bsnm{Schindelin}, \binits{J.}},
\bauthor{\bsnm{Tomancak}, \binits{P.}}:
\batitle{Software for bead-based registration of selective plane illumination
  microscopy data}.
\bjtitle{Nature methods}
\bvolume{7}(\bissue{6}),
\bfpage{418}--\blpage{419}
(\byear{2010})
\end{barticle}
\endbibitem

\bibitem{multiview_2014}
\begin{barticle}
\bauthor{\bsnm{Preibisch}, \binits{S.}},
\bauthor{\bsnm{Amat}, \binits{F.}},
\bauthor{\bsnm{Stamataki}, \binits{E.}},
\bauthor{\bsnm{Sarov}, \binits{M.}},
\bauthor{\bsnm{Singer}, \binits{R.H.}},
\bauthor{\bsnm{Myers}, \binits{E.}},
\bauthor{\bsnm{Tomancak}, \binits{P.}}:
\batitle{Efficient bayesian-based multiview deconvolution}.
\bjtitle{nature methods}
\bvolume{11}(\bissue{6}),
\bfpage{645}--\blpage{648}
(\byear{2014})
\end{barticle}
\endbibitem

\bibitem{science_multiview_1}
\begin{botherref}
\oauthor{\bsnm{Icha}, \binits{J.}},
\oauthor{\bsnm{Schmied}, \binits{C.}},
\oauthor{\bsnm{Sidhaye}, \binits{J.}},
\oauthor{\bsnm{Tomancak}, \binits{P.}},
\oauthor{\bsnm{Preibisch}, \binits{S.}},
\oauthor{\bsnm{Norden}, \binits{C.}}:
Using light sheet fluorescence microscopy to image zebrafish eye development.
Journal of visualized experiments: JoVE
(110)
(2016)
\end{botherref}
\endbibitem

\bibitem{science_multiview_2}
\begin{barticle}
\bauthor{\bsnm{Fame}, \binits{R.M.}},
\bauthor{\bsnm{Chang}, \binits{J.T.}},
\bauthor{\bsnm{Hong}, \binits{A.}},
\bauthor{\bsnm{Aponte-Santiago}, \binits{N.A.}},
\bauthor{\bsnm{Sive}, \binits{H.}}:
\batitle{Directional cerebrospinal fluid movement between brain ventricles in
  larval zebrafish}.
\bjtitle{Fluids and Barriers of the CNS}
\bvolume{13}(\bissue{1}),
\bfpage{11}
(\byear{2016})
\end{barticle}
\endbibitem

\bibitem{sholl_analysis}
\begin{barticle}
\bauthor{\bsnm{Ferreira}, \binits{T.A.}},
\bauthor{\bsnm{Blackman}, \binits{A.V.}},
\bauthor{\bsnm{Oyrer}, \binits{J.}},
\bauthor{\bsnm{Jayabal}, \binits{S.}},
\bauthor{\bsnm{Chung}, \binits{A.J.}},
\bauthor{\bsnm{Watt}, \binits{A.J.}},
\bauthor{\bsnm{Sj{\"o}str{\"o}m}, \binits{P.J.}},
\bauthor{\bsnm{Van~Meyel}, \binits{D.J.}}:
\batitle{Neuronal morphometry directly from bitmap images}.
\bjtitle{Nature methods}
\bvolume{11}(\bissue{10}),
\bfpage{982}--\blpage{984}
(\byear{2014})
\end{barticle}
\endbibitem

\bibitem{simple_neurite_tracer}
\begin{barticle}
\bauthor{\bsnm{Longair}, \binits{M.H.}},
\bauthor{\bsnm{Baker}, \binits{D.A.}},
\bauthor{\bsnm{Armstrong}, \binits{J.D.}}:
\batitle{Simple neurite tracer: open source software for reconstruction,
  visualization and analysis of neuronal processes}.
\bjtitle{Bioinformatics}
\bvolume{27}(\bissue{17}),
\bfpage{2453}--\blpage{2454}
(\byear{2011})
\end{barticle}
\endbibitem

\bibitem{science_sholl_1}
\begin{barticle}
\bauthor{\bsnm{Luczynski}, \binits{P.}},
\bauthor{\bsnm{Whelan}, \binits{S.O.}},
\bauthor{\bsnm{O'sullivan}, \binits{C.}},
\bauthor{\bsnm{Clarke}, \binits{G.}},
\bauthor{\bsnm{Shanahan}, \binits{F.}},
\bauthor{\bsnm{Dinan}, \binits{T.G.}},
\bauthor{\bsnm{Cryan}, \binits{J.F.}}:
\batitle{Adult microbiota-deficient mice have distinct dendritic morphological
  changes: differential effects in the amygdala and hippocampus}.
\bjtitle{European Journal of Neuroscience}
\bvolume{44}(\bissue{9}),
\bfpage{2654}--\blpage{2666}
(\byear{2016})
\end{barticle}
\endbibitem

\bibitem{science_sholl_2}
\begin{barticle}
\bauthor{\bsnm{Li}, \binits{K.}},
\bauthor{\bsnm{Zhong}, \binits{X.}},
\bauthor{\bsnm{Yang}, \binits{S.}},
\bauthor{\bsnm{Luo}, \binits{Z.}},
\bauthor{\bsnm{Li}, \binits{K.}},
\bauthor{\bsnm{Liu}, \binits{Y.}},
\bauthor{\bsnm{Cai}, \binits{S.}},
\bauthor{\bsnm{Gu}, \binits{H.}},
\bauthor{\bsnm{Lu}, \binits{S.}},
\bauthor{\bsnm{Zhang}, \binits{H.}}, \betal:
\batitle{Hipsc-derived retinal ganglion cells grow dendritic arbors and
  functional axons on a tissue-engineered scaffold}.
\bjtitle{Acta Biomaterialia}
\bvolume{54},
\bfpage{117}--\blpage{127}
(\byear{2017})
\end{barticle}
\endbibitem

\bibitem{science_sholl_3}
\begin{barticle}
\bauthor{\bsnm{Valdez}, \binits{C.M.}},
\bauthor{\bsnm{Murphy}, \binits{G.G.}},
\bauthor{\bsnm{Beg}, \binits{A.A.}}:
\batitle{The rac-gap alpha2-chimaerin regulates hippocampal dendrite and spine
  morphogenesis}.
\bjtitle{Molecular and Cellular Neuroscience}
\bvolume{75},
\bfpage{14}--\blpage{26}
(\byear{2016})
\end{barticle}
\endbibitem

\bibitem{cathedral_bazaar}
\begin{bbook}
\bauthor{\bsnm{Raymond}, \binits{E.S.}}:
\bbtitle{The Cathedral \& the Bazaar: Musings on Linux and Open Source by an
  Accidental Revolutionary}.
\bpublisher{" O'Reilly Media, Inc."}, \blocation{???}
(\byear{2001})
\end{bbook}
\endbibitem

\bibitem{imagej_communication}
\begin{botherref}
ImageJ Communication Channels.
\url{https://imagej.net/Communication}
\end{botherref}
\endbibitem

\bibitem{imagej_forum}
\begin{botherref}
ImageJ Forum.
\url{http://forum.imagej.net/}
\end{botherref}
\endbibitem

\bibitem{discourse}
\begin{botherref}
Discourse.
\url{https://www.discourse.org/}
\end{botherref}
\endbibitem

\bibitem{imagej_team}
\begin{botherref}
SciJava Team Roles.
\url{https://imagej.net/Team}
\end{botherref}
\endbibitem

\bibitem{imagej_roadmap}
\begin{botherref}
ImageJ Roadmap.
\url{https://imagej.net/Roadmap}
\end{botherref}
\endbibitem

\bibitem{deep_learning}
\begin{botherref}
Deep Learning.
\url{http://deeplearning.net/}
\end{botherref}
\endbibitem

\bibitem{scikit_image}
\begin{botherref}
Scikit-image: Image Processing in Python.
\url{http://scikit-image.org/}
\end{botherref}
\endbibitem

\bibitem{aws}
\begin{botherref}
Amazon Web Services.
\url{https://aws.amazon.com/}
\end{botherref}
\endbibitem

\bibitem{imagej_user_guide}
\begin{botherref}
\oauthor{\bsnm{Ferreira}, \binits{T.}}:
ImageJ User Guide.
\url{https://imagej.net/docs/guide/}
\end{botherref}
\endbibitem

\bibitem{imagej1_docs}
\begin{botherref}
ImageJ 1.x Documentation.
\url{https://imagej.net/index.html}
\end{botherref}
\endbibitem

\bibitem{imagej_docu}
\begin{botherref}
ImageJ Information and Documentation Portal.
\url{http://imagejdocu.tudor.lu/}
\end{botherref}
\endbibitem

\bibitem{imagej_search}
\begin{botherref}
ImageJ Search.
\url{https://search.imagej.net/}
\end{botherref}
\endbibitem

\bibitem{imagej_funding}
\begin{botherref}
ImageJ Funding.
\url{https://imagej.net/Funding}
\end{botherref}
\endbibitem

\bibitem{imagej_mamut}
\begin{botherref}
MaMuT.
\url{https://imagej.net/MaMuT}
\end{botherref}
\endbibitem

\bibitem{moma_seg}
\begin{bchapter}
\bauthor{\bsnm{Jug}, \binits{F.}},
\bauthor{\bsnm{Pietzsch}, \binits{T.}},
\bauthor{\bsnm{Kainm{\"u}ller}, \binits{D.}},
\bauthor{\bsnm{Funke}, \binits{J.}},
\bauthor{\bsnm{Kaiser}, \binits{M.}},
\bauthor{\bparticle{van} \bsnm{Nimwegen}, \binits{E.}},
\bauthor{\bsnm{Rother}, \binits{C.}},
\bauthor{\bsnm{Myers}, \binits{G.}}:
\bctitle{Optimal joint segmentation and tracking of escherichia coli in the
  mother machine}.
In: \bbtitle{Bayesian and grAphical Models for Biomedical Imaging},
pp. \bfpage{25}--\blpage{36}.
\bpublisher{Springer}, \blocation{???}
(\byear{2014})
\end{bchapter}
\endbibitem

\bibitem{moma_tracking}
\begin{bchapter}
\bauthor{\bsnm{Jug}, \binits{F.}},
\bauthor{\bsnm{Pietzsch}, \binits{T.}},
\bauthor{\bsnm{Kainm{\"u}ller}, \binits{D.}},
\bauthor{\bsnm{Myers}, \binits{G.}}:
\bctitle{Tracking by assignment facilitates data curation}.
In: \bbtitle{MICCAI IMIC Workshop},
vol. \bseriesno{2}
(\byear{2014})
\end{bchapter}
\endbibitem

\bibitem{kymograph}
\begin{botherref}
\oauthor{\bsnm{Mary}, \binits{H.}},
\oauthor{\bsnm{Rueden}, \binits{C.}},
\oauthor{\bsnm{Ferreira}, \binits{T.}}:
KymographBuilder: Release 1.2.4
(2016).
doi:\doiurl{10.5281/zenodo.56702}.
\url{https://doi.org/10.5281/zenodo.56702}
\end{botherref}
\endbibitem

\bibitem{z_spacing}
\begin{bchapter}
\bauthor{\bsnm{Hanslovsky}, \binits{P.}},
\bauthor{\bsnm{Bogovic}, \binits{J.A.}},
\bauthor{\bsnm{Saalfeld}, \binits{S.}}:
\bctitle{Post-acquisition image based compensation for thickness variation in
  microscopy section series}.
In: \bbtitle{Biomedical Imaging (ISBI), 2015 IEEE 12th International Symposium
  On},
pp. \bfpage{507}--\blpage{511}
(\byear{2015}).
\bcomment{IEEE}
\end{bchapter}
\endbibitem

\bibitem{trainable_weka}
\begin{botherref}
\oauthor{\bsnm{Arganda-Carreras}, \binits{I.}},
\oauthor{\bsnm{Kaynig}, \binits{V.}},
\oauthor{\bsnm{Rueden}, \binits{C.}},
\oauthor{\bsnm{Eliceiri}, \binits{K.}},
\oauthor{\bsnm{Schindelin}, \binits{J.}},
\oauthor{\bsnm{Cardona}, \binits{A.}},
\oauthor{\bsnm{Seung}, \binits{H.}}:
Trainable weka segmentation: a machine learning tool for microscopy pixel
  classification.
Bioinformatics (Oxford, England)
(2017)
\end{botherref}
\endbibitem

\bibitem{pendent_drop}
\begin{botherref}
\oauthor{\bsnm{Daerr}, \binits{A.}},
\oauthor{\bsnm{Mogne}, \binits{A.}}:
Pendent\_drop: an imagej plugin to measure the surface tension from an image of
  a pendent drop.
Journal of Open Research Software
\textbf{4}(1)
(2016)
\end{botherref}
\endbibitem

\bibitem{sciview}
\begin{botherref}
SciView.
\url{https://github.com/scenerygraphics/SciView}
\end{botherref}
\endbibitem

\bibitem{image_stitching}
\begin{barticle}
\bauthor{\bsnm{Preibisch}, \binits{S.}},
\bauthor{\bsnm{Saalfeld}, \binits{S.}},
\bauthor{\bsnm{Tomancak}, \binits{P.}}:
\batitle{Globally optimal stitching of tiled 3d microscopic image
  acquisitions}.
\bjtitle{Bioinformatics}
\bvolume{25}(\bissue{11}),
\bfpage{1463}--\blpage{1465}
(\byear{2009})
\end{barticle}
\endbibitem

\bibitem{imagej_coloc_2}
\begin{botherref}
Coloc 2.
\url{https://imagej.net/Coloc\_2}
\end{botherref}
\endbibitem

\end{thebibliography}

\newcommand{\BMCxmlcomment}[1]{}

\BMCxmlcomment{

<refgrp>

<bibl id="B1">
  <title><p>NIH Image to ImageJ: 25 years of image analysis</p></title>
  <aug>
    <au><snm>Schneider</snm><fnm>CA</fnm></au>
    <au><snm>Rasband</snm><fnm>WS</fnm></au>
    <au><snm>Eliceiri</snm><fnm>KW</fnm></au>
    <au><cnm>others</cnm></au>
  </aug>
  <source>Nat methods</source>
  <pubdate>2012</pubdate>
  <volume>9</volume>
  <issue>7</issue>
  <fpage>671</fpage>
  <lpage>-675</lpage>
</bibl>

<bibl id="B2">
  <title><p>Quantitating the cell: turning images into numbers with
  ImageJ</p></title>
  <aug>
    <au><snm>Arena</snm><fnm>ET</fnm></au>
    <au><snm>Rueden</snm><fnm>CT</fnm></au>
    <au><snm>Hiner</snm><fnm>MC</fnm></au>
    <au><snm>Wang</snm><fnm>S</fnm></au>
    <au><snm>Yuan</snm><fnm>M</fnm></au>
    <au><snm>Eliceiri</snm><fnm>KW</fnm></au>
  </aug>
  <source>Wiley Interdisciplinary Reviews: Developmental Biology</source>
  <publisher>Wiley Online Library</publisher>
  <pubdate>2016</pubdate>
</bibl>

<bibl id="B3">
  <title><p>ImageJ Contributors</p></title>
  <url>https://imagej.net/Contributors</url>
</bibl>

<bibl id="B4">
  <title><p>Bioimage informatics: a new area of engineering biology</p></title>
  <aug>
    <au><snm>Peng</snm><fnm>H</fnm></au>
  </aug>
  <source>Bioinformatics</source>
  <publisher>Oxford Univ Press</publisher>
  <pubdate>2008</pubdate>
  <volume>24</volume>
  <issue>17</issue>
  <fpage>1827</fpage>
  <lpage>-1836</lpage>
</bibl>

<bibl id="B5">
  <title><p>Biological imaging software tools</p></title>
  <aug>
    <au><snm>Eliceiri</snm><fnm>KW</fnm></au>
    <au><snm>Berthold</snm><fnm>MR</fnm></au>
    <au><snm>Goldberg</snm><fnm>IG</fnm></au>
    <au><snm>Ib{\'a}{\~n}ez</snm><fnm>L</fnm></au>
    <au><snm>Manjunath</snm><fnm>BS</fnm></au>
    <au><snm>Martone</snm><fnm>ME</fnm></au>
    <au><snm>Murphy</snm><fnm>RF</fnm></au>
    <au><snm>Peng</snm><fnm>H</fnm></au>
    <au><snm>Plant</snm><fnm>AL</fnm></au>
    <au><snm>Roysam</snm><fnm>B</fnm></au>
    <au><cnm>others</cnm></au>
  </aug>
  <source>Nature methods</source>
  <publisher>Nature Publishing Group</publisher>
  <pubdate>2012</pubdate>
  <volume>9</volume>
  <issue>7</issue>
  <fpage>697</fpage>
  <lpage>-710</lpage>
</bibl>

<bibl id="B6">
  <title><p>Open source bioimage informatics for cell biology</p></title>
  <aug>
    <au><snm>Swedlow</snm><fnm>JR</fnm></au>
    <au><snm>Eliceiri</snm><fnm>KW</fnm></au>
  </aug>
  <source>Trends in cell biology</source>
  <publisher>Elsevier</publisher>
  <pubdate>2009</pubdate>
  <volume>19</volume>
  <issue>11</issue>
  <fpage>656</fpage>
  <lpage>-660</lpage>
</bibl>

<bibl id="B7">
  <title><p>Computer control of microscopes using $\mu$Manager</p></title>
  <aug>
    <au><snm>Edelstein</snm><fnm>A</fnm></au>
    <au><snm>Amodaj</snm><fnm>N</fnm></au>
    <au><snm>Hoover</snm><fnm>K</fnm></au>
    <au><snm>Vale</snm><fnm>R</fnm></au>
    <au><snm>Stuurman</snm><fnm>N</fnm></au>
  </aug>
  <source>Current protocols in molecular biology</source>
  <publisher>Wiley Online Library</publisher>
  <pubdate>2010</pubdate>
  <fpage>14</fpage>
  <lpage>-20</lpage>
</bibl>

<bibl id="B8">
  <title><p>Advanced methods of microscope control using $\mu$Manager
  software</p></title>
  <aug>
    <au><snm>Edelstein</snm><fnm>AD</fnm></au>
    <au><snm>Tsuchida</snm><fnm>MA</fnm></au>
    <au><snm>Amodaj</snm><fnm>N</fnm></au>
    <au><snm>Pinkard</snm><fnm>H</fnm></au>
    <au><snm>Vale</snm><fnm>RD</fnm></au>
    <au><snm>Stuurman</snm><fnm>N</fnm></au>
  </aug>
  <source>Journal of biological methods</source>
  <publisher>NIH Public Access</publisher>
  <pubdate>2014</pubdate>
  <volume>1</volume>
  <issue>2</issue>
</bibl>

<bibl id="B9">
  <title><p>Bisque: a platform for bioimage analysis and management</p></title>
  <aug>
    <au><snm>Kvilekval</snm><fnm>K</fnm></au>
    <au><snm>Fedorov</snm><fnm>D</fnm></au>
    <au><snm>Obara</snm><fnm>B</fnm></au>
    <au><snm>Singh</snm><fnm>A</fnm></au>
    <au><snm>Manjunath</snm><fnm>BS</fnm></au>
  </aug>
  <source>Bioinformatics</source>
  <publisher>Oxford Univ Press</publisher>
  <pubdate>2010</pubdate>
  <volume>26</volume>
  <issue>4</issue>
  <fpage>544</fpage>
  <lpage>-552</lpage>
</bibl>

<bibl id="B10">
  <title><p>OMERO: flexible, model-driven data management for experimental
  biology</p></title>
  <aug>
    <au><snm>Allan</snm><fnm>C</fnm></au>
    <au><snm>Burel</snm><fnm>JM</fnm></au>
    <au><snm>Moore</snm><fnm>J</fnm></au>
    <au><snm>Blackburn</snm><fnm>C</fnm></au>
    <au><snm>Linkert</snm><fnm>M</fnm></au>
    <au><snm>Loynton</snm><fnm>S</fnm></au>
    <au><snm>MacDonald</snm><fnm>D</fnm></au>
    <au><snm>Moore</snm><fnm>WJ</fnm></au>
    <au><snm>Neves</snm><fnm>C</fnm></au>
    <au><snm>Patterson</snm><fnm>A</fnm></au>
    <au><cnm>others</cnm></au>
  </aug>
  <source>Nature methods</source>
  <publisher>Nature Publishing Group</publisher>
  <pubdate>2012</pubdate>
  <volume>9</volume>
  <issue>3</issue>
  <fpage>245</fpage>
  <lpage>-253</lpage>
</bibl>

<bibl id="B11">
  <title><p>Icy: an open bioimage informatics platform for extended
  reproducible research</p></title>
  <aug>
    <au><snm>De Chaumont</snm><fnm>F</fnm></au>
    <au><snm>Dallongeville</snm><fnm>S</fnm></au>
    <au><snm>Chenouard</snm><fnm>N</fnm></au>
    <au><snm>Herv{\'e}</snm><fnm>N</fnm></au>
    <au><snm>Pop</snm><fnm>S</fnm></au>
    <au><snm>Provoost</snm><fnm>T</fnm></au>
    <au><snm>Meas Yedid</snm><fnm>V</fnm></au>
    <au><snm>Pankajakshan</snm><fnm>P</fnm></au>
    <au><snm>Lecomte</snm><fnm>T</fnm></au>
    <au><snm>Le Montagner</snm><fnm>Y</fnm></au>
    <au><cnm>others</cnm></au>
  </aug>
  <source>Nature methods</source>
  <publisher>Nature Publishing Group</publisher>
  <pubdate>2012</pubdate>
  <volume>9</volume>
  <issue>7</issue>
  <fpage>690</fpage>
  <lpage>-696</lpage>
</bibl>

<bibl id="B12">
  <title><p>BioImageXD: an open, general-purpose and high-throughput
  image-processing platform</p></title>
  <aug>
    <au><snm>Kankaanp{\"a}{\"a}</snm><fnm>P</fnm></au>
    <au><snm>Paavolainen</snm><fnm>L</fnm></au>
    <au><snm>Tiitta</snm><fnm>S</fnm></au>
    <au><snm>Karjalainen</snm><fnm>M</fnm></au>
    <au><snm>P{\"a}iv{\"a}rinne</snm><fnm>J</fnm></au>
    <au><snm>Nieminen</snm><fnm>J</fnm></au>
    <au><snm>Marjom{\"a}ki</snm><fnm>V</fnm></au>
    <au><snm>Heino</snm><fnm>J</fnm></au>
    <au><snm>White</snm><fnm>DJ</fnm></au>
  </aug>
  <source>Nature methods</source>
  <publisher>Nature Research</publisher>
  <pubdate>2012</pubdate>
  <volume>9</volume>
  <issue>7</issue>
  <fpage>683</fpage>
  <lpage>-689</lpage>
</bibl>

<bibl id="B13">
  <title><p>CellProfiler: image analysis software for identifying and
  quantifying cell phenotypes</p></title>
  <aug>
    <au><snm>Carpenter</snm><fnm>AE</fnm></au>
    <au><snm>Jones</snm><fnm>TR</fnm></au>
    <au><snm>Lamprecht</snm><fnm>MR</fnm></au>
    <au><snm>Clarke</snm><fnm>C</fnm></au>
    <au><snm>Kang</snm><fnm>IH</fnm></au>
    <au><snm>Friman</snm><fnm>O</fnm></au>
    <au><snm>Guertin</snm><fnm>DA</fnm></au>
    <au><snm>Chang</snm><fnm>JH</fnm></au>
    <au><snm>Lindquist</snm><fnm>RA</fnm></au>
    <au><snm>Moffat</snm><fnm>J</fnm></au>
    <au><cnm>others</cnm></au>
  </aug>
  <source>Genome biology</source>
  <publisher>BioMed Central Ltd</publisher>
  <pubdate>2006</pubdate>
  <volume>7</volume>
  <issue>10</issue>
  <fpage>R100</fpage>
</bibl>

<bibl id="B14">
  <title><p>Improved structure, function and compatibility for CellProfiler:
  modular high-throughput image analysis software</p></title>
  <aug>
    <au><snm>Kamentsky</snm><fnm>L</fnm></au>
    <au><snm>Jones</snm><fnm>TR</fnm></au>
    <au><snm>Fraser</snm><fnm>A</fnm></au>
    <au><snm>Bray</snm><fnm>MA</fnm></au>
    <au><snm>Logan</snm><fnm>DJ</fnm></au>
    <au><snm>Madden</snm><fnm>KL</fnm></au>
    <au><snm>Ljosa</snm><fnm>V</fnm></au>
    <au><snm>Rueden</snm><fnm>C</fnm></au>
    <au><snm>Eliceiri</snm><fnm>KW</fnm></au>
    <au><snm>Carpenter</snm><fnm>AE</fnm></au>
  </aug>
  <source>Bioinformatics</source>
  <publisher>Oxford Univ Press</publisher>
  <pubdate>2011</pubdate>
  <volume>27</volume>
  <issue>8</issue>
  <fpage>1179</fpage>
  <lpage>-1180</lpage>
</bibl>

<bibl id="B15">
  <title><p>KNIME: The Konstanz Information Miner</p></title>
  <aug>
    <au><snm>Michael</snm><fnm>B</fnm></au>
    <au><snm>Nicolas</snm><fnm>C</fnm></au>
    <au><snm>Fabian</snm><fnm>D</fnm></au>
    <au><snm>Thomas</snm><fnm>G</fnm></au>
    <au><snm>Tobias</snm><fnm>O</fnm></au>
    <au><snm>Thorsten</snm><fnm>M</fnm></au>
    <au><snm>Peter</snm><fnm>O</fnm></au>
    <au><snm>Christoph</snm><fnm>S</fnm></au>
    <au><snm>Kilian</snm><fnm>T</fnm></au>
    <au><snm>Bernd</snm><fnm>W</fnm></au>
  </aug>
  <source>Studies in Classification, Data Analysis, and Knowledge Organization.
  GfKL</source>
  <pubdate>2007</pubdate>
</bibl>

<bibl id="B16">
  <title><p>KNIME for Open-Source Bioimage Analysis: A Tutorial</p></title>
  <aug>
    <au><snm>Dietz</snm><fnm>C</fnm></au>
    <au><snm>Berthold</snm><fnm>MR</fnm></au>
  </aug>
  <source>Focus on Bio-Image Informatics</source>
  <publisher>Springer</publisher>
  <pubdate>2016</pubdate>
  <fpage>179</fpage>
  <lpage>-197</lpage>
</bibl>

<bibl id="B17">
  <title><p>Scientific workflow systems: Pipeline Pilot and KNIME</p></title>
  <aug>
    <au><snm>Warr</snm><fnm>WA</fnm></au>
  </aug>
  <source>Journal of computer-aided molecular design</source>
  <publisher>Springer</publisher>
  <pubdate>2012</pubdate>
  <volume>26</volume>
  <issue>7</issue>
  <fpage>801</fpage>
  <lpage>-804</lpage>
</bibl>

<bibl id="B18">
  <title><p>FluoRender: an application of 2D image space methods for 3D and 4D
  confocal microscopy data visualization in neurobiology research</p></title>
  <aug>
    <au><snm>Wan</snm><fnm>Y</fnm></au>
    <au><snm>Otsuna</snm><fnm>H</fnm></au>
    <au><snm>Chien</snm><fnm>CB</fnm></au>
    <au><snm>Hansen</snm><fnm>C</fnm></au>
  </aug>
  <source>Pacific Visualization Symposium (PacificVis), 2012 IEEE</source>
  <pubdate>2012</pubdate>
  <fpage>201</fpage>
  <lpage>-208</lpage>
</bibl>

<bibl id="B19">
  <title><p>Extensible visualization and analysis for multidimensional images
  using Vaa3D</p></title>
  <aug>
    <au><snm>Peng</snm><fnm>H</fnm></au>
    <au><snm>Bria</snm><fnm>A</fnm></au>
    <au><snm>Zhou</snm><fnm>Z</fnm></au>
    <au><snm>Iannello</snm><fnm>G</fnm></au>
    <au><snm>Long</snm><fnm>F</fnm></au>
  </aug>
  <source>Nature protocols</source>
  <publisher>Nature Research</publisher>
  <pubdate>2014</pubdate>
  <volume>9</volume>
  <issue>1</issue>
  <fpage>193</fpage>
  <lpage>-208</lpage>
</bibl>

<bibl id="B20">
  <title><p>The ImageJ ecosystem: An open platform for biomedical image
  analysis</p></title>
  <aug>
    <au><snm>Schindelin</snm><fnm>J</fnm></au>
    <au><snm>Rueden</snm><fnm>CT</fnm></au>
    <au><snm>Hiner</snm><fnm>MC</fnm></au>
    <au><snm>Eliceiri</snm><fnm>KW</fnm></au>
  </aug>
  <source>Molecular reproduction and development</source>
  <publisher>Wiley Online Library</publisher>
  <pubdate>2015</pubdate>
  <volume>82</volume>
  <issue>7-8</issue>
  <fpage>518</fpage>
  <lpage>-529</lpage>
</bibl>

<bibl id="B21">
  <title><p>List of ImageJ Update Sites</p></title>
  <url>https://imagej.net/List\_of\_update\_sites</url>
</bibl>

<bibl id="B22">
  <title><p>SciJava</p></title>
  <url>http://www.scijava.org/</url>
</bibl>

<bibl id="B23">
  <title><p>2015 ImageJ conference presentation: survey</p></title>
  <url>https://imagej.github.io/presentations/2015-09-03-imagej2-and-fiji/\#/6</url>
</bibl>

<bibl id="B24">
  <title><p>A call for bioimaging software usability</p></title>
  <aug>
    <au><snm>Carpenter</snm><fnm>AE</fnm></au>
    <au><snm>Kamentsky</snm><fnm>L</fnm></au>
    <au><snm>Eliceiri</snm><fnm>KW</fnm></au>
  </aug>
  <source>Nature methods</source>
  <publisher>NIH Public Access</publisher>
  <pubdate>2012</pubdate>
  <volume>9</volume>
  <issue>7</issue>
  <fpage>666</fpage>
</bibl>

<bibl id="B25">
  <title><p>Hardware is Cheap, Programmers are Expensive</p></title>
  <aug>
    <au><snm>Atwood</snm><fnm>J</fnm></au>
  </aug>
  <pubdate>2008</pubdate>
  <url>https://blog.codinghorror.com/hardware-is-cheap-programmers-are-expensive/</url>
</bibl>

<bibl id="B26">
  <title><p>SciJava</p></title>
  <url>https://imagej.net/SciJava</url>
</bibl>

<bibl id="B27">
  <title><p>ImgLib2---generic image processing in Java</p></title>
  <aug>
    <au><snm>Pietzsch</snm><fnm>T</fnm></au>
    <au><snm>Preibisch</snm><fnm>S</fnm></au>
    <au><snm>Toman{\v{c}}{\'a}k</snm><fnm>P</fnm></au>
    <au><snm>Saalfeld</snm><fnm>S</fnm></au>
  </aug>
  <source>Bioinformatics</source>
  <publisher>Oxford Univ Press</publisher>
  <pubdate>2012</pubdate>
  <volume>28</volume>
  <issue>22</issue>
  <fpage>3009</fpage>
  <lpage>-3011</lpage>
</bibl>

<bibl id="B28">
  <title><p>SCIFIO: an extensible framework to support scientific image
  formats</p></title>
  <aug>
    <au><snm>Hiner</snm><fnm>MC</fnm></au>
    <au><snm>Rueden</snm><fnm>CT</fnm></au>
    <au><snm>Eliceiri</snm><fnm>KW</fnm></au>
  </aug>
  <source>BMC bioinformatics</source>
  <publisher>BioMed Central</publisher>
  <pubdate>2016</pubdate>
  <volume>17</volume>
  <issue>1</issue>
  <fpage>521</fpage>
</bibl>

<bibl id="B29">
  <title><p>ImageJ</p></title>
  <url>https://imagej.net/</url>
</bibl>

<bibl id="B30">
  <title><p>ImageJ Architecture</p></title>
  <url>https://imagej.net/Architecture</url>
</bibl>

<bibl id="B31">
  <title><p>SciJava Common</p></title>
  <url>https://imagej.net/SciJava\_Common</url>
</bibl>

<bibl id="B32">
  <title><p>Spring</p></title>
  <url>https://spring.io/</url>
</bibl>

<bibl id="B33">
  <title><p>Dependency injection</p></title>
  <url>https://en.wikipedia.org/wiki/Dependency\_injection</url>
</bibl>

<bibl id="B34">
  <title><p>Inversion of control</p></title>
  <url>https://en.wikipedia.org/wiki/Inversion\_of\_control</url>
</bibl>

<bibl id="B35">
  <title><p>Daring Fireball: Markdown</p></title>
  <aug>
    <au><snm>Gruber</snm><fnm>J</fnm></au>
  </aug>
  <url>https://daringfireball.net/projects/markdown/</url>
</bibl>

<bibl id="B36">
  <title><p>ImageJ Common</p></title>
  <url>https://imagej.net/ImageJ\_Common</url>
</bibl>

<bibl id="B37">
  <title><p>Groovy</p></title>
  <url>http://groovy-lang.org/</url>
</bibl>

<bibl id="B38">
  <title><p>ImageJ-MATLAB: a bidirectional framework for scientific image
  analysis interoperability</p></title>
  <aug>
    <au><snm>Hiner</snm><fnm>MC</fnm></au>
    <au><snm>Rueden</snm><fnm>CT</fnm></au>
    <au><snm>Eliceiri</snm><fnm>KW</fnm></au>
  </aug>
  <source>Bioinformatics</source>
  <publisher>Oxford Univ Press</publisher>
  <pubdate>2016</pubdate>
  <fpage>btw681</fpage>
</bibl>

<bibl id="B39">
  <title><p>Engineering and algorithm design for an image processing API: a
  technical report on ITK-the insight toolkit</p></title>
  <aug>
    <au><snm>Yoo</snm><fnm>TS</fnm></au>
    <au><snm>Ackerman</snm><fnm>MJ</fnm></au>
    <au><snm>Lorensen</snm><fnm>WE</fnm></au>
    <au><snm>Schroeder</snm><fnm>W</fnm></au>
    <au><snm>Chalana</snm><fnm>V</fnm></au>
    <au><snm>Aylward</snm><fnm>S</fnm></au>
    <au><snm>Metaxas</snm><fnm>D</fnm></au>
    <au><snm>Whitaker</snm><fnm>R</fnm></au>
  </aug>
  <source>Studies in health technology and informatics</source>
  <publisher>IOS Press; 1999</publisher>
  <pubdate>2002</pubdate>
  <fpage>586</fpage>
  <lpage>-592</lpage>
</bibl>

<bibl id="B40">
  <title><p>ImageJ-ITK</p></title>
  <url>https://imagej.net/ITK</url>
</bibl>

<bibl id="B41">
  <title><p>ImageJ-OMERO</p></title>
  <url>https://github.com/imagej/imagej-omero</url>
</bibl>

<bibl id="B42">
  <title><p>The Open Microscopy Environment (OME) Data Model and XML file: open
  tools for informatics and quantitative analysis in biological
  imaging</p></title>
  <aug>
    <au><snm>Goldberg</snm><fnm>IG</fnm></au>
    <au><snm>Allan</snm><fnm>C</fnm></au>
    <au><snm>Burel</snm><fnm>JM</fnm></au>
    <au><snm>Creager</snm><fnm>D</fnm></au>
    <au><snm>Falconi</snm><fnm>A</fnm></au>
    <au><snm>Hochheiser</snm><fnm>H</fnm></au>
    <au><snm>Johnston</snm><fnm>J</fnm></au>
    <au><snm>Mellen</snm><fnm>J</fnm></au>
    <au><snm>Sorger</snm><fnm>PK</fnm></au>
    <au><snm>Swedlow</snm><fnm>JR</fnm></au>
  </aug>
  <source>Genome biology</source>
  <publisher>BioMed Central</publisher>
  <pubdate>2005</pubdate>
  <volume>6</volume>
  <issue>5</issue>
  <fpage>R47</fpage>
</bibl>

<bibl id="B43">
  <title><p>ImageJ tutorial notebooks</p></title>
  <url>https://imagej.github.io/tutorials/</url>
</bibl>

<bibl id="B44">
  <title><p>Bat Cochlea Volume</p></title>
  <aug>
    <au><snm>Keating</snm><fnm>A</fnm></au>
  </aug>
  <url>https://imagej.net/images/bat-cochlea-volume.txt</url>
</bibl>

<bibl id="B45">
  <title><p>Marching cubes: A high resolution 3D surface construction
  algorithm</p></title>
  <aug>
    <au><snm>Lorensen</snm><fnm>WE</fnm></au>
    <au><snm>Cline</snm><fnm>HE</fnm></au>
  </aug>
  <source>ACM siggraph computer graphics</source>
  <pubdate>1987</pubdate>
  <volume>21</volume>
  <issue>4</issue>
  <fpage>163</fpage>
  <lpage>-169</lpage>
</bibl>

<bibl id="B46">
  <title><p>Meshlab: an open-source mesh processing tool.</p></title>
  <aug>
    <au><snm>Cignoni</snm><fnm>P</fnm></au>
    <au><snm>Callieri</snm><fnm>M</fnm></au>
    <au><snm>Corsini</snm><fnm>M</fnm></au>
    <au><snm>Dellepiane</snm><fnm>M</fnm></au>
    <au><snm>Ganovelli</snm><fnm>F</fnm></au>
    <au><snm>Ranzuglia</snm><fnm>G</fnm></au>
  </aug>
  <source>Eurographics Italian Chapter Conference</source>
  <pubdate>2008</pubdate>
  <volume>2008</volume>
  <fpage>129</fpage>
  <lpage>-136</lpage>
</bibl>

<bibl id="B47">
  <title><p>Richardson--Lucy algorithm with total variation regularization for
  3D confocal microscope deconvolution</p></title>
  <aug>
    <au><snm>Dey</snm><fnm>N</fnm></au>
    <au><snm>Blanc Feraud</snm><fnm>L</fnm></au>
    <au><snm>Zimmer</snm><fnm>C</fnm></au>
    <au><snm>Roux</snm><fnm>P</fnm></au>
    <au><snm>Kam</snm><fnm>Z</fnm></au>
    <au><snm>Olivo Marin</snm><fnm>JC</fnm></au>
    <au><snm>Zerubia</snm><fnm>J</fnm></au>
  </aug>
  <source>Microscopy research and technique</source>
  <publisher>Wiley Online Library</publisher>
  <pubdate>2006</pubdate>
  <volume>69</volume>
  <issue>4</issue>
  <fpage>260</fpage>
  <lpage>-266</lpage>
</bibl>

<bibl id="B48">
  <title><p>Leica microscope GPU deconvolution Stellaris FISH dataset
  \#1</p></title>
  <aug>
    <au><snm>McNamara</snm><fnm>G</fnm></au>
  </aug>
  <url>https://works.bepress.com/gmcnamara/31/</url>
</bibl>

<bibl id="B49">
  <title><p>Flexible Deconvolution using ImageJ Ops</p></title>
  <aug>
    <au><snm>Northan</snm><fnm>B</fnm></au>
  </aug>
  <url>https://imagej.github.io/presentations/2015-09-04-imagej2-deconvolution/</url>
</bibl>

<bibl id="B50">
  <title><p>OpenCL</p></title>
  <url>https://www.khronos.org/opencl/</url>
</bibl>

<bibl id="B51">
  <title><p>CUDA</p></title>
  <url>http://www.nvidia.com/object/cuda\_home\_new.html</url>
</bibl>

<bibl id="B52">
  <title><p>Apache Spark</p></title>
  <url>https://spark.apache.org/</url>
</bibl>

<bibl id="B53">
  <title><p>An easy-to-use toolkit for efficient Java bytecode
  translators</p></title>
  <aug>
    <au><snm>Chiba</snm><fnm>S</fnm></au>
    <au><snm>Nishizawa</snm><fnm>M</fnm></au>
  </aug>
  <source>International Conference on Generative Programming and Component
  Engineering</source>
  <pubdate>2003</pubdate>
  <fpage>364</fpage>
  <lpage>-376</lpage>
</bibl>

<bibl id="B54">
  <title><p>Working effectively with legacy code</p></title>
  <aug>
    <au><snm>Feathers</snm><fnm>M</fnm></au>
  </aug>
  <publisher>Prentice Hall Professional</publisher>
  <pubdate>2004</pubdate>
</bibl>

<bibl id="B55">
  <title><p>Fiji: an open-source platform for biological-image
  analysis</p></title>
  <aug>
    <au><snm>Schindelin</snm><fnm>J</fnm></au>
    <au><snm>Arganda Carreras</snm><fnm>I</fnm></au>
    <au><snm>Frise</snm><fnm>E</fnm></au>
    <au><snm>Kaynig</snm><fnm>V</fnm></au>
    <au><snm>Longair</snm><fnm>M</fnm></au>
    <au><snm>Pietzsch</snm><fnm>T</fnm></au>
    <au><snm>Preibisch</snm><fnm>S</fnm></au>
    <au><snm>Rueden</snm><fnm>C</fnm></au>
    <au><snm>Saalfeld</snm><fnm>S</fnm></au>
    <au><snm>Schmid</snm><fnm>B</fnm></au>
    <au><cnm>others</cnm></au>
  </aug>
  <source>Nature methods</source>
  <publisher>Nature Publishing Group</publisher>
  <pubdate>2012</pubdate>
  <volume>9</volume>
  <issue>7</issue>
  <fpage>676</fpage>
  <lpage>-682</lpage>
</bibl>

<bibl id="B56">
  <title><p>MorphoLibJ: integrated library and plugins for mathematical
  morphology with ImageJ</p></title>
  <aug>
    <au><snm>Legland</snm><fnm>D</fnm></au>
    <au><snm>Arganda Carreras</snm><fnm>I</fnm></au>
    <au><snm>Andrey</snm><fnm>P</fnm></au>
  </aug>
  <source>Bioinformatics</source>
  <publisher>Oxford Univ Press</publisher>
  <pubdate>2016</pubdate>
  <volume>32</volume>
  <issue>22</issue>
  <fpage>3532</fpage>
  <lpage>-3534</lpage>
</bibl>

<bibl id="B57">
  <title><p>Apache Maven</p></title>
  <url>https://maven.apache.org/</url>
</bibl>

<bibl id="B58">
  <title><p>Metadata matters: access to image data in the real
  world</p></title>
  <aug>
    <au><snm>Linkert</snm><fnm>M</fnm></au>
    <au><snm>Rueden</snm><fnm>CT</fnm></au>
    <au><snm>Allan</snm><fnm>C</fnm></au>
    <au><snm>Burel</snm><fnm>JM</fnm></au>
    <au><snm>Moore</snm><fnm>W</fnm></au>
    <au><snm>Patterson</snm><fnm>A</fnm></au>
    <au><snm>Loranger</snm><fnm>B</fnm></au>
    <au><snm>Moore</snm><fnm>J</fnm></au>
    <au><snm>Neves</snm><fnm>C</fnm></au>
    <au><snm>MacDonald</snm><fnm>D</fnm></au>
    <au><cnm>others</cnm></au>
  </aug>
  <source>The Journal of cell biology</source>
  <publisher>Rockefeller Univ Press</publisher>
  <pubdate>2010</pubdate>
  <volume>189</volume>
  <issue>5</issue>
  <fpage>777</fpage>
  <lpage>-782</lpage>
</bibl>

<bibl id="B59">
  <title><p>A Graphical User Interface for R in a Rich Client Platform for
  Ecological Modeling</p></title>
  <aug>
    <au><snm>Austenfeld</snm><fnm>M</fnm></au>
    <au><snm>Beyschlag</snm><fnm>W</fnm></au>
  </aug>
  <source>Journal of Statistical Software</source>
  <publisher>Foundation for Open Access Statistics</publisher>
  <pubdate>2012</pubdate>
  <volume>49</volume>
  <issue>4</issue>
  <fpage>1</fpage>
  <lpage>-19</lpage>
</bibl>

<bibl id="B60">
  <title><p>ImageJFX - an enhanced interface for ImageJ</p></title>
  <aug>
    <au><snm>Mongis</snm><fnm>C</fnm></au>
  </aug>
  <url>http://www.imagejfx.net/</url>
</bibl>

<bibl id="B61">
  <title><p>SciJava Scripting: Groovy</p></title>
  <url>https://github.com/scijava/scripting-groovy</url>
</bibl>

<bibl id="B62">
  <title><p>BeanShell: Lightweight Scripting for Java</p></title>
  <url>http://beanshell.org/</url>
</bibl>

<bibl id="B63">
  <title><p>SciJava Scripting: BeanShell</p></title>
  <url>https://github.com/scijava/scripting-beanshell</url>
</bibl>

<bibl id="B64">
  <title><p>scifio-bf-compat</p></title>
  <url>https://github.com/scifio/scifio-bf-compat</url>
</bibl>

<bibl id="B65">
  <title><p>SCIFIO OME-XML support</p></title>
  <url>https://github.com/scifio/scifio-ome-xml</url>
</bibl>

<bibl id="B66">
  <title><p>Eclipse</p></title>
  <url>https://eclipse.org/</url>
</bibl>

<bibl id="B67">
  <title><p>ImageJ Server</p></title>
  <url>https://github.com/imagej/imagej-server</url>
</bibl>

<bibl id="B68">
  <title><p>ImageJ Legacy</p></title>
  <url>https://github.com/imagej/imagej-legacy</url>
</bibl>

<bibl id="B69">
  <title><p>ImageJ 1.x Patcher</p></title>
  <url>https://github.com/imagej/ij1-patcher</url>
</bibl>

<bibl id="B70">
  <title><p>SimpleITK</p></title>
  <url>https://simpleitk.org/</url>
</bibl>

<bibl id="B71">
  <title><p>JavaScript</p></title>
  <url>https://developer.mozilla.org/en-US/docs/Web/JavaScript</url>
</bibl>

<bibl id="B72">
  <title><p>SciJava Scripting: JavaScript</p></title>
  <url>https://github.com/scijava/scripting-javascript</url>
</bibl>

<bibl id="B73">
  <title><p>Project Nashorn</p></title>
  <url>http://openjdk.java.net/projects/nashorn/</url>
</bibl>

<bibl id="B74">
  <title><p>Rhino JavaScript implementation</p></title>
  <url>https://developer.mozilla.org/en-US/docs/Mozilla/Projects/Rhino</url>
</bibl>

<bibl id="B75">
  <title><p>Project Jupyter</p></title>
  <url>https://jupyter.org/</url>
</bibl>

<bibl id="B76">
  <title><p>SciJava Jupyter Kernel</p></title>
  <url>https://github.com/scijava/scijava-jupyter-kernel</url>
</bibl>

<bibl id="B77">
  <title><p>Beaker Extensions for Jupyter Notebook</p></title>
  <url>https://github.com/twosigma/beakerx</url>
</bibl>

<bibl id="B78">
  <title><p>Kotlin</p></title>
  <url>https://kotlinlang.org/</url>
</bibl>

<bibl id="B79">
  <title><p>SciJava Scripting: Kotlin</p></title>
  <url>https://github.com/scijava/scripting-kotlin</url>
</bibl>

<bibl id="B80">
  <title><p>Lisp (programming language)</p></title>
  <url>https://en.wikipedia.org/wiki/Lisp\_(programming\_language)</url>
</bibl>

<bibl id="B81">
  <title><p>SciJava Scripting: Clojure</p></title>
  <url>https://github.com/scijava/scripting-clojure</url>
</bibl>

<bibl id="B82">
  <title><p>The Clojure Programming Language</p></title>
  <url>https://clojure.org/</url>
</bibl>

<bibl id="B83">
  <title><p>MATLAB: The Language of Technical Computing</p></title>
  <url>https://www.mathworks.com/products/matlab.html</url>
</bibl>

<bibl id="B84">
  <title><p>SciJava Scripting: MATLAB</p></title>
  <url>https://github.com/scijava/scripting-matlab</url>
</bibl>

<bibl id="B85">
  <title><p>matlabcontrol</p></title>
  <url>https://code.google.com/archive/p/matlabcontrol/</url>
</bibl>

<bibl id="B86">
  <title><p>MiToBo-A Toolbox for Image Processing and Analysis</p></title>
  <aug>
    <au><snm>M{\"o}ller</snm><fnm>B</fnm></au>
    <au><snm>Gla{\ss}</snm><fnm>M</fnm></au>
    <au><snm>Misiak</snm><fnm>D</fnm></au>
    <au><snm>Posch</snm><fnm>S</fnm></au>
  </aug>
  <source>Journal of Open Research Software</source>
  <publisher>Ubiquity Press</publisher>
  <pubdate>2016</pubdate>
  <volume>4</volume>
  <issue>1</issue>
</bibl>

<bibl id="B87">
  <title><p>Alida</p></title>
  <url>http://www.informatik.uni-halle.de/alida/</url>
</bibl>

<bibl id="B88">
  <title><p>OpenCV: Open Source Computer Vision</p></title>
  <url>http://opencv.org/</url>
</bibl>

<bibl id="B89">
  <title><p>IJ-OpenCV: Combining ImageJ and OpenCV for processing images in
  biomedicine</p></title>
  <aug>
    <au><snm>Domínguez</snm><fnm>C</fnm></au>
    <au><snm>Heras</snm><fnm>J</fnm></au>
    <au><snm>Pascual</snm><fnm>V</fnm></au>
  </aug>
  <source>Computers in Biology and Medicine</source>
  <publisher>Pergamon Press, Inc.</publisher>
  <pubdate>2017</pubdate>
  <volume>84</volume>
  <issue>C</issue>
  <fpage>189</fpage>
  <lpage>-194</lpage>
</bibl>

<bibl id="B90">
  <title><p>JavaCV: Java interface to OpenCV and more</p></title>
  <url>https://github.com/bytedeco/javacv</url>
</bibl>

<bibl id="B91">
  <title><p>Python</p></title>
  <url>https://python.org/</url>
</bibl>

<bibl id="B92">
  <title><p>imglib2-imglyb</p></title>
  <url>https://github.com/hanslovsky/imglib2-imglyb</url>
</bibl>

<bibl id="B93">
  <title><p>PyJNIus: Access Java classes from Python</p></title>
  <url>https://github.com/kivy/pyjnius</url>
</bibl>

<bibl id="B94">
  <title><p>Jython: Python for the Java Platform</p></title>
  <url>http://jython.org/</url>
</bibl>

<bibl id="B95">
  <title><p>JyNI -- Jython Native Interface</p></title>
  <url>https://jyni.org/</url>
</bibl>

<bibl id="B96">
  <title><p>imagey: ImageJ with CPython REPL</p></title>
  <url>https://github.com/hanslovsky/imagey</url>
</bibl>

<bibl id="B97">
  <title><p>SciJava Scripting: CPython</p></title>
  <url>https://github.com/scijava/scripting-cpython</url>
</bibl>

<bibl id="B98">
  <title><p>python-javabridge: Python wrapper for the Java Native
  Interface</p></title>
  <url>https://github.com/LeeKamentsky/python-javabridge</url>
</bibl>

<bibl id="B99">
  <title><p>SciJava Scripting: Jython</p></title>
  <url>https://github.com/scijava/scripting-jython</url>
</bibl>

<bibl id="B100">
  <title><p>The R Project for Statistical Computing</p></title>
  <url>https://r-project.org/</url>
</bibl>

<bibl id="B101">
  <title><p>SciJava Scripting: Renjin</p></title>
  <url>https://github.com/scijava/scripting-renjin</url>
</bibl>

<bibl id="B102">
  <title><p>Renjin</p></title>
  <url>http://renjin.org/</url>
</bibl>

<bibl id="B103">
  <title><p>Representational state transfer</p></title>
  <url>https://en.wikipedia.org/wiki/Representational\_state\_transfer</url>
</bibl>

<bibl id="B104">
  <title><p>Dropwizard</p></title>
  <url>http://dropwizard.io/</url>
</bibl>

<bibl id="B105">
  <title><p>Ruby Programming Language</p></title>
  <url>https://www.ruby-lang.org/</url>
</bibl>

<bibl id="B106">
  <title><p>SciJava Scripting: JRuby</p></title>
  <url>https://github.com/scijava/scripting-jruby</url>
</bibl>

<bibl id="B107">
  <title><p>The Scala Programming Language</p></title>
  <url>https://scala-lang.org/</url>
</bibl>

<bibl id="B108">
  <title><p>SciJava Scripting: Scala</p></title>
  <url>https://github.com/scijava/scripting-scala</url>
</bibl>

<bibl id="B109">
  <title><p>TensorFlow: An open-source software library for Machine
  Intelligence</p></title>
  <url>https://www.tensorflow.org/</url>
</bibl>

<bibl id="B110">
  <title><p>ImageJ-TensorFlow</p></title>
  <url>https://github.com/imagej/imagej-tensorflow</url>
</bibl>

<bibl id="B111">
  <title><p>The GitHub Effect: Forking Your Way to Better Code (FOWA Vegas
  2011)</p></title>
  <aug>
    <au><snm>Preston Werner</snm><fnm>T</fnm></au>
  </aug>
  <pubdate>2011</pubdate>
  <url>http://lanyrd.com/2011/fowa-vegas/sfxcw/</url>
</bibl>

<bibl id="B112">
  <title><p>ImageJ source code</p></title>
  <url>https://imagej.net/Source\_code</url>
</bibl>

<bibl id="B113">
  <title><p>ImageJ Licensing</p></title>
  <url>https://imagej.net/Licensing</url>
</bibl>

<bibl id="B114">
  <title><p>ImageJ Javadocs</p></title>
  <url>https://javadoc.imagej.net/</url>
</bibl>

<bibl id="B115">
  <title><p>ImageJ Tutorials</p></title>
  <url>https://imagej.net/Tutorials</url>
</bibl>

<bibl id="B116">
  <title><p>ImageJ Issue Management</p></title>
  <url>https://imagej.net/Issues</url>
</bibl>

<bibl id="B117">
  <title><p>Contributing to ImageJ</p></title>
  <url>https://imagej.net/Contributing</url>
</bibl>

<bibl id="B118">
  <title><p>Travis CI</p></title>
  <url>https://travis-ci.org/</url>
</bibl>

<bibl id="B119">
  <title><p>Uber-JARs</p></title>
  <url>https://imagej.net/Uber-JAR</url>
</bibl>

<bibl id="B120">
  <title><p>Personal Update Sites</p></title>
  <url>https://sites.imagej.net/</url>
</bibl>

<bibl id="B121">
  <title><p>The fallacy of premature optimization</p></title>
  <aug>
    <au><snm>Hyde</snm><fnm>R</fnm></au>
  </aug>
  <source>Ubiquity</source>
  <publisher>ACM</publisher>
  <pubdate>2009</pubdate>
  <volume>2009</volume>
  <issue>February</issue>
  <fpage>1</fpage>
</bibl>

<bibl id="B122">
  <title><p>ImgLib2 Benchmarks</p></title>
  <url>https://imagej.net/ImgLib2\_Benchmarks</url>
</bibl>

<bibl id="B123">
  <title><p>BigDataViewer: visualization and processing for large image data
  sets</p></title>
  <aug>
    <au><snm>Pietzsch</snm><fnm>T</fnm></au>
    <au><snm>Saalfeld</snm><fnm>S</fnm></au>
    <au><snm>Preibisch</snm><fnm>S</fnm></au>
    <au><snm>Tomancak</snm><fnm>P</fnm></au>
  </aug>
  <source>Nature methods</source>
  <publisher>Nature Publishing Group</publisher>
  <pubdate>2015</pubdate>
  <volume>12</volume>
  <issue>6</issue>
  <fpage>481</fpage>
  <lpage>-483</lpage>
</bibl>

<bibl id="B124">
  <title><p>TrackMate: An open and extensible platform for single-particle
  tracking</p></title>
  <aug>
    <au><snm>Tinevez</snm><fnm>JY</fnm></au>
    <au><snm>Perry</snm><fnm>N</fnm></au>
    <au><snm>Schindelin</snm><fnm>J</fnm></au>
    <au><snm>Hoopes</snm><fnm>GM</fnm></au>
    <au><snm>Reynolds</snm><fnm>GD</fnm></au>
    <au><snm>Laplantine</snm><fnm>E</fnm></au>
    <au><snm>Bednarek</snm><fnm>SY</fnm></au>
    <au><snm>Shorte</snm><fnm>SL</fnm></au>
    <au><snm>Eliceiri</snm><fnm>KW</fnm></au>
  </aug>
  <source>Methods</source>
  <publisher>Elsevier</publisher>
  <pubdate>2017</pubdate>
  <volume>115</volume>
  <fpage>80</fpage>
  <lpage>-90</lpage>
</bibl>

<bibl id="B125">
  <title><p>In vivo imaging of DNA double-strand break induced telomere
  mobility during alternative lengthening of telomeres</p></title>
  <aug>
    <au><snm>Cho</snm><fnm>NW</fnm></au>
    <au><snm>Lampson</snm><fnm>MA</fnm></au>
    <au><snm>Greenberg</snm><fnm>RA</fnm></au>
  </aug>
  <source>Methods</source>
  <publisher>Elsevier</publisher>
  <pubdate>2017</pubdate>
  <volume>114</volume>
  <fpage>54</fpage>
  <lpage>-59</lpage>
</bibl>

<bibl id="B126">
  <title><p>Reconstruction of cell lineages and behaviors underlying arthropod
  limb outgrowth with multi-view light-sheet imaging and tracking</p></title>
  <aug>
    <au><snm>Wolff</snm><fnm>C</fnm></au>
    <au><snm>Tinevez</snm><fnm>JY</fnm></au>
    <au><snm>Pietzsch</snm><fnm>T</fnm></au>
    <au><snm>Stamataki</snm><fnm>E</fnm></au>
    <au><snm>Harich</snm><fnm>B</fnm></au>
    <au><snm>Preibisch</snm><fnm>S</fnm></au>
    <au><snm>Shorte</snm><fnm>S</fnm></au>
    <au><snm>Keller</snm><fnm>PJ</fnm></au>
    <au><snm>Tomancak</snm><fnm>P</fnm></au>
    <au><snm>Pavlopoulos</snm><fnm>A</fnm></au>
  </aug>
  <source>bioRxiv</source>
  <publisher>Cold Spring Harbor Labs Journals</publisher>
  <pubdate>2017</pubdate>
  <fpage>112623</fpage>
</bibl>

<bibl id="B127">
  <title><p>Software for bead-based registration of selective plane
  illumination microscopy data</p></title>
  <aug>
    <au><snm>Preibisch</snm><fnm>S</fnm></au>
    <au><snm>Saalfeld</snm><fnm>S</fnm></au>
    <au><snm>Schindelin</snm><fnm>J</fnm></au>
    <au><snm>Tomancak</snm><fnm>P</fnm></au>
  </aug>
  <source>Nature methods</source>
  <publisher>Nature Publishing Group</publisher>
  <pubdate>2010</pubdate>
  <volume>7</volume>
  <issue>6</issue>
  <fpage>418</fpage>
  <lpage>-419</lpage>
</bibl>

<bibl id="B128">
  <title><p>Efficient Bayesian-based multiview deconvolution</p></title>
  <aug>
    <au><snm>Preibisch</snm><fnm>S</fnm></au>
    <au><snm>Amat</snm><fnm>F</fnm></au>
    <au><snm>Stamataki</snm><fnm>E</fnm></au>
    <au><snm>Sarov</snm><fnm>M</fnm></au>
    <au><snm>Singer</snm><fnm>RH</fnm></au>
    <au><snm>Myers</snm><fnm>E</fnm></au>
    <au><snm>Tomancak</snm><fnm>P</fnm></au>
  </aug>
  <source>nature methods</source>
  <publisher>Nature Research</publisher>
  <pubdate>2014</pubdate>
  <volume>11</volume>
  <issue>6</issue>
  <fpage>645</fpage>
  <lpage>-648</lpage>
</bibl>

<bibl id="B129">
  <title><p>Using light sheet fluorescence microscopy to image zebrafish eye
  development</p></title>
  <aug>
    <au><snm>Icha</snm><fnm>J</fnm></au>
    <au><snm>Schmied</snm><fnm>C</fnm></au>
    <au><snm>Sidhaye</snm><fnm>J</fnm></au>
    <au><snm>Tomancak</snm><fnm>P</fnm></au>
    <au><snm>Preibisch</snm><fnm>S</fnm></au>
    <au><snm>Norden</snm><fnm>C</fnm></au>
  </aug>
  <source>Journal of visualized experiments: JoVE</source>
  <publisher>MyJoVE Corporation</publisher>
  <pubdate>2016</pubdate>
  <issue>110</issue>
</bibl>

<bibl id="B130">
  <title><p>Directional cerebrospinal fluid movement between brain ventricles
  in larval zebrafish</p></title>
  <aug>
    <au><snm>Fame</snm><fnm>RM</fnm></au>
    <au><snm>Chang</snm><fnm>JT</fnm></au>
    <au><snm>Hong</snm><fnm>A</fnm></au>
    <au><snm>Aponte Santiago</snm><fnm>NA</fnm></au>
    <au><snm>Sive</snm><fnm>H</fnm></au>
  </aug>
  <source>Fluids and Barriers of the CNS</source>
  <publisher>BioMed Central</publisher>
  <pubdate>2016</pubdate>
  <volume>13</volume>
  <issue>1</issue>
  <fpage>11</fpage>
</bibl>

<bibl id="B131">
  <title><p>Neuronal morphometry directly from bitmap images</p></title>
  <aug>
    <au><snm>Ferreira</snm><fnm>TA</fnm></au>
    <au><snm>Blackman</snm><fnm>AV</fnm></au>
    <au><snm>Oyrer</snm><fnm>J</fnm></au>
    <au><snm>Jayabal</snm><fnm>S</fnm></au>
    <au><snm>Chung</snm><fnm>AJ</fnm></au>
    <au><snm>Watt</snm><fnm>AJ</fnm></au>
    <au><snm>Sj{\"o}str{\"o}m</snm><fnm>PJ</fnm></au>
    <au><snm>Van Meyel</snm><fnm>DJ</fnm></au>
  </aug>
  <source>Nature methods</source>
  <publisher>Nature Research</publisher>
  <pubdate>2014</pubdate>
  <volume>11</volume>
  <issue>10</issue>
  <fpage>982</fpage>
  <lpage>-984</lpage>
</bibl>

<bibl id="B132">
  <title><p>Simple Neurite Tracer: open source software for reconstruction,
  visualization and analysis of neuronal processes</p></title>
  <aug>
    <au><snm>Longair</snm><fnm>MH</fnm></au>
    <au><snm>Baker</snm><fnm>DA</fnm></au>
    <au><snm>Armstrong</snm><fnm>JD</fnm></au>
  </aug>
  <source>Bioinformatics</source>
  <publisher>Oxford University Press</publisher>
  <pubdate>2011</pubdate>
  <volume>27</volume>
  <issue>17</issue>
  <fpage>2453</fpage>
  <lpage>-2454</lpage>
</bibl>

<bibl id="B133">
  <title><p>Adult microbiota-deficient mice have distinct dendritic
  morphological changes: differential effects in the amygdala and
  hippocampus</p></title>
  <aug>
    <au><snm>Luczynski</snm><fnm>P</fnm></au>
    <au><snm>Whelan</snm><fnm>SO</fnm></au>
    <au><snm>O'sullivan</snm><fnm>C</fnm></au>
    <au><snm>Clarke</snm><fnm>G</fnm></au>
    <au><snm>Shanahan</snm><fnm>F</fnm></au>
    <au><snm>Dinan</snm><fnm>TG</fnm></au>
    <au><snm>Cryan</snm><fnm>JF</fnm></au>
  </aug>
  <source>European Journal of Neuroscience</source>
  <publisher>Wiley Online Library</publisher>
  <pubdate>2016</pubdate>
  <volume>44</volume>
  <issue>9</issue>
  <fpage>2654</fpage>
  <lpage>-2666</lpage>
</bibl>

<bibl id="B134">
  <title><p>HiPSC-derived retinal ganglion cells grow dendritic arbors and
  functional axons on a tissue-engineered scaffold</p></title>
  <aug>
    <au><snm>Li</snm><fnm>K</fnm></au>
    <au><snm>Zhong</snm><fnm>X</fnm></au>
    <au><snm>Yang</snm><fnm>S</fnm></au>
    <au><snm>Luo</snm><fnm>Z</fnm></au>
    <au><snm>Li</snm><fnm>K</fnm></au>
    <au><snm>Liu</snm><fnm>Y</fnm></au>
    <au><snm>Cai</snm><fnm>S</fnm></au>
    <au><snm>Gu</snm><fnm>H</fnm></au>
    <au><snm>Lu</snm><fnm>S</fnm></au>
    <au><snm>Zhang</snm><fnm>H</fnm></au>
    <au><cnm>others</cnm></au>
  </aug>
  <source>Acta Biomaterialia</source>
  <publisher>Elsevier</publisher>
  <pubdate>2017</pubdate>
  <volume>54</volume>
  <fpage>117</fpage>
  <lpage>-127</lpage>
</bibl>

<bibl id="B135">
  <title><p>The Rac-GAP alpha2-chimaerin regulates hippocampal dendrite and
  spine morphogenesis</p></title>
  <aug>
    <au><snm>Valdez</snm><fnm>CM</fnm></au>
    <au><snm>Murphy</snm><fnm>GG</fnm></au>
    <au><snm>Beg</snm><fnm>AA</fnm></au>
  </aug>
  <source>Molecular and Cellular Neuroscience</source>
  <publisher>Elsevier</publisher>
  <pubdate>2016</pubdate>
  <volume>75</volume>
  <fpage>14</fpage>
  <lpage>-26</lpage>
</bibl>

<bibl id="B136">
  <title><p>The Cathedral &amp the Bazaar: Musings on linux and open source by
  an accidental revolutionary</p></title>
  <aug>
    <au><snm>Raymond</snm><fnm>ES</fnm></au>
  </aug>
  <publisher>" O'Reilly Media, Inc."</publisher>
  <pubdate>2001</pubdate>
</bibl>

<bibl id="B137">
  <title><p>ImageJ communication channels</p></title>
  <url>https://imagej.net/Communication</url>
</bibl>

<bibl id="B138">
  <title><p>ImageJ Forum</p></title>
  <url>http://forum.imagej.net/</url>
</bibl>

<bibl id="B139">
  <title><p>Discourse</p></title>
  <url>https://www.discourse.org/</url>
</bibl>

<bibl id="B140">
  <title><p>SciJava team roles</p></title>
  <url>https://imagej.net/Team</url>
</bibl>

<bibl id="B141">
  <title><p>ImageJ Roadmap</p></title>
  <url>https://imagej.net/Roadmap</url>
</bibl>

<bibl id="B142">
  <title><p>Deep Learning</p></title>
  <url>http://deeplearning.net/</url>
</bibl>

<bibl id="B143">
  <title><p>scikit-image: Image Processing in Python</p></title>
  <url>http://scikit-image.org/</url>
</bibl>

<bibl id="B144">
  <title><p>Amazon Web Services</p></title>
  <url>https://aws.amazon.com/</url>
</bibl>

<bibl id="B145">
  <title><p>ImageJ User Guide</p></title>
  <aug>
    <au><snm>Ferreira</snm><fnm>T</fnm></au>
  </aug>
  <url>https://imagej.net/docs/guide/</url>
</bibl>

<bibl id="B146">
  <title><p>ImageJ 1.x documentation</p></title>
  <url>https://imagej.net/index.html</url>
</bibl>

<bibl id="B147">
  <title><p>ImageJ Information and Documentation Portal</p></title>
  <url>http://imagejdocu.tudor.lu/</url>
</bibl>

<bibl id="B148">
  <title><p>ImageJ Search</p></title>
  <url>https://search.imagej.net/</url>
</bibl>

<bibl id="B149">
  <title><p>ImageJ Funding</p></title>
  <url>https://imagej.net/Funding</url>
</bibl>

<bibl id="B150">
  <title><p>MaMuT</p></title>
  <url>https://imagej.net/MaMuT</url>
</bibl>

<bibl id="B151">
  <title><p>Optimal joint segmentation and tracking of Escherichia coli in the
  mother machine</p></title>
  <aug>
    <au><snm>Jug</snm><fnm>F</fnm></au>
    <au><snm>Pietzsch</snm><fnm>T</fnm></au>
    <au><snm>Kainm{\"u}ller</snm><fnm>D</fnm></au>
    <au><snm>Funke</snm><fnm>J</fnm></au>
    <au><snm>Kaiser</snm><fnm>M</fnm></au>
    <au><snm>Nimwegen</snm><fnm>E</fnm></au>
    <au><snm>Rother</snm><fnm>C</fnm></au>
    <au><snm>Myers</snm><fnm>G</fnm></au>
  </aug>
  <source>Bayesian and grAphical Models for Biomedical Imaging</source>
  <publisher>Springer</publisher>
  <pubdate>2014</pubdate>
  <fpage>25</fpage>
  <lpage>-36</lpage>
</bibl>

<bibl id="B152">
  <title><p>Tracking by assignment facilitates data curation</p></title>
  <aug>
    <au><snm>Jug</snm><fnm>F</fnm></au>
    <au><snm>Pietzsch</snm><fnm>T</fnm></au>
    <au><snm>Kainm{\"u}ller</snm><fnm>D</fnm></au>
    <au><snm>Myers</snm><fnm>G</fnm></au>
  </aug>
  <source>MICCAI IMIC Workshop</source>
  <pubdate>2014</pubdate>
  <volume>2</volume>
</bibl>

<bibl id="B153">
  <title><p>KymographBuilder: Release 1.2.4</p></title>
  <aug>
    <au><snm>Mary</snm><fnm>H</fnm></au>
    <au><snm>Rueden</snm><fnm>C</fnm></au>
    <au><snm>Ferreira</snm><fnm>T</fnm></au>
  </aug>
  <pubdate>2016</pubdate>
  <url>https://doi.org/10.5281/zenodo.56702</url>
</bibl>

<bibl id="B154">
  <title><p>Post-acquisition image based compensation for thickness variation
  in microscopy section series</p></title>
  <aug>
    <au><snm>Hanslovsky</snm><fnm>P</fnm></au>
    <au><snm>Bogovic</snm><fnm>JA</fnm></au>
    <au><snm>Saalfeld</snm><fnm>S</fnm></au>
  </aug>
  <source>Biomedical Imaging (ISBI), 2015 IEEE 12th International Symposium
  on</source>
  <pubdate>2015</pubdate>
  <fpage>507</fpage>
  <lpage>-511</lpage>
</bibl>

<bibl id="B155">
  <title><p>Trainable Weka Segmentation: a machine learning tool for microscopy
  pixel classification.</p></title>
  <aug>
    <au><snm>Arganda Carreras</snm><fnm>I</fnm></au>
    <au><snm>Kaynig</snm><fnm>V</fnm></au>
    <au><snm>Rueden</snm><fnm>C</fnm></au>
    <au><snm>Eliceiri</snm><fnm>KW</fnm></au>
    <au><snm>Schindelin</snm><fnm>J</fnm></au>
    <au><snm>Cardona</snm><fnm>A</fnm></au>
    <au><snm>Seung</snm><fnm>HS</fnm></au>
  </aug>
  <source>Bioinformatics (Oxford, England)</source>
  <pubdate>2017</pubdate>
</bibl>

<bibl id="B156">
  <title><p>Pendent\_Drop: an ImageJ plugin to measure the surface tension from
  an image of a pendent drop</p></title>
  <aug>
    <au><snm>Daerr</snm><fnm>A</fnm></au>
    <au><snm>Mogne</snm><fnm>A</fnm></au>
  </aug>
  <source>Journal of Open Research Software</source>
  <publisher>Ubiquity Press</publisher>
  <pubdate>2016</pubdate>
  <volume>4</volume>
  <issue>1</issue>
</bibl>

<bibl id="B157">
  <title><p>SciView</p></title>
  <url>https://github.com/scenerygraphics/SciView</url>
</bibl>

<bibl id="B158">
  <title><p>Globally optimal stitching of tiled 3D microscopic image
  acquisitions</p></title>
  <aug>
    <au><snm>Preibisch</snm><fnm>S</fnm></au>
    <au><snm>Saalfeld</snm><fnm>S</fnm></au>
    <au><snm>Tomancak</snm><fnm>P</fnm></au>
  </aug>
  <source>Bioinformatics</source>
  <publisher>Oxford University Press</publisher>
  <pubdate>2009</pubdate>
  <volume>25</volume>
  <issue>11</issue>
  <fpage>1463</fpage>
  <lpage>-1465</lpage>
</bibl>

<bibl id="B159">
  <title><p>Coloc 2</p></title>
  <url>https://imagej.net/Coloc\_2</url>
</bibl>

</refgrp>
} 


\section*{Supplemental material}

\subsection*{Figures and illustrations}
  \renewcommand\thefigure{S.\arabic{figure}}
  \setcounter{figure}{0}

  \begin{figure}[h!]
    \caption{Module execution in different contexts.}
    \includegraphics[width=4.75in]{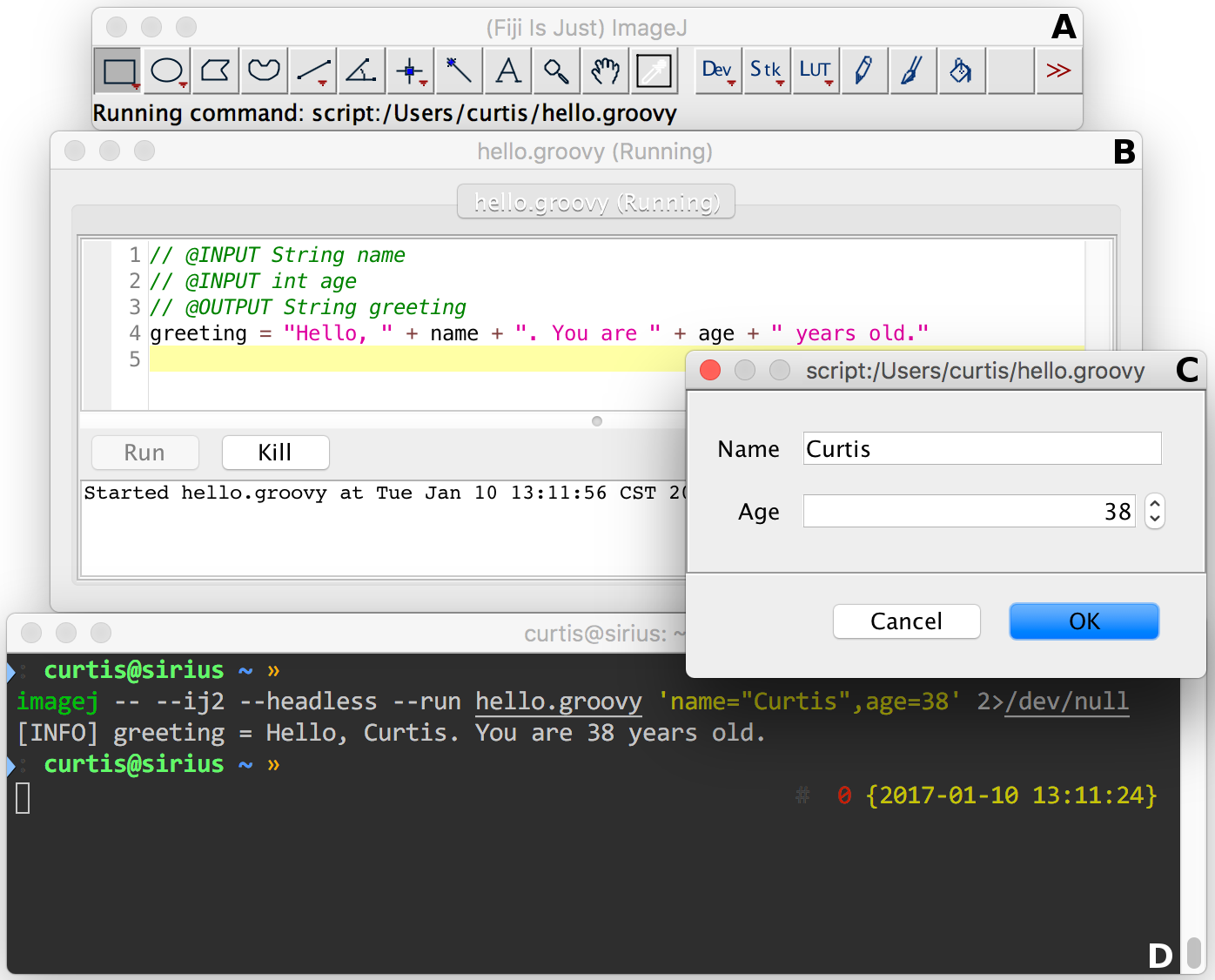}
    \begin{flushleft}
      When running a parameterized script (panel B) from the ImageJ user
      interface (panel A), a pop-up dialog box (panel C) enables the user to
      enter the name and age values; when running the script headless from the
      command line (panel D), input values are passed as arguments and output
      values echoed to the standard output stream.
    \end{flushleft}
  \end{figure}

  \begin{figure}[h!]
    \caption{Comparison of time performance across ImageJ 1.x and ImgLib2 data
    structures.}
    \includegraphics[width=4.75in]{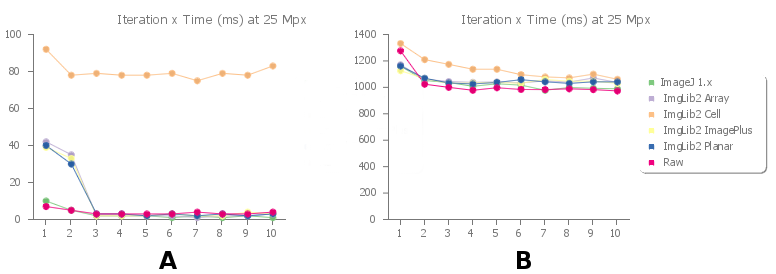}
    \begin{flushleft}
      For ten iterations, we ran a ``cheap'' per-pixel operation and an
      ``expensive'' operation on a 25 Mpx image stored in the ImageJ 1.x
      container, various ImgLib2 containers, and raw byte arrays. Panel A
      (left) shows the time (ms) it took to complete a ``cheap'' operations
      versus the loop iteration for each container. Panel B (right) shows the
      same information but for the time (ms) it took to complete the expensive
      operation.
    \end{flushleft}
  \end{figure}

  \begin{figure}[h!]
    \caption{Sample ImageJ plugin usage of ImageJ 1.x and ImageJ2.}
    \includegraphics[width=4.75in]{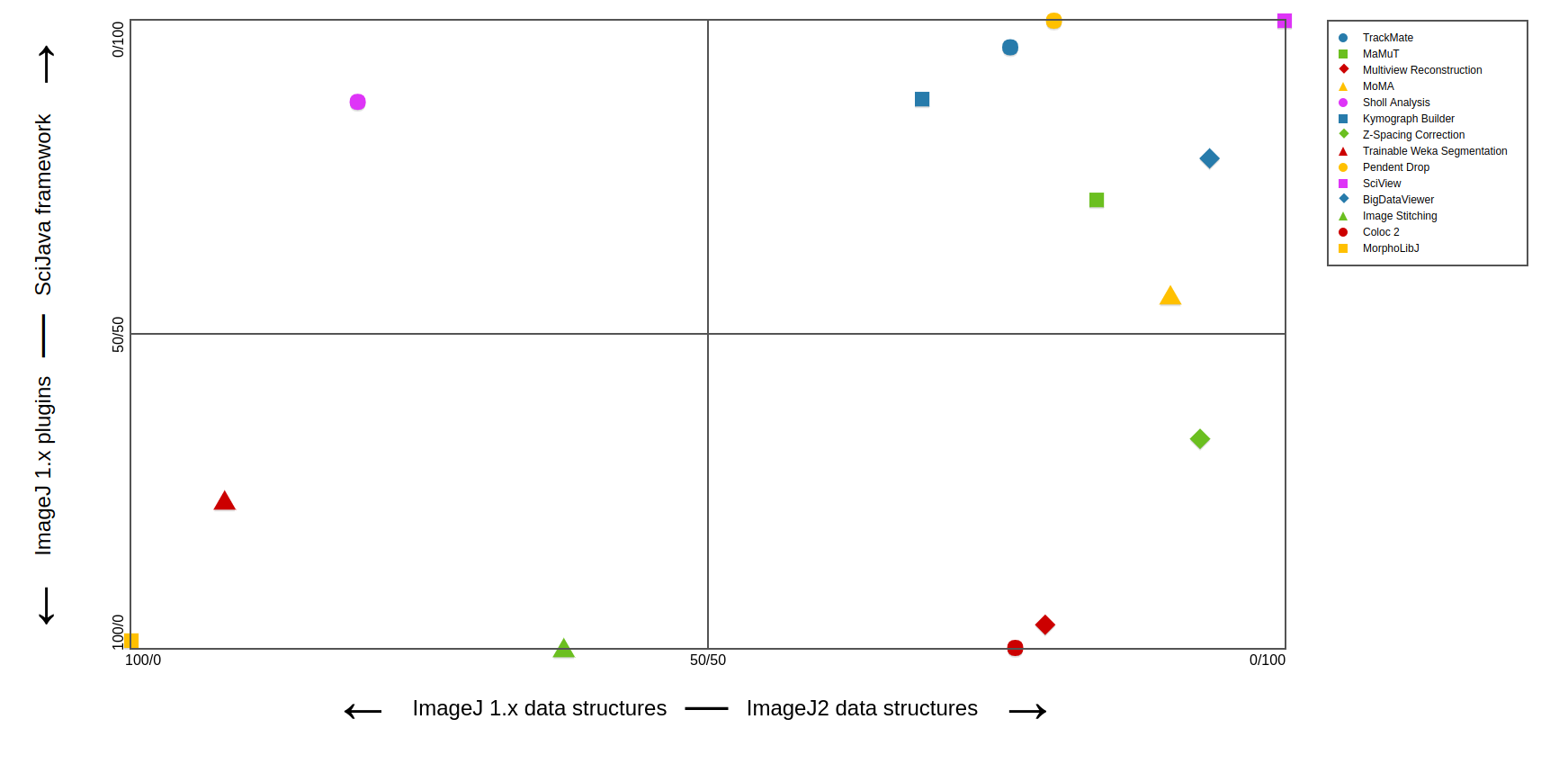}
    \begin{flushleft}
      This plot displays a select few ImageJ plugins in varying stages of
      transition, from ImageJ 1.x to ImageJ2, as of 11 Aug 2017 2:35 PM CDT.
      The ratio of ImageJ 1.x to ImageJ2 usage was computed by counting the
      number of imports each plugin uses from relevant Java packages:
      "ImageJ 1.x plugins" is \texttt{ij.plugin.*}, "ImageJ 1.x data
      structures" is \texttt{ij.*} excluding the \texttt{plugin} subpackage,
      "SciJava framework" is \texttt{org.scijava.*}, and "ImageJ2 data
      structures" is \texttt{net.imagej.*} and \texttt{net.imglib2.*}.
      References for plugins shown:
      TrackMate                    \cite{trackmate},
      \acrshort{mamut}             \cite{imagej_mamut},
      Multiview Reconstruction     \cite{multiview_2010, multiview_2014},
      \acrfull{moma}               \cite{moma_seg, moma_tracking},
      Sholl Analysis               \cite{sholl_analysis},
      Kymograph Builder            \cite{kymograph},
      Z-Spacing Correction         \cite{z_spacing},
      Trainable Weka Segmentation  \cite{trainable_weka},
      Pendent Drop                 \cite{pendent_drop},
      SciView                      \cite{sciview},
      BigDataViewer                \cite{bigdataviewer},
      Image Stitching              \cite{image_stitching},
      Coloc 2                      \cite{imagej_coloc_2},
      MorphoLibJ                   \cite{morpholibj}.
    \end{flushleft}
  \end{figure}

\FloatBarrier

\subsection*{Tables and captions}
  \renewcommand\thetable{S.\arabic{table}}
  \setcounter{table}{0}

  \begin{table}[h!]
    \caption{Built-in SciJava input widgets}
    \begin{tabular}{| l | l |}
      \hline
      \textbf{Java data type}                                                                                                            & \textbf{Widget type}                   \\ \hline
      \texttt{boolean} $\vert$ \texttt{Boolean}                                                                                          & checkbox                               \\ \hline
      \texttt{byte} $\vert$ \texttt{short} $\vert$ \texttt{int}     $\vert$ \texttt{long} $\vert$ \texttt{float} $\vert$ \texttt{double} & numeric field                          \\ \hline
      \texttt{Byte} $\vert$ \texttt{Short} $\vert$ \texttt{Integer} $\vert$ \texttt{Long} $\vert$ \texttt{Float} $\vert$ \texttt{Double} & numeric field                          \\ \hline
      \texttt{BigInteger} $\vert$ \texttt{BigDecimal}                                                                                    & numeric field                          \\ \hline
      \texttt{char} $\vert$ \texttt{Character} $\vert$ \texttt{String}                                                                   & text field                             \\ \hline
      \texttt{Dataset} $\vert$ \texttt{ImagePlus}                                                                                        & (\textgreater=2 images) drop-down list \\ \hline
      \texttt{ColorRGB}                                                                                                                  & color chooser                          \\ \hline
      \texttt{Date}                                                                                                                      & date chooser                           \\ \hline
      \texttt{File}                                                                                                                      & file chooser                           \\ \hline
    \end{tabular}
  \end{table}

  \begin{table}[h!]
    \caption{Kinds and arities of special ops}
    \renewcommand{\arraystretch}{1.5}
    \begin{tabular}{| l | l | c | l | p{1.4in} |}
      \hline
      \textbf{Kind}              & \textbf{Stipulations} & \textbf{Arity} & \textbf{Interface}         & \textbf{Methods}                    \\ \hline
      \multirow{3}{*}{computer}  & \multirow{3}{1.8in}{
                                   \begin{itemize}[leftmargin=*]
                                     \renewcommand{\labelitemi}{-}
                                     \item Mutating the inputs is forbidden.\\
                                     \item The output and input references must differ (i.e., computers do not work in-place).\\
                                     \item The output's initial contents must not affect the value of the result.
                                   \end{itemize}
                                   }                     & 0              & \texttt{NullaryComputerOp} & \texttt{void compute(O)}            \\ \cline{3-5}
                                 &                       & 1              & \texttt{UnaryComputerOp}   & \texttt{void compute(I, O)}         \\ \cline{3-5}
                                 &                       & 2              & \texttt{BinaryComputerOp}  & \texttt{void compute(I1, I2, O)}    \\[0.45in] \hline
      \multirow{3}{*}{function}  & \multirow{3}{1.8in}{
                                   Mutating the inputs is forbidden.
                                   }                     & 0              & \texttt{NullaryFunctionOp} & \texttt{O calculate()}              \\ \cline{3-5}
                                 &                       & 1              & \texttt{UnaryFunctionOp}   & \texttt{O calculate(I)}             \\ \cline{3-5}
                                 &                       & 2              & \texttt{BinaryFunctionOp}  & \texttt{O calculate(I1, I2)}        \\ \hline
      \multirow{3}{*}{inplace}   & \multirow{3}{1.8in}{
                                   -
                                   }                     & 1              & \texttt{UnaryInplaceOp}    & \texttt{void mutate(O)}             \\ \cline{3-5}
                                 &                       & 1              & \texttt{BinaryInplace1Op}  & \texttt{void mutate1(O, I2)}        \\ \cline{3-5}
                                 &                       & 2              & \texttt{BinaryInplaceOp}   & \parbox[t]{2in}{
                                                                                                         \texttt{void mutate1(O, I2)}\\
                                                                                                         \texttt{void mutate2(I1, O)}
                                                                                                         }                                   \\[0.15in] \hline
      \multirow{3}{*}{hybrid CF} & \multirow{3}{1.8in}{
                                   Same as \textit{computer} and \textit{function} respectively.
                                   }                     & 0              & \texttt{NullaryHybridCF}   & \parbox[t]{2in}{
                                                                                                         \texttt{void compute(O)}\\
                                                                                                         \texttt{O calculate()}
                                                                                                         }                                   \\[0.15in] \cline{3-5}
                                 &                       & 1              & \texttt{UnaryHybridCF}     & \parbox[t]{2in}{
                                                                                                         \texttt{void compute(I, O)}\\
                                                                                                         \texttt{O calculate(I)}
                                                                                                         }                                   \\[0.15in] \cline{3-5}
                                 &                       & 2              & \texttt{BinaryHybridCF}    & \parbox[t]{2in}{
                                                                                                         \texttt{void compute(I1, I2, O)}\\
                                                                                                         \texttt{O calculate(I1, I2)}
                                                                                                         }                                   \\[0.15in] \hline
      \multirow{3}{*}{hybrid CI} & \multirow{3}{1.8in}{
                                   Same as \textit{computer} and \textit{inplace} respectively.
                                   }                     & 1              & \texttt{UnaryHybridCI}     & \parbox[t]{2in}{
                                                                                                         \texttt{void compute(I, O)}\\
                                                                                                         \texttt{void mutate(O)}
                                                                                                         }                                   \\[0.15in] \cline{3-5}
                                 &                       & 2              & \texttt{BinaryHybridCI1}   & \parbox[t]{2in}{
                                                                                                         \texttt{void compute(I1, I2, O)}\\
                                                                                                         \texttt{void mutate1(O, I2)}
                                                                                                         }                                   \\[0.15in] \cline{3-5}
                                 &                       & 2              & \texttt{BinaryHybridCI}    & \parbox[t]{2in}{
                                                                                                         \texttt{void compute(I1, I2, O)}\\
                                                                                                         \texttt{void mutate1(O, I2)}\\
                                                                                                         \texttt{void mutate2(I1, O)}
                                                                                                         }                                   \\[0.25in] \hline
      \multirow{3}{*}{hybrid CFI}& \multirow{3}{1.8in}{
                                   Same as \textit{computer}, \textit{function} and \textit{inplace} respectively.
                                   }                     & 1              & \texttt{UnaryHybridCFI}    & \parbox[t]{2in}{
                                                                                                         \texttt{void compute(I, O)}\\
                                                                                                         \texttt{O calculate(I)}\\
                                                                                                         \texttt{void mutate(O)}
                                                                                                         }                                   \\[0.25in] \cline{3-5}
                                 &                       & 2              & \texttt{BinaryHybridCFI1}  & \parbox[t]{2in}{
                                                                                                         \texttt{void compute(I1, I2, O)}\\
                                                                                                         \texttt{O calculate(I1, I2)}\\
                                                                                                         \texttt{void mutate1(O, I2)}
                                                                                                         }                                   \\[0.25in] \cline{3-5}
                                 &                       & 2              & \texttt{BinaryHybridCFI}   & \parbox[t]{2in}{
                                                                                                         \texttt{void compute(I1, I2, O)}\\
                                                                                                         \texttt{O calculate(O, I1, I2)}\\
                                                                                                         \texttt{void mutate1(O, I2)}\\
                                                                                                         \texttt{void mutate2(I1, O)}
                                                                                                         }                                   \\[0.4in] \hline
    \end{tabular}
  \end{table}

  \begin{table}[h!]
    \caption{Image types supported by ImageJ}
    \begin{tabular}{| l | l | l | p{0.4in} | p{0.8in} | p{0.8in} | l | l |}
      \hline
      \textbf{Name}            & \textbf{Bit depth} & \textbf{Signedness} & \textbf{Values} & \textbf{Min. Value}     & \textbf{Max. Value}    & \textbf{ImageJ 1.x} & \textbf{ImageJ2}  \\ \hline
      \texttt{bool}            & 1-bit              & N/A                 & boolean         & $false$                 & $true$                 & no                  & yes               \\ \hline
      \texttt{bit}             & 1-bit              & unsigned            & binary          & $0$                     & $1$                    & no                  & yes               \\ \hline
      \texttt{uint2}           & 2-bit              & unsigned            & integer         & $0$                     & $3$                    & no                  & yes               \\ \hline
      \texttt{uint4}           & 4-bit              & unsigned            & integer         & $0$                     & $15$                   & no                  & yes               \\ \hline
      \texttt{uint8}           & 8-bit              & unsigned            & integer         & $0$                     & $255$                  & yes                 & yes               \\ \hline
      \texttt{uint12}          & 12-bit             & unsigned            & integer         & $0$                     & $4,095$                & no                  & yes               \\ \hline
      \texttt{uint16}          & 16-bit             & unsigned            & integer         & $0$                     & $65,535$               & yes                 & yes               \\ \hline
      \texttt{uint32}          & 32-bit             & unsigned            & integer         & $0$                     & $2^{32} - 1$           & no                  & yes               \\ \hline
      \texttt{uint64}          & 64-bit             & unsigned            & integer         & $0$                     & $2^{64} - 1$           & no                  & yes               \\ \hline
      \texttt{uint128}         & 128-bit            & unsigned            & integer         & $0$                     & $2^{128} - 1$          & no                  & yes               \\ \hline
      \texttt{int8}            & 8-bit              & signed              & integer         & $-128$                  & $127$                  & partial*            & yes               \\ \hline
      \texttt{int16}           & 16-bit             & signed              & integer         & $-32,768$               & $32,767$               & partial*            & yes               \\ \hline
      \texttt{int32}           & 32-bit             & signed              & integer         & $-2^{31}$               & $2^{31} - 1$           & no                  & yes               \\ \hline
      \texttt{float32}         & 32-bit             & signed              & floating point  & \parbox[t]{0.8in}{
                                                                                              approx.\\
                                                                                              $-3.4 \times 10^{38}$
                                                                                              }                       & \parbox[t]{0.8in}{
                                                                                                                        approx.\\
                                                                                                                        $3.4 \times 10^{38}$
                                                                                                                        }                      & yes                 & yes               \\ \hline
      \texttt{float64}         & 64-bit             & signed              & floating point  & \parbox[t]{0.8in}{
                                                                                              approx.\\
                                                                                              $-1.8 \times 10^{308}$
                                                                                              }                       & \parbox[t]{0.8in}{
                                                                                                                        approx.\\
                                                                                                                        $1.8 \times 10^{308}$
                                                                                                                        }                      & no                  & yes               \\ \hline
      \texttt{cfloat32}        & 2 $\times$ 32-bit  & signed              & floating point  & \parbox[t]{0.8in}{
                                                                                              approx.\\
                                                                                              $(-3.4 \times 10^{38},\\
                                                                                              -3.4 \times 10^{38})$
                                                                                              }                       & \parbox[t]{0.8in}{
                                                                                                                        approx.\\
                                                                                                                        $(3.4 \times 10^{38},\\
                                                                                                                        3.4 \times 10^{38})$
                                                                                                                        }                      & no                  & yes               \\ \hline
      \texttt{cfloat64}        & 2 $\times$ 64-bit  & signed              & floating point  & \parbox[t]{0.8in}{
                                                                                              approx.\\
                                                                                              $(-1.8 \times 10^{308},\\
                                                                                              -1.8 \times 10^{308})$
                                                                                              }                       & \parbox[t]{0.8in}{
                                                                                                                        approx.\\
                                                                                                                        $(1.8 \times 10^{308},\\
                                                                                                                        1.8 \times 10^{308})$
                                                                                                                        }                      & no                  & yes               \\ \hline
      \texttt{bigint}          & unlimited          & signed              & integer         & none                    & none                   & no                  & yes               \\ \hline
      \texttt{bigdec}          & arbitrary          & signed              & decimal         & none                    & none                   & no                  & yes               \\ \hline
      \acrshort{rgb}$^\dagger$ & 3 $\times$ 8-bit   & unsigned            & integer         & $(0, 0, 0)$             & $(255, 255, 255)$      & yes                 & legacy$^\ddagger$ \\ \hline
      8-bit color              & 8-bit              & indexed             & integer         & $(0, 0, 0)$             & $(255, 255, 255)$      & yes                 & legacy$^\ddagger$ \\ \hline
      custom                   & any                & any                 & any             & any                     & any                    & no                  & yes               \\ \hline
    \end{tabular}
    \begin{flushleft}
      * ImageJ 1.x partially supports \texttt{int8} and \texttt{int16} types
      via an ``image calibration'' feature.

      $^\dagger$ \acrfull{rgb}.

      $^\ddagger$ ImageJ2 supports these image types only for backwards
      compatibility with ImageJ 1.x, since composite color mode with one
      \acrshort{lut} per channel achieves the same results in a more flexible
      way.
    \end{flushleft}
  \end{table}

\end{backmatter}
\end{document}